\documentclass[aps,reprint,nofootinbib,onecolumn,notitlepage,preprintnumbers]{revtex4-1}
\pdfoutput=1

\usepackage{amsmath}
\usepackage{amssymb}
\usepackage{bbm}
\usepackage{braket}
\usepackage{graphicx}
\usepackage{slashed}
\usepackage[dvinames]{xcolor}
\usepackage{xspace} 
\usepackage{siunitx}

\DeclareMathOperator{\tr}{Tr}

\newcommand{\qq}{\ensuremath{q^{2}}\xspace}
\newcommand{\thetal}{\ensuremath{\theta_{l}}\xspace}
\newcommand{\costhetal}{\ensuremath{\cos\thetal}\xspace}
\newcommand{\dd}{\text{d}}
\newcommand{\eps}{\varepsilon}
\newcommand{\refapp}[1]{appendix~\ref{app:#1}}
\newcommand{\refeq}[1]{eq.~(\ref{eq:#1})}
\newcommand{\refeqs}[2]{eqs.~(\ref{eq:#1})~--~(\ref{eq:#2})}

\newcommand{\refsec}[1]{section~\ref{sec:#1}}

\newcommand{\LamB}{\ensuremath{\Lambda_b^{0}}\xspace}
\newcommand{\LamC}{\ensuremath{\Lambda_c^{+}}\xspace}
\newcommand{\LamCstBoth}{\ensuremath{\Lambda_c^{*+}}\xspace}
\newcommand{\LamCst}[1][]{\ensuremath{\Lambda_{c}(#1)^{+}}}
\newcommand{\GoodDecay}{\ensuremath{\LamB\to\LamCst[2625]\mu^{-}\bar{\nu}}\xspace}
\newcommand{\BadDecay}{\ensuremath{\LamB\to\LamCst[2595]\mu^{-}\bar{\nu}}\xspace}

\newcommand{\SignalDecayBoth}{\ensuremath{\LamB\to\LamCstBoth\mu^{-}\bar{\nu}}\xspace}
\newcommand{\mLamB}{m_{\Lambda_b}\xspace}
\newcommand{\mLamCst}{m_{\Lambda_c^*}}

\newcommand{\cev}[1]{\reflectbox{\ensuremath{\vec{\reflectbox{\ensuremath{#1}}}}}}
\newcommand{\mum}{\si{\micro\meter}\xspace}

\newcommand{\GeV}{\si{\giga\electronvolt}\xspace}
\newcommand{\TeV}{\si{\tera\electronvolt}\xspace}

\newcommand{\mb}[1]{\textcolor{red}{#1}}

\graphicspath {{../}{./figs/}}

\begin{document}
\title{Testing lepton flavour universality in semileptonic $\Lambda_b \to \Lambda_c^*$ decays}

\author{Philipp B\"oer}

\preprint{TUM-HEP-1118/17}
\preprint{SI-HEP-2017-24}
\preprint{QFET-2017-21}
\preprint{ZU-TH-04/18}

\email{boeer@physik.uni-siegen.de}
\affiliation{Theoretische Physik 1, Naturwissenschaftlich-Technische Fakult\"at,
Universit\"at Siegen, Walter-Flex-Stra\ss{}e 3, D-57068 Siegen, Germany}

\author{Marzia Bordone}
\email{mbordone@physik.uzh.ch}
\affiliation{Physik Institut, Universit\"at Z\"urich, Winterthurerstrasse 190, CH-8057 Z\"urich, Switzerland}

\author{Elena Graverini}
\email{elena.graverini@cern.ch}
\affiliation{Physik Institut, Universit\"at Z\"urich, Winterthurerstrasse 190, CH-8057 Z\"urich, Switzerland}

\author{Patrick Owen}
\email{powen@physik.uzh.ch}
\affiliation{Physik Institut, Universit\"at Z\"urich, Winterthurerstrasse 190, CH-8057 Z\"urich, Switzerland}


\author{Marcello Rotondo}
\email{marcello.rotondo@cern.ch}
\affiliation{Laboratori Nazionali dell'INFN di Frascati, Frascati, Italy}

\author{Danny van Dyk}
\email{danny.van.dyk@gmail.com}
\affiliation{Physik Institut, Universit\"at Z\"urich, Winterthurerstrasse 190, CH-8057 Z\"urich, Switzerland}
\affiliation{Physik Department, Technische Universit\"at M\"unchen, James-Franck-Stra\ss{}e 1, D-85748 Garching, Germany}

\begin{abstract}
    Lepton Flavour Universality tests with semileptonic $\Lambda_b \to \Lambda_c^*$ decays
    are important to corroborate the present anomalies in the similar ratios $R_{D^{(*)}}$,
    and can provide complementary constraints on possible origins of these anomalies beyond the Standard Model.
    In this paper we provide -- for the first time -- all the necessary theoretical ingredients to perform
    and interpret measurements of $R_{\Lambda_c^*}$ at the LHCb experiment. For this, we revisit
    the heavy-quark expansion of the relevant hadronic matrix elements, and provide their
    expressions to order $\alpha_s$ and $1/m$ accuracy.
    Moreover, we study the sensitivity to the form factor parameters given the projected
    size and purity of upcoming and future LHCb datasets of $\Lambda_b\to \Lambda_c^*\mu\bar\nu$ decays.
    We demonstrate explicitly the need to perform a simultaneous fit to both $\Lambda_c^*$ final states.
    Finally, we provide projections for the uncertainty of $R_{\Lambda_c^*}$ based on the form factors
    analysis from semimuonic decays and theoretical relations based on the heavy-quark expansion.
\end{abstract}

\maketitle

\section{Introduction}

Tests of lepton flavour universality in semileptonic decays of $b$ quarks are presently in focus
of both experimental as well as theoretical particle physics. This interest has been sparked
by deviations between Standard Model (SM) estimates and measurements in both 
charged-current~\cite{Lees:2012xj,Lees:2013uzd,Huschle:2015rga,Aaij:2015yra,Sato:2016svk,Hirose:2016wfn,Aaij:2017uff,LHCb-PAPER-2017-027,LHCb-PAPER-2017-035} and neutral-current~\cite{Aaij:2014ora,Aaij:2017vbb} semileptonic $b$ quark decays.
Deviations in both sectors are at the level of three to four standard deviations, which is at
present intriguing but does not yet provide conclusive evidence for particles beyond the SM. It is therefore 
important to extend the current tests to new decay modes to provide measurements with 
orthogonal experimental and theoretical systematic uncertainties as well as a complementary 
sensitivity to new physics.\\

In this paper we will concentrate on Lepton Flavour Universality (LFU) in $b\to
c\tau \bar\nu$ versus $b\to c \mu \bar\nu$ decays, in particular for \LamB decays. 
At the LHC, \LamB baryons are copiously produced, at approximately half the rate of $B^{0}$ mesons~\cite{LHCb-PAPER-2011-018,LHCb-PAPER-2014-004}. 
The decay involving the ground state charmed baryon, $\LamB \to \LamC \ell^{-} \bar{\nu}$ has been 
studied in lattice QCD in Ref.~\cite{Detmold:2015aaa} and precise predictions for the LFU ratio $R_{\Lambda_{c}}$ are provided in the SM and beyond \cite{Detmold:2015aaa,Datta:2017aue}.
In addition, the LHCb collaboration has recently measured the slope of the leading order Isgur-Wise (IW) function of the 
decay $\LamB \to \LamC \mu^{-} \bar{\nu}$~\cite{LHCb-PAPER-2017-016}. 
While studying backgrounds to this decay, large samples of \LamCst[2595] and \LamCst[2625] candidates were reconstructed as background, 
which demonstrates the potential of precise LFU tests in these decays. Therefore, we propose to
investigate the LFU ratios
\begin{equation}
    R_{\Lambda_c^*} \equiv \frac{\mathcal{B}(\Lambda_b^0 \to \Lambda_c^{*+} \tau^- \bar{\nu})}{\mathcal{B}(\Lambda_b^0 \to \Lambda_c^{*+} \mu^- \bar{\nu})}
\end{equation}
where $\Lambda_c^{*+}$ denotes either the $\LamCst[2595]$ (with $J^P =
1/2^{-}$) or the $\LamCst[2625]$ (with $J^P = 3/2^{-}$) charmed
baryon.

 The challenge in
exploiting these modes for LFU tests is controlling uncertainties
related to the hadronic matrix elements, which are genuinely non-perturbative objects.
As a consequence of both baryons forming a
doublet under Heavy Quark Spin Symmetry (HQSS), the hadronic matrix elements
for the $\Lambda_b \to \Lambda_c^*$ transitions can be expressed -- in the
infinite mass limit -- through a single IW function
$\zeta$ \cite{Falk:1991nq} at leading power in $1/m$. The power suppressed contributions at the $1/m$ level -- where
$m=m_b, m_c$ -- have been previously calculated in~\cite{Leibovich:1997az}.\\

The purpose of this paper is to provide for the first time all the
necessary ingredients to carry out a LFU study of these decays. In
\refsec{form_factors}, we first revisit the definition of the hadronic form
factors, and provide a helicity decomposition that is convenient
for the description of the decay observables. Subsequently,
we provide formulae for these hadronic form factors in the Heavy Quark
Expansion (HQE) up to order $\alpha_s$ and $1/m$, beyond what has been done in the literature so far. Continuing in \refsec{pheno},
we model the kinematic dependence of the leading and subleading
IW functions, and then provide a set of benchmark points based
on inputs from non-perturbative approaches. Afterwards, we calculate the
differential decay width, including the finite lepton-mass
contributions that are necessary for testing LFU. The following \refsec{exp_prospects} shows the impact of using LHCb data
for constraining the relevant form factor parameters, and control the theory uncertainties for the prediction of the LFU ratios.
We conclude in \refsec{conclusion}.

\section{Form factors for $\Lambda_b\to \Lambda_c^*$ transitions}
\label{sec:form_factors}

In the following we investigate form factors for the transitions
\begin{equation}
    \LamB(p, s_b) \to
    \begin{cases}
        \LamCst[2595](k, J_z \equiv s_c)                    & \text{with } J^P = 1/2^-\\
        \LamCst[2625](k, J_z \equiv s_c +\lambda_c)         & \text{with } J^P = 3/2^-
    \end{cases}\,,
\end{equation}
where $p$ and $k$ denote the four momenta of the initial and final state respectively,
and $J^P$ indicates both angular momentum and parity eigenvalues of the $\Lambda_c^{*+}$ states.
The states' rest-frame helicities are denoted as $s_b$ and $J_z$. Note that, for
the $J^P=3/2^-$ state, $J_z$ can be decomposed into the rest-frame helicity of a $1/2^+$ spinor ($s_c$),
and the polarisation of a polarisation vector $\eta \equiv \eta(\lambda_c)$.
For later use we also define the momentum transfer to the leptons $q^\mu \equiv p^\mu - k^\mu$.

\subsection{Helicity form factors}
\label{sec:helicity_form_factors}

We define the hadronic matrix elements for vector and axialvector transitions to the $\LamCst[2595]$ state
as:
\begin{equation}
\label{eq:FF-12-def}
\begin{aligned}
    \bra{\LamCst[2595](k, \eta(\lambda_c), s_c)} \bar{c}\gamma^\mu b \ket{\LamB(p, s_b)}
        & = +\bar{u}_{\alpha}^{(1/2)}(k, \eta(\lambda_c), s_c)
            \left[\sum_{i} f_i(q^2) \Gamma^{\alpha \mu}_{V,i} \right] u(p, s_b)\,,\\
    \bra{\LamCst[2595](k, \eta(\lambda_c), s_c)} \bar{c}\gamma^\mu \gamma_5 b \ket{\LamB(p, s_b)}
        & = -\bar{u}_{\alpha}^{(1/2)}(k, \eta(\lambda_c), s_c)
            \left[\sum_{i} g_i(q^2) \gamma_5 \Gamma^{\alpha \mu}_{A,i} \right] u(p, s_b)\,,
\end{aligned}
\end{equation}
where $\bar{u}^{(1/2)}_\alpha$ is the spin $1/2$ projection of a Rarita-Schwinger object $u_\alpha^\text{RS}(k, \eta, s) \equiv \eta_\alpha(k) u(k, s)$ (see Appendix
\ref{app:RSspinor}). For the hadronic matrix element of the vector and axialvector transitions to
the $\LamCst[2625]$ state we use:
\begin{equation}
\label{eq:FF-32-def}
\begin{aligned}
    \bra{\LamCst[2625](k, \eta(\lambda_c), s_c)} \bar{c}\gamma^\mu b \ket{\LamB(p, s_b)}
        & = +\bar{u}_\alpha^{(3/2)}(k, \eta(\lambda_c), s_c)
            \left[\sum_{i} F_i(q^2) \Gamma^{\alpha \mu}_{V,i} \right] u(p, s_b)\,,\\
    \bra{\LamCst[2625](k, \eta(\lambda_c), s_c)} \bar{c}\gamma^\mu \gamma_5 b \ket{\LamB(p, s_b)}
        & = -\bar{u}_\alpha^{(3/2)}(k, \eta(\lambda_c), s_c)
            \left[\sum_{i} G_i(q^2) \gamma_5 \Gamma^{\alpha \mu}_{A,i} \right] u(p, s_b)\,,
\end{aligned}
\end{equation}
where $\bar{u}^{(3/2)}_\alpha$ is the spin $3/2$ projection of a Rarita-Schwinger object; see also Appendix \ref{app:RSspinor}.
A possible basis of Dirac structures for the vector current is given in \cite{Meinel:2016cxo}.
We choose a different basis for both vector and axialvector currents. We compile the list of all
Dirac structures $\Gamma_{V(A),i}^{\alpha\mu}$ in \refapp{ff-details}.\\

We define the helicity amplitudes for the two currents $\Gamma^\mu =\gamma^\mu, \gamma^\mu\gamma_5$
as
\begin{equation}
    \label{eq:def:hel-amp}
    \mathcal{A}_\Gamma(s_b, s_c, \lambda_c, \lambda_q)
        \equiv \bra{\Lambda_c^*(s_c, \eta(\lambda_c))}\bar{c} \  \Gamma^\mu \eps_\mu^*(\lambda_q)b\ket{\Lambda_b(s_b)}\,,
\end{equation}
where the $\eps_\mu^*(\lambda_q)$ are a basis of polarisation vectors for the
virtual $W$ exchange with the polarisation states \mbox{$\lambda_q \in \lbrace t, 0,
+1, -1\rbrace$}; see \refapp{kin}. Due to the fact that the angular momentum configurations
$\lambda_c$ and $s_c$ in \refeq{def:hel-amp} can be independently chosen, there are more
possible combinations of $\lambda_c$ and $s_c$ than physically permitted.
We identify the helicity amplitudes with total angular moment $J=1/2$ as
\begin{equation}
\label{eq:def:hel-amp12}
\begin{aligned}
    \mathcal{A}^{(1/2)}_\Gamma(+1/2, +1/2,  0) & \equiv -\sqrt{\frac{1}{3}} \mathcal{A}_\Gamma(+1/2, +1/2, 0, 0) + \sqrt{\frac{2}{3}} \mathcal{A}_\Gamma(+1/2, -1/2, +1, 0)\,,\\
    \mathcal{A}^{(1/2)}_\Gamma(+1/2, +1/2,  t) & \equiv -\sqrt{\frac{1}{3}} \mathcal{A}_\Gamma(+1/2, +1/2, 0, t) + \sqrt{\frac{2}{3}} \mathcal{A}_\Gamma(+1/2, -1/2, +1, t)\,,\\
    \mathcal{A}^{(1/2)}_\Gamma(+1/2, -1/2, -1) & \equiv \sqrt{\frac{1}{3}} \mathcal{A}_\Gamma(+1/2, -1/2, 0, -1) - \sqrt{\frac{2}{3}} \mathcal{A}_\Gamma(+1/2, +1/2, -1, -1)\,.
\end{aligned}
\end{equation}
The complementary set of $J=3/2$ amplitudes reads
\begin{equation}
\label{eq:def:hel-amp32}
\begin{aligned}
    \mathcal{A}^{(3/2)}_\Gamma(+1/2, +3/2, +1) & \equiv \mathcal{A}_\Gamma(+1/2, +1/2, +1, +1)\,,\\
    \mathcal{A}^{(3/2)}_\Gamma(+1/2, +1/2,  0) & \equiv \sqrt{\frac{2}{3}} \mathcal{A}_\Gamma(+1/2, +1/2, 0, 0) + \sqrt{\frac{1}{3}} \mathcal{A}^{(3/2)}_\Gamma(+1/2, -1/2, +1, 0)\,,\\
    \mathcal{A}^{(3/2)}_\Gamma(+1/2, +1/2,  t) & \equiv \sqrt{\frac{2}{3}} \mathcal{A}_\Gamma(+1/2, +1/2, 0, t) + \sqrt{\frac{1}{3}} \mathcal{A}^{(3/2)}_\Gamma(+1/2, -1/2, +1, t)\,,\\
    \mathcal{A}^{(3/2)}_\Gamma(+1/2, -1/2, -1) & \equiv \sqrt{\frac{2}{3}} \mathcal{A}_\Gamma(+1/2, -1/2, 0,-1) + \sqrt{\frac{1}{3}} \mathcal{A}^{(3/2)}_\Gamma(+1/2, +1/2, -1,-1)\,.
\end{aligned}
\end{equation}
For transitions to $J=1/2$ the set of amplitudes in \refeq{def:hel-amp32} is required to vanish identically,
and similarly for transitions to $J=3/2$ the set in \refeq{def:hel-amp12} needs to be zero.
We explicitly verify this to be the case for the structures listed in \refapp{ff-details}.\\

Our Dirac structures $\Gamma^{\alpha\mu}_{V(A),i}$ have been chosen such that the form factors $F_{1/2,\lambda_q}$ and
$G_{1/2,\lambda_q}$, $\lambda_q \in \lbrace t,0,\perp\rbrace$, correspond to transitions into
$\LamCst[2595]$ states with $|J_z| = 1/2$, while the $\LamCst[2625]$ states
with $|J_z| = 3/2$ are only produced via the form factors $F_{3/2,\perp}$ and
$G_{3/2,\perp}$. Note that all helicity amplitudes depend only on one single form factor;
see \refeqs{helamp12-vector-first}{helamp12-vector-last}, \refeqs{helamp12-axialvector-first}{helamp12-axialvector-last}, \refeqs{helamp32-vector-first}{helamp32-vector-last}, and \refeqs{helamp32-axialvector-first}{helamp32-axialvector-last}.
We have therefore achieved a decomposition of the (axial)vector hadronic matrix
elements in terms of helicity form factors as inspired by
\cite{Feldmann:2011xf}. We note that our definitions of the form factors differ
from the one adopted in \cite{Leibovich:1997az}, where
the decomposition of the vector and axial vector hadronic matrix elements do not yield
form factors for transitions with well-defined angular momentum of the final states. In particular
in the conventions of \cite{Leibovich:1997az} the time-like polarisation, which is
relevant for the LFU ratio $R_{\Lambda_c^*}$, depends on linear combinations of multiple form
factors instead of one form factor per current.

\subsection{Heavy-quark expansion}
\label{sec:heavy_quark_expansion}

In Ref.~\cite{Leibovich:1997az}, the usual basis of form factors has been studied in the HQE up to
$1/m$ contributions. We cross-check their results, and adapt them to our choice of a helicity basis
for the form factors. In particular, we study the hadronic matrix elements in and beyond
the heavy quark limit $m_b \to \infty$, $m_c \to \infty$ with $m_c / m_b = \text{const}$. Following
\cite{Falk:1991nq}, we use that the transition matrix elements can be written at leading power in
the expansion as
\begin{equation}
\label{eq:hqet_expansion}
    \bra{\Lambda_c^{*}(k, \eta, s_c)} \bar{c} \ \Gamma^\mu b \ket{\LamB(p, s_b)}
    = \sqrt{4} \bar{u}_\alpha(\mLamCst v', \eta, s_c) \Gamma^{\mu} u(\mLamB v, s_b) \zeta^\alpha(w)\,,
\end{equation}
where $w \equiv v\cdot v'= (\mLamB^2+\mLamCst^2-q^2)/(2 \mLamB\mLamCst)$,
$v$ and $v'$ are the four-velocities of the initial and final states, respectively,
and $\Gamma$ denotes a Dirac structure.
Here the most general decomposition of the light-state transition
amplitude $\zeta$ reads
\begin{equation}
    \zeta^\alpha(w) = \zeta(w) (v - v')^\alpha\,.
\end{equation}
As a consequence, at leading power all form factors can be expressed in terms of the single amplitude $\zeta(w)$,
which must vanish at the zero hadronic recoil $w = 1$, which corresponds to $q^2 = (\mLamB - \mLamCst)^2$.
In order to include also $1/m$ and $\alpha_s$ corrections, we use for the vector current (and similarly for the axialvector current)
\begin{equation}
\label{eq:hqet_vect_current}
\gamma^{\mu}\mapsto J^{\mu}_V=C_1(\bar w) \gamma^{\mu} + C_2(\bar w) v^\mu + C_3(\bar w) v'^\mu +\Delta J_V^{\mu}\big\vert_{\mathcal{O}_1}+\Delta J_V^{\mu}\big\vert_{\mathcal{O}_8}
+\mathcal{O}(\alpha_s / m, 1/m^2)\,,
\end{equation}
with perturbative coefficients $C_i$ and power corrections $\Delta J_V^\mu$.\\

The perturbative functions $C_i$ are the Wilson coefficients arising in the matching of HQET onto QCD.
Their argument $\bar{w}$
is the recoil parameter as experienced by the heavy quarks within the hadrons.
Note that for a decay to orbitally excited hadrons $\bar w$ is not the same as defined for transitions
among ground-state baryons. Instead, we use
\begin{equation}
    \bar{w} \equiv w \left(1 + \frac{\bar{\Lambda}}{m_b} + \frac{\bar{\Lambda}'}{m_c}\right) - \left(\frac{\bar\Lambda}{m_c} + \frac{\bar{\Lambda}'}{m_b}\right)\,,
\end{equation}
where $\bar{\Lambda}$ and $\bar{\Lambda}'$ are the usual HQET parameters in the infinite mass limit.
The above yields the product of heavy-quark velocities as defined in \cite{Neubert:1993mb} in the limit
$\bar{\Lambda}' \to \bar{\Lambda}$. 
We use the matching coefficients $C_i$ to order $\alpha_s$, which are given in eq. (3.111) of \cite{Neubert:1993mb}.
At the precision that we aim for, we do not require the renormalization-group improved matching
coefficients, which can be extracted from \cite{Neubert:1993mb}, eq. (3.121).\\

In \refeq{hqet_vect_current} we use only power corrections $\Delta J_V^{\mu}\big\vert_{\mathcal{O}_1}$ and $\Delta J_V^{\mu}\big\vert_{\mathcal{O}_8}$, arising from the local operators $\mathcal{O}_1$ and $\mathcal{O}_8$ as defined in
\cite{Neubert:1993mb}, respectively. The remaining local operators only contribute at the order $\alpha_s / m$
and are therefore beyond the precision we aim for. The hadronic matrix elements of $\mathcal{O}_{1}$ and $\mathcal{O}_8$
can be parametrised as:
\begin{equation}
 \bra{\Lambda_c^{*}(k, \eta, s_c)}\Delta J_{V\mu}\big\vert_{\mathcal{O}_{1(8)}} \ket{\LamB(p, s_b)}=\sqrt{4} \bar{u}_\alpha(\mLamCst v', \eta, s_c)\big[\mathcal{O}_{1(8)}\big]_{\mu\beta}u(\mLamB v, s_b)  \zeta_{b(c)}^{\alpha\beta}(w)\,,
\end{equation}
where
\begin{equation}
\zeta^{\alpha\beta}_{(q)}(w)=(v-v')^{\alpha}\left[\zeta_{1}^{(q)}(w)v^{\beta}+\zeta^{(q)}_{2}(w)v'^{\beta}\right]+g^{\alpha\beta}\zeta^{(q)}_3(w)\,.
\end{equation}
and $[\mathcal{O}_1]_{\mu\beta}=\gamma_{\mu}\gamma_{\beta}$, $[\mathcal{O}_8]_{\mu\beta}=\gamma_{\beta}\gamma_{\mu}$. \\
After some algebra, we obtain the following for the contributions from $\Delta J_{V\mu}\big\vert_{\mathcal{O}_1}$ and $\Delta J_{V\mu}\big\vert_{\mathcal{O}_8}$:
\begin{equation}
\begin{aligned}
 \bra{\Lambda_c^{*}(k, \eta, s_c)}    \Delta J_{V\mu}\big\vert_{\mathcal{O}_1} \ket{\LamB(p, s_b)}
         = \frac{1}{2m_b}\bigg[&2\bar{u}_\alpha(\mLamCst v', \eta, s_c)\gamma_{\mu}u(\mLamB v, s_b) v^\alpha\left(\zeta^{(b)}_{1}(w)-\zeta^{(b)}_{2}(w)\right) \\
        +&4\bar{u}_\alpha(\mLamCst v', \eta, s_c) u(\mLamB v, s_b)v^{\alpha}v'^{\mu}\zeta^{(b)}_{2}(w) \\
         +&2\bar{u}_{\alpha}(\mLamCst v', \eta, s_c)\gamma_\mu\gamma^\alpha u(\mLamB v, s_b)\zeta^{(b)}_3(w)\bigg]\,, \\
 \bra{\Lambda_c^{*}(k, \eta, s_c)}    \Delta J_{V\mu}\big\vert_{\mathcal{O}_8} \ket{\LamB(p, s_b)}
        = \frac{1}{2m_c}\bigg[&2\bar{u}_\alpha(\mLamCst v', \eta, s_c)\gamma_{\mu}u(\mLamB v, s_b) v^\alpha \left(\zeta^{(c)}_{2}(w)-\zeta^{(c)}_{1}(w)\right) \\
         +&4\bar{u}_\alpha(\mLamCst v', \eta, s_c) u(\mLamB v, s_b)v^{\alpha}v^{\mu}\zeta^{(c)}_{1}(w) \\
          +&2\bar{u}_\alpha(\mLamCst v', \eta, s_c) \gamma^\alpha\gamma_\mu u(\mLamB v, s_b)\zeta^{(c)}_{3}(w)
          \bigg]\,.
\end{aligned}
\end{equation}
We can follow the very same steps also with the axial vector current. In this case we have:
\begin{equation}
\label{eq:hqet_axial_current}
\gamma^{\mu}\gamma_5\mapsto    J^{\mu}_A=C_1^{(5)}(\bar w)\gamma^{\mu}\gamma^5+ C_2^{(5)}(\bar w) v^\mu\gamma^5 + C_3^{(5)}(\bar w) v'^\mu\gamma^5 +\Delta J_A^{\mu}\big\vert_{\mathcal{O}^{\tiny A}_1}+\Delta J_A^{\mu}\big\vert_{\mathcal{O}^{A}_8}+\mathcal{O}(\alpha_s / m, 1/m^2)\,,
\end{equation}
where the subleading contributions $\Delta J_A^{\mu}\big\vert_{\mathcal{O}^{\tiny A}_1}$ and $\Delta J_A^{\mu}\big\vert_{\mathcal{O}^{A}_8}$ can by computed from
\begin{equation}
 \bra{\Lambda_c^{*}(k, \eta, s_c)} \Delta J_{A\mu}\big\vert_{\mathcal{O}^{A}_{1(8)}} \ket{\LamB(p, s_b)}= \sqrt{4} \bar{u}_\alpha(\mLamCst v', \eta, s_c) [\mathcal{O}^A_{1(8)}]_{\mu\beta} u(\mLamB v, s_b)\zeta_{b(c)}^{\alpha\beta}(w)\,,
\end{equation}
and  $[\mathcal{O}^{A}_1]_{\mu\beta}=\gamma_{\mu}\gamma^{5}\gamma_{\beta}$, $[\mathcal{O}^{A}_8]_{\mu\beta}=\gamma_{\beta}\gamma_{\mu}\gamma^{5}$. From this we obtain:
\begin{equation}
\begin{aligned}
\bra{\Lambda_c^{*}(k, \eta, s_c)} \Delta J_{A\mu}\big\vert_{\mathcal{O}^{A}_1} \ket{\LamB(p, s_b)}=\frac{1}{2m_b}\bigg[&2\bar{u}_\alpha(\mLamCst v', \eta, s_c)\gamma_{\mu}u(\mLamB v, s_b)v^\alpha\left(\zeta^{(b)}_{1}(w)+\zeta^{(b)}_{2}(w)\right) \\
 -&4\bar{u}_\alpha(\mLamCst v', \eta, s_c) u(\mLamB v, s_b)v^{\alpha}v'^{\mu}\zeta^{(b)}_{2}(w) \\
 +&2\bar{u}_{\alpha}(\mLamCst v', \eta, s_c)\gamma^{\mu}\gamma^5\gamma^{\alpha}u(\mLamB v, s_b)\zeta^{(b)}_3(w)\bigg]\,, \\
\bra{\Lambda_c^{*}(k, \eta, s_c)} \Delta J_{A\mu}\big\vert_{\mathcal{O}^{A}_8} \ket{\LamB(p, s_b)}=\frac{1}{2m_c}\bigg[&2\bar{u}_\alpha(\mLamCst v', \eta, s_c)\gamma_{\mu}u(\mLamB v, s_b)v^\alpha\left(\zeta^{(c)}_{1}(w)+\zeta^{(c)}_{2}(w)\right) \\
 +&4\bar{u}_\alpha(\mLamCst v', \eta, s_c) u(\mLamB v, s_b)v^{\alpha}v^{\mu}\zeta^{(c)}_{1}(w) \bigg]\,, \\
  +&2\bar{u}_{\alpha}(\mLamCst v', \eta, s_c)\gamma^{\alpha}\gamma^{\mu}\gamma^5u(\mLamB v, s_b)\zeta^{(b)}_3(w)\bigg]\,. 
\end{aligned}
\end{equation}
The subleading IW functions are related by the equations of motion.
In particular we have that $v_{\beta}\zeta^{\alpha\beta}_{(b)}=0$, and
$v'_{\beta}\zeta^{\alpha\beta}_{(c)}=0$. This leads to the following relations:
\begin{align}
\zeta_1^{(b)}(w)+w\zeta_{2}^{(b)}(w)+\zeta_{3}^{(b)}(w)&=0 \,,\\
w\zeta_1^{(c)}(w)+\zeta_{2}^{(c)}(w)&=0 \,.
\end{align}

Furthermore we know that $i\partial_\alpha[\bar{h}_c(v')\Gamma h_b(v)]=\bar{h}_c(v')i\cev{D}_{\alpha}\Gamma h_b(v)+\bar{h}_c(v')\Gamma iD_{\alpha}h_b(v)$,
where we denote $h_{b(c)}$ as the usual HQET fields. This identity allows us to write the following relations:
\begin{align}
\zeta_{1}^{(b)}(w)+\zeta_{1}^{(c)}(w)&=\bar{\Lambda}\zeta(w) \,, \\
\zeta_{2}^{(b)}(w)+\zeta_{2}^{(c)}(w)&=-\bar{\Lambda}'\zeta(w) \,, \\
\zeta_{3}^{(b)}(w)+\zeta_{3}^{(c)}(w)&=0\,.
\end{align}
With these $5$ relations we can reduce the initial $6$ subleading IW functions
to one independent subleading IW function. We find it convenient to use $\zeta_3^{(b)}$:
\begin{equation}
    \begin{aligned}
        \zeta_1^{(b)} & = -\frac{\zeta_3^{(b)}}{1 - w^2} + \frac{w \zeta}{1 - w^2}\left(\bar{\Lambda}' - \bar{\Lambda} w\right)\,,  \ & \   \zeta_2^{(b)} & = +\frac{w \zeta_3^{(b)}}{1 - w^2} - \frac{\zeta}{1 - w^2}\left(\bar{\Lambda}' - \bar{\Lambda} w\right)\,,\\
        \zeta_1^{(c)} & = +\frac{\zeta_3^{(b)}}{1 - w^2} - \frac{\zeta}{1 - w^2}\left(w \bar{\Lambda}' - \bar{\Lambda}\right)\,, \  & \ \zeta_2^{(c)} & = -\frac{w \zeta_3^{(b)}}{1 - w^2} + \frac{w \zeta}{1 - w^2}\left(w \bar{\Lambda}' - \bar{\Lambda}\right)\,.
    \end{aligned}
\end{equation}
From this point on we identify $\zeta_{\text{SL}}\equiv\zeta_3^{(b)} = - \zeta_3^{(c)}$.\\

Beside the effects on local operators, we also need to consider effects from
non-local insertions of the HQET Lagrangian at power $1/m$. Following the
discussion in \cite{Neubert:1993mb,Leibovich:1997az}, non-local insertions of
the kinetic operator give rise to an $w$-dependent shift $\eta_\text{kin}(w)$
to the leading-power IW function $\zeta(w)$. We can absorb this shift into the
definition of $\zeta$:
\begin{equation}
    \zeta(w) + \frac{1}{2 m_b m_c} \left[m_b + m_c\right] \eta_\text{kin}(w)
    \mapsto \zeta(w)\,.
\end{equation}
The $w$-dependent shift due to the chromomagnetic operator is more delicate.
The two contributions are:
\begin{align}
\label{eq:eta_mag_c}
    \eta_\text{mag}^{(c)}(w) & : \left[g_{\mu \alpha} v_\nu\right] \bar{u}_{J}^\alpha(\mLamCst v', \eta, s_c) \  i \sigma^{\mu\nu} \frac{1 + \slashed{v'}}{2} \Gamma u(\mLamB v, s_b)\\
\label{eq:eta_mag_b}
    \eta_\text{mag}^{(b)}(w) & : \left[g_{\mu \alpha} v'_\nu\right] \bar{u}_{J}^\alpha(\mLamCst v', \eta, s_c) \   \Gamma \frac{1 + \slashed{v}}{2} \  i \sigma^{\mu\nu} u(\mLamB v, s_b)\,.
\end{align}
In \cite{Leibovich:1997az}, it is argued that the two functions $\eta_\text{mag}^{(q)}(w)$ must vanish at zero recoil, and
are expected to be small compared to the size of $\Lambda_\text{QCD}$.
We follow this argument, and therefore choose to not consider contributions from either
$\eta_\text{mag}^{(q)}(w)$ from this point on.\\

If we want now to express the form factors in terms of the leading and
subleading IW functions we need to match the HQE expansion of the
helicity amplitudes onto the direct calculation presented in
Sec.~\ref{sec:helicity_form_factors}. Concerning the $\LamCst[2595]$ final
state, the comparison between
eqs.~(\ref{eq:hqe_hel_ampl_vect_first12})--(\ref{eq:hqe_hel_ampl_vect_last12})
and eqs.~(\ref{eq:helamp12-vector-first})--(\ref{eq:helamp12-vector-last})
leads to
\begin{align}
f_{1/2,0}=&\frac{\sqrt{s_+}}{2(\mLamB\mLamCst)^{3/2}} \bigg\{\bigg[s_{-}\left(C_1(\bar w)+\frac{s_+(C_2(\bar w)\mLamCst+C_3(\bar w)\mLamB)}{2\mLamB\mLamCst(\mLamB+\mLamCst)}\right) \notag \\ 
&+\frac{(\mLamB-\mLamCst)}{\mLamB+\mLamCst}\bigg(\frac{\mLamB^2-\mLamCst^2+q^2}{2\mLamB}\bar\Lambda-\frac{\mLamB^2-\mLamCst^2-q^2}{2\mLamCst}\bar\Lambda'\bigg)\bigg]\zeta-2(\mLamB-\mLamCst)\zeta_{\text{SL}}\bigg\} \,,\\
f_{1/2,t}=&\frac{\sqrt{s_-}}{2(\mLamB\mLamCst)^{3/2}} \bigg\{\bigg[C_1(\bar w)s_{+}+\frac{\mLamB+\mLamCst}{\mLamB-\mLamCst}\bigg(\frac{\mLamB^2-\mLamCst^2+q^2}{2\mLamB}\left(\bar\Lambda+\frac{C_2(\bar w)s_+}{\mLamB+\mLamCst}\right) \notag \\
&-\frac{\mLamB^2-\mLamCst^2-q^2}{2\mLamCst}\left(\bar\Lambda'-\frac{C_3(\bar w)s_+}{\mLamB+\mLamCst}\right)\bigg)\bigg]\zeta-2\frac{(\mLamB +\mLamCst)^2}{\mLamB-\mLamCst}\zeta_{\text{SL}}\bigg\}  \,,\\
f_{1/2,\perp}=& \frac{\sqrt{s_+}}{2(\mLamB\mLamCst)^{3/2}}\bigg\{\bigg[C_1(\bar w)s_{-}+\frac{3\mLamB^2+\mLamCst^2-q^2}{2\mLamB}\bar\Lambda-\frac{\mLamB^2+3\mLamCst^2-q^2}{2\mLamCst}\bar\Lambda'\bigg]\zeta-2\mLamB\zeta_{\text{SL}}\bigg\}\,,
\end{align}
for the vector form factors, while for the axial-vector form factors the matching of eqs.~(\ref{eq:hqe_hel_ampl_axial_first12})--(\ref{eq:hqe_hel_ampl_axial_last12}) onto eqs.~(\ref{eq:helamp12-axialvector-first})--(\ref{eq:helamp12-axialvector-last}) gives
\begin{align}
g_{1/2,0}=&\frac{\sqrt{s_-}}{2(\mLamB\mLamCst)^{3/2}} \bigg\{\bigg[s_{+}\left(C_1(\bar w)-\frac{s_-(C_2(\bar w)\mLamCst+C_3(\bar w)\mLamB)}{2\mLamB\mLamCst(\mLamB-\mLamCst)}\right) \notag  \\
&+\frac{\mLamB+\mLamCst}{\mLamB-\mLamCst}\bigg(\frac{\mLamB^2-\mLamCst^2+q^2}{2\mLamB}\bar\Lambda-\frac{\mLamB^2-\mLamCst^2-q^2}{2\mLamCst}\bar\Lambda'\bigg)\bigg]\zeta-2(\mLamB+\mLamCst)\zeta_{\text{SL}}\bigg\}\,,\\
g_{1/2,t}=&\frac{\sqrt{s_+}}{2(\mLamB\mLamCst)^{3/2}} \bigg\{\bigg[C_1(\bar w)s_{-}+\frac{\mLamB-\mLamCst}{\mLamB+\mLamCst}\bigg(\frac{\mLamB^2-\mLamCst^2+q^2}{2\mLamB}\left(\bar\Lambda-\frac{C_2(\bar w)s_-}{\mLamB-\mLamCst}\right) \notag  \\
&-\frac{\mLamB^2-\mLamCst^2-q^2}{2\mLamCst}\left(\bar\Lambda'+\frac{C_3(\bar w)s_-}{\mLamB-\mLamCst}\right)\bigg)\bigg]\zeta-2\frac{(\mLamB-\mLamCst)^2}{\mLamB+\mLamCst}\zeta_{\text{SL}}\bigg\}  \,,\\
g_{1/2,\perp}=& \frac{\sqrt{s_-}}{2(\mLamB\mLamCst)^{3/2}}\bigg\{\bigg[C_1(\bar w)s_{+}+\bar\Lambda\frac{3\mLamB^2+\mLamCst^2-q^2}{2\mLamB}-\bar\Lambda'\frac{\mLamB^2+3\mLamCst^2-q^2}{2\mLamCst}\bigg]\zeta-2\mLamB\zeta_{\text{SL}}\bigg\} \,.
\end{align}
Here and in the following we denote $s_\pm \equiv (\mLamB \pm \mLamCst)^2 -
q^2$.  Concerning the $\LamCst[2625]$ final state, the vector form factors are
obtained by matching
eqs.~(\ref{eq:hqe_hel_ampl_vect_first32})--(\ref{eq:hqe_hel_ampl_vect_last32})
with eqs.~(\ref{eq:helamp32-vector-first})--(\ref{eq:helamp32-vector-last})
\begin{align}
    F_{1/2,\perp} =&   \frac{\sqrt{s_{+}}}{2 (\mLamB\mLamCst)^{3/2}}\left\{\left[C_1(\bar w)s_{-}+\frac{3\mLamB^2+\mLamCst^2-q^2}{2\mLamB}\bar\Lambda-\frac{\mLamB^2+3\mLamCst^2-q^2}{2\mLamCst}\bar\Lambda'\right]\zeta+\mLamB \zeta_{\text{SL}}\right\}\, , \\
    F_{1/2,t} = & \frac{\sqrt{s_{-}}}{2 (\mLamB\mLamCst)^{3/2}}\bigg\{\bigg[C_1(\bar w)s_{+}+\frac{\mLamB+\mLamCst}{\mLamB-\mLamCst}\bigg(\frac{\mLamB^2-\mLamCst^2+q^2}{2\mLamB}\left(\bar\Lambda+\frac{C_2(\bar w)s_+}{\mLamB+\mLamCst}\right) \notag \\
    &-\frac{\mLamB^2-\mLamCst^2-q^2}{2\mLamCst}\left(\bar\Lambda'-\frac{C_3(\bar w)s_+}{\mLamB+\mLamCst}\right)\bigg)\bigg]\zeta 
  +  \frac{(\mLamB + \mLamCst)^2}{\mLamB-\mLamCst}\zeta_{\text{SL}}\bigg\} \, , \\
    F_{1/2,0} = &  \frac{\sqrt{s_{+}}}{2 (\mLamB\mLamCst)^{3/2}}\bigg\{\bigg[s_{-}\left(C_1(\bar w)+\frac{s_+(C_2(\bar w)\mLamCst+C_3(\bar w)\mLamB)}{2\mLamB\mLamCst(\mLamB+\mLamCst)}\right) \notag \\
     &+\frac{\mLamB-\mLamCst}{\mLamB+\mLamCst}\left(\frac{\mLamB^2-\mLamCst^2+q^2}{2\mLamB}\bar\Lambda-\frac{\mLamB^2-\mLamCst^2-q^2}{2\mLamCst}\bar\Lambda'\right)\bigg]\zeta+(\mLamB-\mLamCst)\zeta_{\text{SL}}\bigg\} \, , \\
 F_{3/2,\perp} =& -\frac{\sqrt{s_+}}{2 \mLamB^{3/2}\mLamCst^{1/2}}\zeta_{\text{SL}} \,,
\end{align}
while for the axial-vector form factor the comparison of eqs.~(\ref{eq:hqe_hel_ampl_axial_first32})--(\ref{eq:hqe_hel_ampl_axial_last32}) and eqs.~(\ref{eq:helamp32-axialvector-first})--(\ref{eq:helamp32-axialvector-last}) yields
\begin{align}
    G_{1/2,\perp} =& \frac{\sqrt{s_{-}}}{2 (\mLamB\mLamCst)^{3/2}}\left\{\left[C_1(\bar w)s_{+}+\frac{3\mLamB^2+\mLamCst^2-q^2}{2\mLamB}\bar\Lambda-\frac{\mLamB^2+3\mLamCst^2-q^2}{2\mLamCst}\bar\Lambda'\right]\zeta+\mLamB\zeta_{\text{SL}}\right\}\, , \\
    G_{1/2,t} =& \frac{\sqrt{s_{+}}}{2 (\mLamB\mLamCst)^{3/2}}\bigg\{\bigg[C_1(\bar w)s_{-}+\frac{\mLamB-\mLamCst}{\mLamB+\mLamCst}\bigg(\frac{\mLamB^2-\mLamCst^2+q^2}{2\mLamB}\left(\bar\Lambda-\frac{C_2(\bar w)s_-}{\mLamB-\mLamCst}\right) \notag \\
    &-\frac{\mLamB^2-\mLamCst^2-q^2}{2\mLamCst}\left(\bar\Lambda'+\frac{C_3(\bar w)s_-}{\mLamB-\mLamCst}\right)\bigg)\bigg] \zeta 
    +\frac{(\mLamB - \mLamCst)^2}{\mLamB+\mLamCst}\zeta_{\text{SL}}\bigg\}\, , \\
     G_{1/2,0} =&  \frac{\sqrt{s_{-}}}{2 (\mLamB\mLamCst)^{3/2}}\bigg\{\bigg[s_{+}\left(C_1(\bar w)-\frac{s_-(C_2(\bar w)\mLamCst+C_3(\bar w)\mLamB)}{2\mLamB\mLamCst(\mLamB-\mLamCst)}\right) \notag \\
     &+\frac{\mLamB+\mLamCst}{\mLamB-\mLamCst}\left(\frac{\mLamB^2-\mLamCst^2+q^2}{2\mLamB}\bar\Lambda-\frac{\mLamB^2-\mLamCst^2-q^2}{2\mLamCst}\bar\Lambda'\right)\bigg]\zeta
     +(\mLamB+\mLamCst)\zeta_{\text{SL}}\bigg\}\, , \\
 G_{3/2,\perp} =& -\frac{\sqrt{s_-}}{2 \mLamB^{3/2}\mLamCst^{1/2}}\zeta_{\text{SL}} \,.
\end{align}

Thus, at leading power in $1/m$ only the $(J,J_z) = (3/2, \pm 1/2)$ form factors receive contributions
from the leading-power IW function. As a consequence, the sum rule at zero recoil ($w = 1$ or $s_- = 0$)
as discussed later will be less sensitive to the contributions from the $J=3/2$ amplitudes.

We note in passing that our results for the HQE of the form factors fulfil the relations
\begin{equation}
\label{eq:form factors relations}
\begin{aligned}
    \frac{f_{1/2,t}(0)}{f_{1/2,0}(0)} & \equiv \frac{\mLamB + \mLamCst}{\mLamB - \mLamCst} \, \,, &
    \frac{g_{1/2,t}(0)}{g_{1/2,0}(0)} & \equiv \frac{\mLamB - \mLamCst}{\mLamB + \mLamCst} \, \,, \\
    \frac{F_{1/2,t}(0)}{F_{1/2,0}(0)} & \equiv \frac{\mLamB + \mLamCst}{\mLamB - \mLamCst} \, \,, &
    \frac{G_{1/2,t}(0)}{G_{1/2,0}(0)} & \equiv \frac{\mLamB - \mLamCst}{\mLamB + \mLamCst} \, \,,
\end{aligned}
\end{equation}
as required by analyticity; i.e., any spurious poles of the hadronic matrix elements in the limit $q^2 \to 0$
do not correspond to any physical states with quantum numbers $B = -C = 1$, and therefore must
be cancelled due to the above relations.

\section{Phenomenology}
\label{sec:pheno}

\subsection{Parametrisation of the Isgur-Wise functions}
\label{sec:parametrization}

Determining the parameters of the leading and subleading IW functions is
a crucial point to evaluate the form factors.  Unfortunately, there are no
first principles in HQET which allow us to estimate the $q^{2}$ dependence of
the IW functions. In light of this, we need to infer a functional form
for $\zeta(q^2)$ and $\zeta_{\text{SL}}(q^2)$ through some other means. For the ground-state
transition $\Lambda_b \to \Lambda_c$ and in the large $N_c$ limit, it has been motivated
in \cite{Jenkins:1992se} to express the IW functions as exponential functions.
Inspired by this, one of the models we consider here for the parametrisation of the leading and
subleading IW function $\zeta(q^2)$ and $\zeta_{\text{SL}}(q^2)$ is
\begin{equation}
\begin{aligned}
\label{eq:IW_functions_exp}
    \zeta(q^2)\bigg|_{\text{exp}}
        & \equiv \zeta(q^2_\text{max}) \exp\left[ \rho \left(\frac{q^2}{q^2_\text{max}} - 1\right)\right]\,,\\
    \zeta_{\text{SL}}(q^2)\bigg|_{\text{exp}}
        & \equiv \zeta(q^2_\text{max}) \delta_{\text{SL}} \exp\left[\frac{\rho_{\text{SL}}}{\delta_{\text{SL}} } \left(\frac{q^2}{q^2_\text{max}} - 1\right)\right]\,,
\end{aligned}
\end{equation}
where the normalisation $ \zeta(q^2_\text{max})$, the relative normalisation $\delta_{\text{SL}}$ and the two shape parameters $\rho$ and  $\rho_{\text{SL}}$ are  to be determined.\\

We can also use a Taylor expansion of $\zeta(q^2)$ and $\zeta_{\text{SL}}(q^2)$
around $q^2\simeq q^2_\text{max}$.  For our purposes we use an
expansion up to the first order in $q^2$:
\begin{equation}
\begin{aligned}
\label{eq:IW_functions}
    \zeta(q^2)\bigg|_{\text{lin}}
        & \equiv \zeta(q^2_\text{max}) \left[1 + \rho \left(\frac{q^2}{q^2_\text{max}} - 1\right)\right] \,,\\
    \zeta_{\text{SL}}(q^2)\bigg|_{\text{lin}}
        & \equiv \zeta(q^2_\text{max}) \left[\delta_{\text{SL}} + \rho_{\text{SL}} \left(\frac{q^2}{q^2_\text{max}} - 1\right)\right]\,.
\end{aligned}
\end{equation}
In the following we will refer to \refeq{IW_functions} as the nominal parametrisation. \\

Both parametrisations have been chosen such that they share their complete parameter
set, and such that both the leading and the subleading IW functions
have a common normalisation $\zeta(q^2_\text{max})$.

\subsection{Benchmarking the form factors' parameters from Zero Recoil Sum Rules}
\label{sec:benchmark}

The kinematic point of zero hadronic recoil is a special one for bottom-to-charm
transitions. In this point the hadronic form factors for $\Lambda_b \to X_c$ transitions,
where $X_c$ denotes a singly-charmed baryonic state, are minimally sensitive to the dynamics
of the light degrees of freedom within the respective hadrons; see e.g.~\cite{Manohar:2000dt}.
As a consequence, the inclusive spectral
density for the forward matrix elements of two bi-local insertions of the weak current can be expressed
in terms of $\Lambda_b \to X_c$ form factors. Inference of weighted sum of squares for the form factor
normalisations follows in what is known as a Zero Recoil Sum Rule (ZRSR)~\cite{Shifman:1994jh,Bigi:1994ga}.
This is only possible
since the spectral density consists of a sum of positive-definite exclusive terms.\\

The ZRSR is well established for $B\to D$ and $B\to D^*$ transitions, with OPE contributions known
up to order $\alpha_s^2$~\cite{Czarnecki:1997wy}. After the first
lattice QCD results for the $\Lambda_b\to \Lambda_c$ form factors appeared \cite{Detmold:2015aaa}, they were
scrutinised in the ZRSR framework \cite{Mannel:2015osa}. The conclusion of the latter analysis is as follows.
Given our present knowledge of the $\Lambda_b$ forward matrix elements, and given the lack of mixed $\alpha_s / m$
results for the ZRSR, the lattice results for $\Lambda_b\to \Lambda_c$ transition lead to a negative contribution
from non-ground state transitions. As mentioned above, negative contributions to the spectral density are
not possible by construction. Hence, either the inclusive calculation of the spectral density yields too small
a value, or the lattice results are too large.\\

For the discussion at hand, we will assume that the inclusive calculation underestimates the magnitude of the
spectral density. Specifically, we assume that $1/m^4$ and $1/m^5$ terms in the Heavy-Quark-Expansion, which
have not been taken into account due to lack of information on the relevant hadronic matrix elements, will
increase the magnitude. A priori it is not intuitive
that terms at order $1/m^4$ or beyond can make a qualitative difference to the ZRSR. However, there is precedent
for numerically relevant shifts in the case of $B\to D^*$ \cite{Gambino:2012rd}. In the latter study, it was observed
that -- based on rather precise knowledge of the HQE parameters for $B$ mesons -- the sum of $1/m^4$ and $1/m^5$
terms yields roughly a third of the $1/m^2$ and $1/m^3$ terms.\\

In the absence of further information on the
$\Lambda_b$ forward matrix elements, we will therefore proceed as follows. We will rescale the estimate of
the $1/m^2$ and $1/m^3$ terms by a factor of $1.33$, thereby copying the situation in $B\to D^*$ decays\footnote{%
    We stress that this rescaling, and the corresponding shift to the
    inclusive upper bound on the form factor normalisations, is based
    on a supposition rather than data, and will only be used for the purpose
    of benchmarking the experimental sensitivity. Ultimately, only improved
    knowledge of the hadronic matrix elements will settle the discrepancy between
    the ZRSR and lattice results.
}.
The corresponding shift can now accommodate fully the lattice results for the $\Lambda_b\to \Lambda_c$
form factors, as well as form factors for $\Lambda_b$ decays to excited charm baryons. The setup of the
ZRSR involves an upper bound on the excitation energies $\eps \equiv M_{X_c} - M_{\Lambda_c}$ of the
contributing charm baryons.
For the analysis at hand, $\eps \leq 0.7\,\text{GeV}$. Based on the known spectrum of charmed baryons \cite[Ch. 109 Charmed Baryons]{Patrignani:2016xqp},
the ZRSR covers -- beside the ground state -- form factors for $\Lambda_b$ decays into $\Sigma_c(2455)$,
$\Sigma_c(2520)$, $\Lambda_c(2595)$, $\Lambda_c(2625)$, and $\Sigma_c(2800)$\footnote{%
    We do not consider here the states of roughly $2.8$ GeV to $2.9$ GeV for which there exists no definite
    assignment as either a $\Lambda_c$, or a $\Sigma_c$ state, or as a kinematical artifact in
    the $\Lambda_c \pi \pi$ spectrum. A recent LHCb analysis of $\Lambda_b\to \Lambda_c\ell\nu$ \cite{LHCb-PAPER-2017-016}
    suggests that the yield of $\Lambda_c\pi\pi$ background stemming from this kinematic
    region corresponds to roughly $10\%$ of the first orbitally excited $\Lambda_c^*$ states.
    Given the overall accuracy of our analysis, this further supports our decision not to consider these states.
}.
The $\Sigma_c$ states form an isospin triplet and therefore carry isospin $I=1$. Consequently, the transitions
$\Lambda_b\to \Sigma_c$ violate isospin conservation, and we will assume them to be further suppressed
with respect to the $\Lambda_b \to \Lambda_c^*$ transitions. This supposition is corroborated by the
non-observation of $\Lambda_b\to \Sigma_c\ell\nu$ decays in the recent LHCb study~\cite{LHCb-PAPER-2017-016}.
Under the above assumptions, the inelastic parts of the ZRSR can be recast as matrix elements involving
only $\Lambda_b\to \Lambda_c^*$ transitions.\\

Following the definitions and analysis of Ref.~\cite{Mannel:2015osa}, applying the assumptions
above we arrive at the following constraints at zero recoil:
\begin{equation}
\label{eq:Finel+Ginel}
\begin{aligned}
    F_\text{inel} & = 0.011^{+0.061}_{-0.055} \approx F_{\text{inel},1/2} + F_{\text{inel},3/2}\,,\\
    G_\text{inel} & = 0.040^{+0.049}_{-0.052} \approx G_{\text{inel},1/2} + G_{\text{inel},3/2}\,.
\end{aligned}
\end{equation}
The individual contributions from the orbitally-excited $\Lambda_c^*$ states for the vector current read:
\begin{align}
    F_{\text{inel},1/2}
        & \equiv \frac{1}{N_V}\, \sum_\text{$\Lambda_c^*$ spin} \bra{\LamB(v, s_b)}\bar{b}\gamma^{\mu}c\ket{\LamCst[2595](v)}\bra{\LamCst[2595](v)}\bar{c}\gamma_{\mu}b\ket{\LamB(v, s_b)}\\
        & = \frac{1}{3} \left[|f_{t,1/2}|^2 + |f_{0,1/2}|^2 \frac{(\mLamB + \mLamCst)^2}{(\mLamB - \mLamCst)^2} + 2 |f_{\perp,1/2}|^2\right]_{\mathrm{zero \ recoil}}\,,
\end{align}
and
\begin{align}
    F_{\text{inel},3/2}
        & \equiv \frac{1}{N_V}\, \sum_\text{$\Lambda_c^*$ spin} \bra{\LamB(v, s_b)}\bar{b}\gamma^{\mu}c\ket{\LamCst[2625](v)}\bra{\LamCst[2625](v)}\bar{c}\gamma_{\mu}b\ket{\LamB(v, s_b)}\\
        & = \frac{2}{3} \left[|F_{t,1/2}|^2 + |F_{0,1/2}|^2 \frac{(\mLamB + \mLamCst)^2}{(\mLamB - \mLamCst)^2} + 2 |F_{\perp,1/2}|^2 + 6 |F_{\perp,3/2}|^2\right] \,,
\end{align}
where $N_V = 1$.
For the axialvector current, including the normalisation factor $N_A = 3$, the individual contributions read:
\begin{align}
    G_{\text{inel},1/2}
        & \equiv  \frac{1}{N_A}\,\sum_\text{$\Lambda_c^*$ spin} \bra{\LamB(v, s_b)}\bar{b}\gamma^{\mu}\gamma_{5}c\ket{\LamCst[2595](v)}\bra{\LamCst[2595](v)}\bar{c}\gamma_{\mu}\gamma_{5}b\ket{\LamB(v, s_b)}\\
        & = \frac{1}{9} \left[|g_{0,1/2}|^2 + |g_{t,1/2}|^2 \frac{(\mLamB + \mLamCst)^2}{(\mLamB - \mLamCst)^2} + 2 |g_{\perp,1/2}|^2\right]_{\text{zero recoil}} \,,
\end{align}
and
\begin{align}
    G_{\text{inel},1/2}
        & \equiv \frac{1}{N_V}\, \sum_\text{$\Lambda_c^*$ spin} \bra{\LamB(v, s_{b})}\bar{b}\gamma^{\mu}\gamma_{5}c\ket{\LamCst[2625](v)}\bra{\LamCst[2625](v)}\bar{c}\gamma_{\mu}\gamma_{5}b\ket{\LamB(p,s_{b})}\\
        & = \frac{2}{9} \left[|G_{0,1/2}|^2 + |G_{t,1/2}|^2 \frac{(\mLamB + \mLamCst)^2}{(\mLamB - \mLamCst)^2} + 2 |G_{\perp,1/2}|^2 + 6 |G_{\perp,3/2}|^2\right]_{\text{zero recoil}} \,.
\end{align}

In the zero-recoil point, both parametrisation~\refeq{IW_functions_exp} and
\refeq{IW_functions} yield the same expressions, involving only the parameters
$\zeta(q^2_\text{max})$ and $\delta_\text{SL}$.\\

Using two uncorrelated gaussian distributions for $F_\text{inel}$ and $G_\text{inel}$
and using symmetrised $68\%$ intervals based on \refeq{Finel+Ginel} we obtain correlated
distributions for $\zeta(q^2_\text{max})$ and $\delta_\text{SL}$. The $\zeta(q^2_\text{max})$ distribution
is highly non-gaussian, and due to the large set of assumptions on which our results are founded,
both distributions are not instructive for physics analyses. However, they can be used to define
a benchmark point for further phenomenological analyses, in particular for the sensitivity study later on
in this article. For later applications, we define the normalisation parameters of our benchmark point
to be compatible with these distributions:
\begin{equation}
\label{eq:ZRSR-benchmark}
\begin{aligned}
    \zeta(q^2_\text{max}) & =  0.25\,, &
    \delta_\text{SL}      & = -0.14\,\GeV\,,
\end{aligned}
\end{equation}
corresponding to a subleading contribution of $14\%$ of the leading-power IW function. This is fully in line
with naive power-counting expectations for the subleading-power IW function. \\

Since the ZRSR cannot provide us with any information on the slopes of either IW function, we have to draw
inspiration from elsewhere. Given the lower bound on the slope of the leading-power IW function for $B\to D^{(*)}$
transitions, we assume $\rho,\rho_\text{SL} \gtrsim 0.25$. On the other hand, in order to avoid
unphysical zero crossings of the IW functions in the semileptonic region in the nominal parametrisation,
we need to impose $\rho,\rho_\text{SL} \lesssim 0.75$. We choose to use the boundaries to define the
slope parameters of our benchmark points as:
\begin{align}
    \rho & = 0.25 & \rho_\text{SL} & = 0.25\,\GeV\,, \label{eq:bp1}\\
    \rho & = 0.25 & \rho_\text{SL} & = 0.75\,\GeV\,, \\
    \rho & = 0.75 & \rho_\text{SL} & = 0.75\,\GeV\,, \\
    \rho & = 0.75 & \rho_\text{SL} & = 0.25\,\GeV\,.
\end{align}
We emphasise again that these values are not viable for any physics analysis, and are merely used when
studying the sensitivity to the IW function parameters for upcoming LHCb analyses.

\subsection{Observables}
\label{sec:observables}

The fully differential decay rate of an unpolarised $\Lambda_b$ to a $\Lambda_c^{*}$ with total angular momentum $J$ can be written as
\begin{align}
\frac{1}{\Gamma^{(\ell)}_{0}}  \frac{\dd^2 \Gamma^{(\ell)}_{J}}{\dd q^2\,\dd \cos\theta_\ell}= &\left(a_{\ell}^{(J)} + b_{\ell}^{(J)} \cos\theta_{\ell}+c_{\ell}^{(J)}\cos^2\theta_{\ell}\right) \,, &  \frac{1}{\Gamma^{(\ell)}_{0}}  \frac{\dd \Gamma^{(\ell)}_{J}}{\dd q^2}=& \  2\left(a_{\ell}^{(J)}+\frac{1}{3} c^{(J)}_{\ell}\right) \,,
\end{align}
with coefficient functions $a^{(J)}_\ell(q^2)$, $b^{(J)}_\ell(q^2)$,
$c^{(J)}_\ell(q^2)$ for the specific final-state lepton flavour $\ell \in
\lbrace e, \mu, \tau\rbrace$.  The momentum transfer $q^2$ is defined as the
invariant mass of the leptons in the final state, and $\theta_\ell$ is the
helicity angle of the charged lepton with the $\ell$-$\nu_\ell$ momentum in the
$\Lambda_b$ rest frame. Our choice of normalisation reads
\begin{equation}
    \Gamma^{(\ell)}_{0}(q^2) = \frac{G_{F}^2V_{cb}^2\sqrt{s_{+}s_{-}}\mLamCst}{96 \pi^3\mLamB^2}\left(1-\frac{m_{\ell}^2}{q^2}\right)^2\,,
\end{equation}
which should not be confused with the total decay width
\begin{equation}
\label{eq:decay-rate}
    \Gamma^{(\ell)}_J
    = 2 \int_{m_\ell^2}^{(\mLamB - \mLamCst)^2} \dd{q^2}\,
        \Gamma_0^{(\ell)}(q^2)\,\left(a_\ell^{(J)}(q^ 2) + \frac13 c_\ell^{(J)}(q^2)\right)\,.
\end{equation}
From the double-differential rate, we can construct two angular observables in addition to the
$q^2$-differential decay rate: first, the forward-backward asymmetry
\begin{equation}
\label{eq:def:A_FB}
\begin{aligned}
    A_\text{FB}(q^2)
        & \equiv \frac{1}{\dd{\Gamma_J^{(\ell)}} / \dd{q^2}}\,
        \int_{-1}^{+1} \dd{\cos\theta_\ell}\, \left[\omega_{A_\text{FB}}(\cos \theta_\ell)\, \frac{\dd^2 \Gamma^{(\ell)}_{J}}{\dd q^2\,\dd \cos\theta_\ell}\right]\\
        & = \frac{1}{\dd{\Gamma_J^{(\ell)}} / \dd{q^2}}\, \Gamma_0^{(\ell)}(q^2) b^{(J)}_\ell(q^2)\,,
\end{aligned}
\end{equation}
which arises from the term linear in $\cos\theta_\ell$. And secondly, the flat term
\begin{equation}
\label{eq:def:F_H}
\begin{aligned}
    F_\text{H}(q^2)
        & \equiv \frac{1}{\dd{\Gamma_J^{(\ell)}} / \dd{q^2}}\,
        \int_{-1}^{+1} \dd{\cos\theta_\ell}\, \left[\omega_{F_\text{H}}(\cos \theta_\ell)\, \frac{\dd^2 \Gamma^{(\ell)}_{J}}{\dd q^2\,\dd \cos\theta_\ell}\right]\\
        & = \frac{1}{\dd{\Gamma_J^{(\ell)}} / \dd{q^2}}\, 2 \Gamma_0^{(\ell)}(q^2) \left[a^{(J)}_\ell(q^2) + c^{(J)}_\ell(q^2)\right]\,,
\end{aligned}
\end{equation}
which arises from a linear combination of the coefficients $a_\ell^{(J)}$ and
$c_\ell^{(J)}$ that differs from the one comprising the decay rate
\refeq{decay-rate}. The weight functions for both observables read:
\begin{equation}
\begin{aligned}
    \omega_{A_{\text{FB}}}(\cos\theta_\ell) & =  \ \frac{3}{2} P_1(\cos\theta_\ell)\,,       &
    \omega_{F_{\text{H}}}(\cos\theta_\ell)  & =  \ 5 P_2(\cos\theta_\ell) + P_0 (\cos\theta_\ell)\,.
\end{aligned}
\end{equation}
In the above, $P_{n}$ denotes the $n$th Legendre polynomial.\\

Note that the definition of the flat term $F_H$ in \refeq{def:F_H} is similar
to the one proposed for e.g. the decay $B\to K\ell^+\ell^-$; see
Ref.~\cite{Bobeth:2007dw}. However, contrary to what happens in the mesonic decays
in the limit $m_{\ell}\rightarrow 0$, the baryonic $F_H$ does not vanish in the
SM. This is due to the fact that the $\Lambda_b\to\Lambda_c^*$ transitions are
also mediated by perpendicular polarisation states of the virtual $W$, which is
impossible in the mesonic transitions.

For the decay to the $J=1/2$ final state the coefficients are
\begin{align}
2a^{(1/2)}_{\ell}=&\bigg[\vert f_{1/2,t}\vert^2\frac{m_{\ell}^2}{q^2}(\mLamB-\mLamCst)^2 +\vert f_{1/2,0}\vert^2(\mLamB+\mLamCst)^2+\vert f_{1/2,\perp}\vert^2(m_{\ell}^2+q^2) \notag\\
 &+ \vert g_{1/2,t}\vert^2\frac{m_{\ell}^2}{q^2}(\mLamB+\mLamCst)^2 +\vert g_{1/2,0}\vert^2(\mLamB-\mLamCst)^2+\vert g_{1/2,\perp}\vert^2(m_{\ell}^2+q^2)\bigg]\, ,\label{eq:a_12}\\
2b^{(1/2)}_{\ell}=&  \ 2 \left[f_{1/2,t}f_{1/2,0}+g_{1/2,t}g_{1/2,0}\right]\frac{m_\ell^2}{q^2}(\mLamB^2-\mLamCst^2) - 4 \ q^2 f_{1/2,\perp}g_{1/2,\perp}\, ,\label{eq:b_12}\\
2c^{(1/2)}_{\ell}=&-\left(1-\frac{m_{\ell}^2}{q^2}\right) \bigg[\vert f_{1/2,0}\vert^2 (\mLamB+\mLamCst)^2-q^2\vert f_{1/2,\perp}\vert^2+\vert g_{1/2,0}\vert^2 (\mLamB-\mLamCst)^2-q^2\vert g_{1/2,\perp}\vert^2\bigg]  \,.\label{eq:c_12}
\end{align}
For the $J=3/2$ we have
\begin{align}
a^{(3/2)}_{\ell}=&\bigg[\vert F_{1/2,t}\vert^2\frac{m_{\ell}^2}{q^2}(\mLamB-\mLamCst)^2+\vert F_{1/2,0}\vert^2(\mLamB+\mLamCst)^2+(\vert F_{1/2,\perp}\vert^2+3 \vert F_{3/2,\perp}\vert^2)(m_{\ell}^2+q^2)\notag\\
 &+ \vert G_{1/2,t}\vert^2\frac{m_{\ell}^2}{q^2}(\mLamB+\mLamCst)^2 +\vert G_{1/2,0}\vert^2(\mLamB-\mLamCst)^2+(\vert G_{1/2,\perp}\vert^2+3 \vert G_{3/2,\perp}\vert^2)(m_{\ell}^2+q^2)\bigg]\, ,\label{eq:a_32}\\
b^{(3/2)}_{\ell}=&  \ 2 \left[F_{1/2,t}F_{1/2,0}+G_{1/2,t}G_{1/2,0}\right]\frac{m_\ell^2}{q^2}(\mLamB^2-\mLamCst^2)
- 4 \ q^2 \left[F_{1/2,\perp}G_{1/2,\perp}+3F_{3/2,\perp}G_{3/2,\perp}\right]\, ,\label{eq:b_32}\\
c^{(3/2)}_{\ell}=&-\left(1-\frac{m_{\ell}^2}{q^2}\right) \bigg[\vert F_{1/2,0}\vert^2 (\mLamB+\mLamCst)^2 -q^2(\vert F_{1/2,\perp}\vert^2+3 \vert F_{3/2,\perp}\vert^2)\notag \\
&+\vert G_{1/2,0}\vert^2 (\mLamB-\mLamCst)^2-q^2(\vert G_{1/2,\perp}\vert^2+3 \vert G_{3/2,\perp}\vert^2)\bigg]  \, .\label{eq:c_32}
\end{align}
Our results for the angular coefficients in eqs.~\eqref{eq:a_12}--\eqref{eq:c_12} and eqs.~\eqref{eq:a_32}--\eqref{eq:c_32}
include the full $m_\ell$ dependence. We can compare them to the results for the fully differential decay
rate in the limit $m_\ell \to 0$ as presented in \cite{Leibovich:1997az}. We find complete agreement between our
limit and the results of \cite{Leibovich:1997az} when converting to the different basis of form factors as shown in \refeq{conversion-LS}.

\section{Prospects for the determination of the $\Lambda_{b}^{0}\to\Lambda_{c}^{*+}$ form factors using LHCb data}
\label{sec:exp_prospects}

Similarly to the mesonic $B\to D^{(*)}$ transitions, the most precise SM prediction for $R_{\Lambda_c^{*}}$ will arise from a combination of theoretical and experimental input. 
In this section, we investigate the sensitivity to the IW parameters from the decay $\Lambda_{b}^{0}\to\Lambda_{c}^{*+}\mu^{-}\bar{\nu}$ in the present
and future LHCb datasets when assuming a SM-like distribution\footnote{%
    Note that a popular NP explanation for the present $R_{D^{(*)}}$ anomalies is a rescaling of the
    coupling associated with effective operator
    $\sim [\bar{c} \gamma^\mu (1 - \gamma_5) b]\,[\bar{\nu} \gamma_\mu (1 - \gamma_5) \ell]$.
    Such a rescaling would leave the angular distribution of $b\to c\ell\bar\nu$ decays used here invariant.
}.
To achieve this, we first produce a series of toy ensembles and subsequently fit the decay distribution to
the simulated pseudo events. Estimates for the theoretical uncertainty on $R_{\Lambda_c^{*}}$ within the SM
are then produced based on our fits.

\subsection{Experimental situation}

Two aspects of the experimental situation are needed to assess the experimental sensitivity. The reconstructed and selected signal yields of the decays \GoodDecay and \BadDecay and the resolution in \qq and \costhetal.
We estimate the expected signal yields for a given luminosity by extrapolating from the numerical values quoted in Ref.~\cite{LHCb-PAPER-2017-016}, taking into the account the increased $b\bar{b}$ cross-section at 13~\TeV~\cite{LHCb-PAPER-2016-031}. We explore the sensitivity to parameters of interest as a function of the luminosity, starting from the current LHCb dataset, up to the luminosity expected at the end of the first LHCb upgrade~\cite{Bediaga:1443882}. 
 
 A key factor which limits the precision of the experimental measurements is the resolution in \qq and \costhetal, induced by the unreconstructed neutrino. The resolution determines how finely the data is binned and introduces a statistical correlation between adjacent bins. At a hadron collider, the momentum of the neutrino can be deduced using the information of the \LamB flight direction and its mass, up to a two-fold ambiguity. The dominant effects on the resulting resolution originate from the measurement of the primary $pp$ collision and \LamB vertices, as well the effect of choosing the wrong kinematic solution from the two available. In order to approximate the resolution of the LHCb detector, a sample of \SignalDecayBoth candidates are simulated using Pythia at $13~\TeV$~\cite{Sjostrand:2006za,Sjostrand:2007gs}, with a required pseudo-rapidity of $2 < \eta < 5$, approximately corresponding to the LHCb acceptance. The vertices of the $pp$ collision and \LamB decay are varied according to a resolution inspired from Ref.~\cite{Barbosa-Marinho:504321} and used in Ref.~\cite{Ciezarek:2016lqu}. The resolutions of $\pm20~\mum$ in the $x$ and $y$ directions and $\pm200~\mum$ in the $z$ direction (defined as the direction aligned with the LHC beam line) is used for the \LamB vertex. For the $pp$ collision vertex, a resolution of $\pm13~\mum$ in $x$ and $y$ and $\pm70~\mum$ in $z$ is assumed. With these new vertex positions the two kinematic solutions for the neutrino are then calculated, and one is chosen randomly.

\begin{figure}[tb]
\centering
\includegraphics[width=0.45\linewidth]{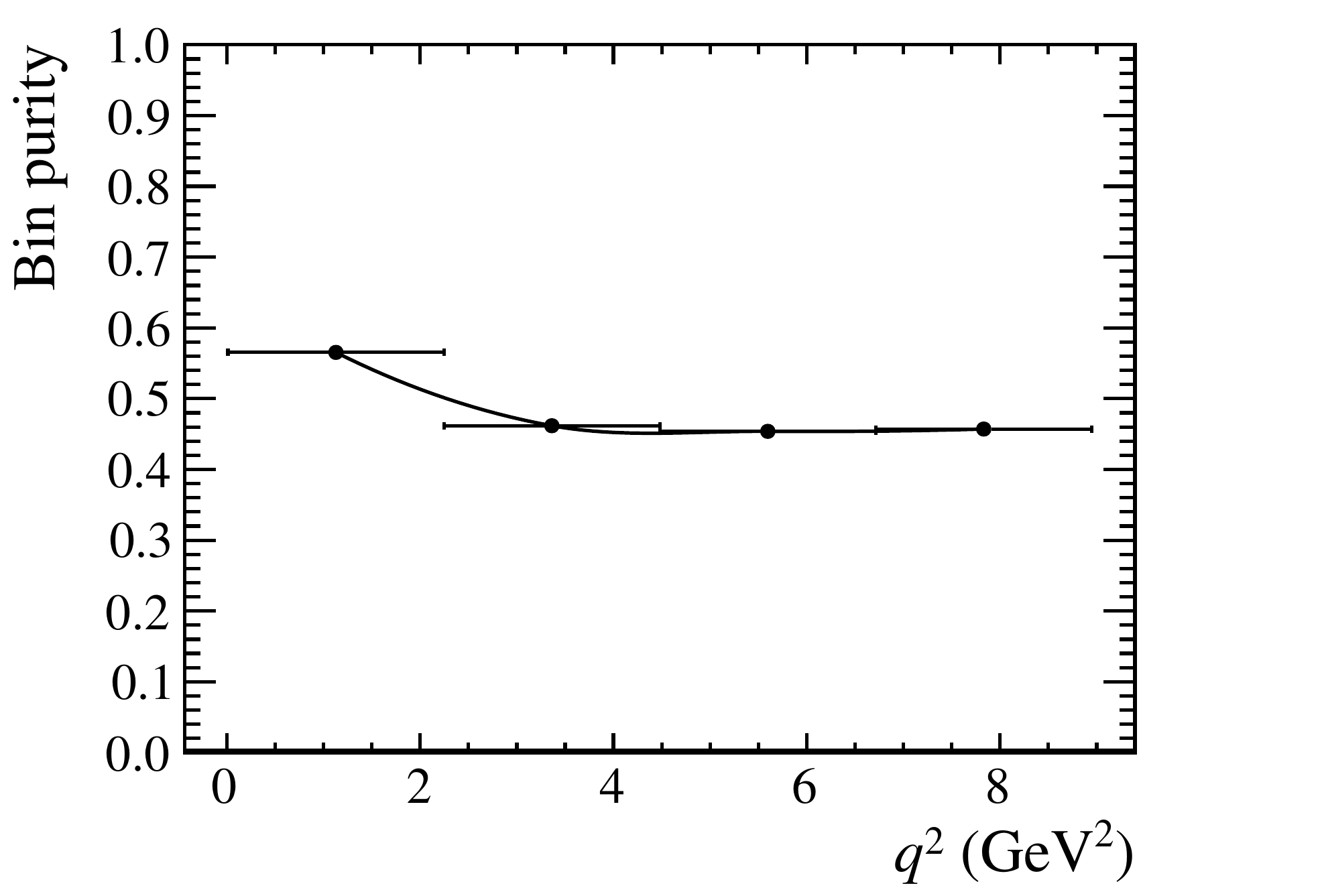}
\includegraphics[width=0.45\linewidth]{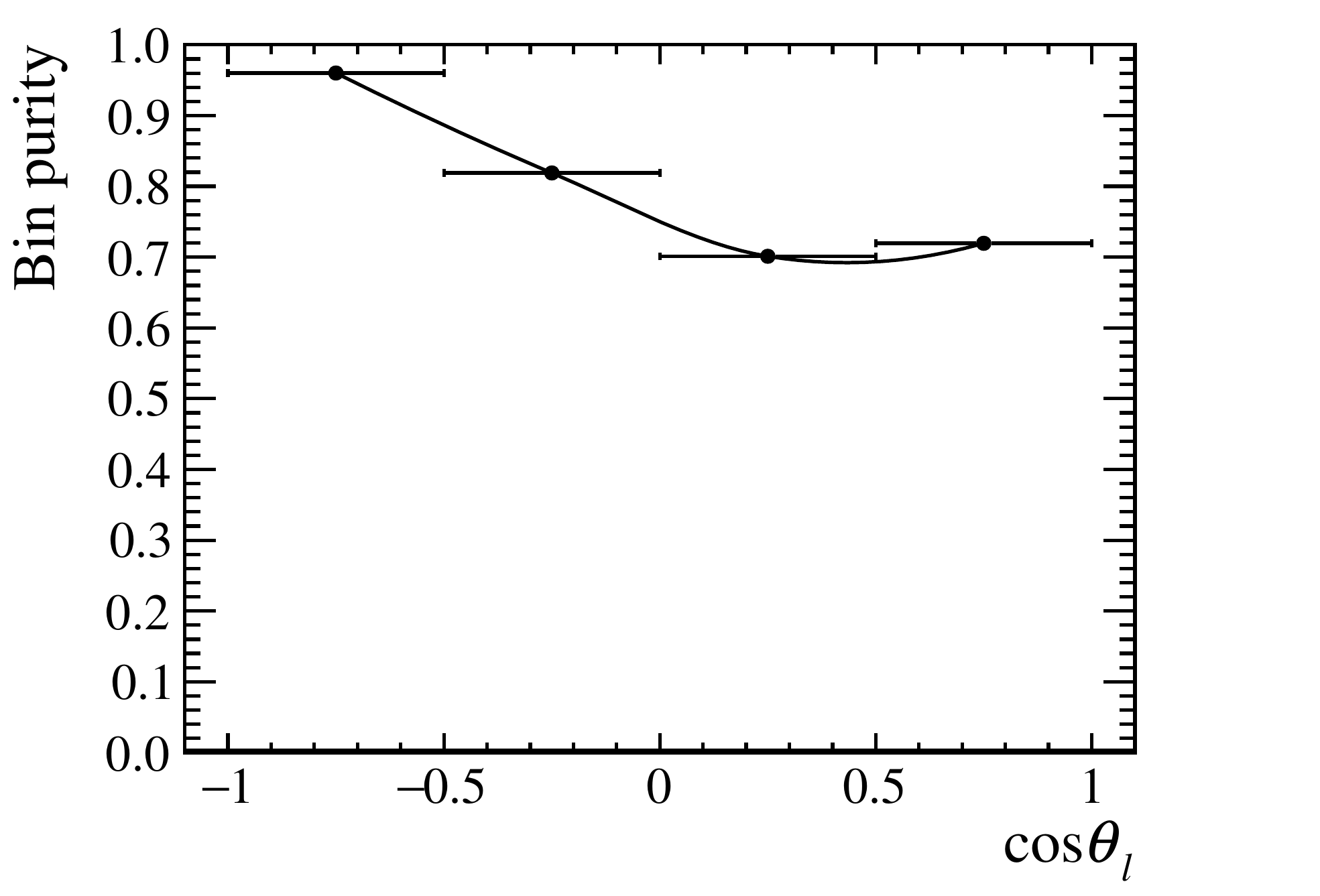}
\caption{
Purity as a function of \qq and \costhetal, defined as the fraction of candidates which belong in a particular kinematic bin. The purity for \costhetal is better than for \qq due to the better resolution.
\label{fig:purities}
}
\end{figure}

The resulting purities with 4 \qq bins and 4 \costhetal bins are shown in Fig.~\ref{fig:purities}, where the purity is defined as the fraction of the number of candidates reconstructed correctly for a given \qq bin. There is a better purity at negative \costhetal, which is due to the interplay between \qq and \costhetal: at high \qq the \costhetal resolution is poor, and in this region there is a positive \costhetal distribution. The resolution limits the number of bins and induces a statistical correlation between neighbouring bins, which is calculated based on the number of candidates which migrate between those two bins. In the $4\times 4$ bins configuration, this correlation is around 10-30\% in both \qq and \costhetal.

In addition to the above, precision measurements of $b$-hadrons branching fractions at the LHC require a well-measured normalisation channel to cancel the uncertainties related to the production. In principle one could normalise to a well measured $B$ meson decay and take the ratio of production fractions. However, this method would inherit substantial systematic uncertainties, and therefore for this study the decay rate is normalised and only the shape information is used to determine the parameters of interest. This means that the absolute normalisation of the form factors cannot be constrained experimentally. As a consequence we do not report any sensitivity for the form factor parameter $\zeta(q^2_\text{max})$, which corresponds to this absolute normalisation.

\subsection{Fits to the differential decay rate}

\begin{figure}[tb]
	\centering
	\includegraphics[width=0.32\linewidth]{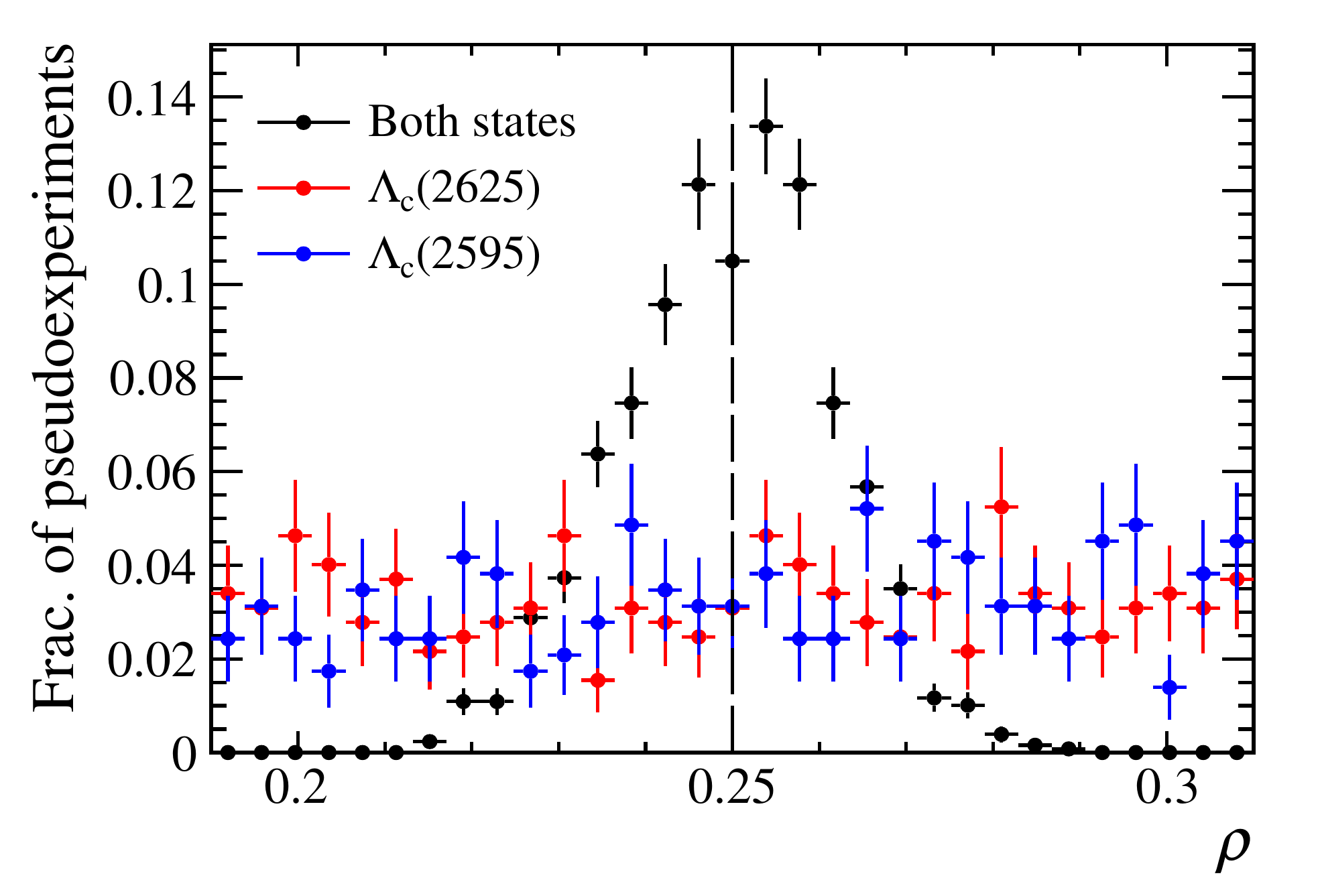}
	\includegraphics[width=0.32\linewidth]{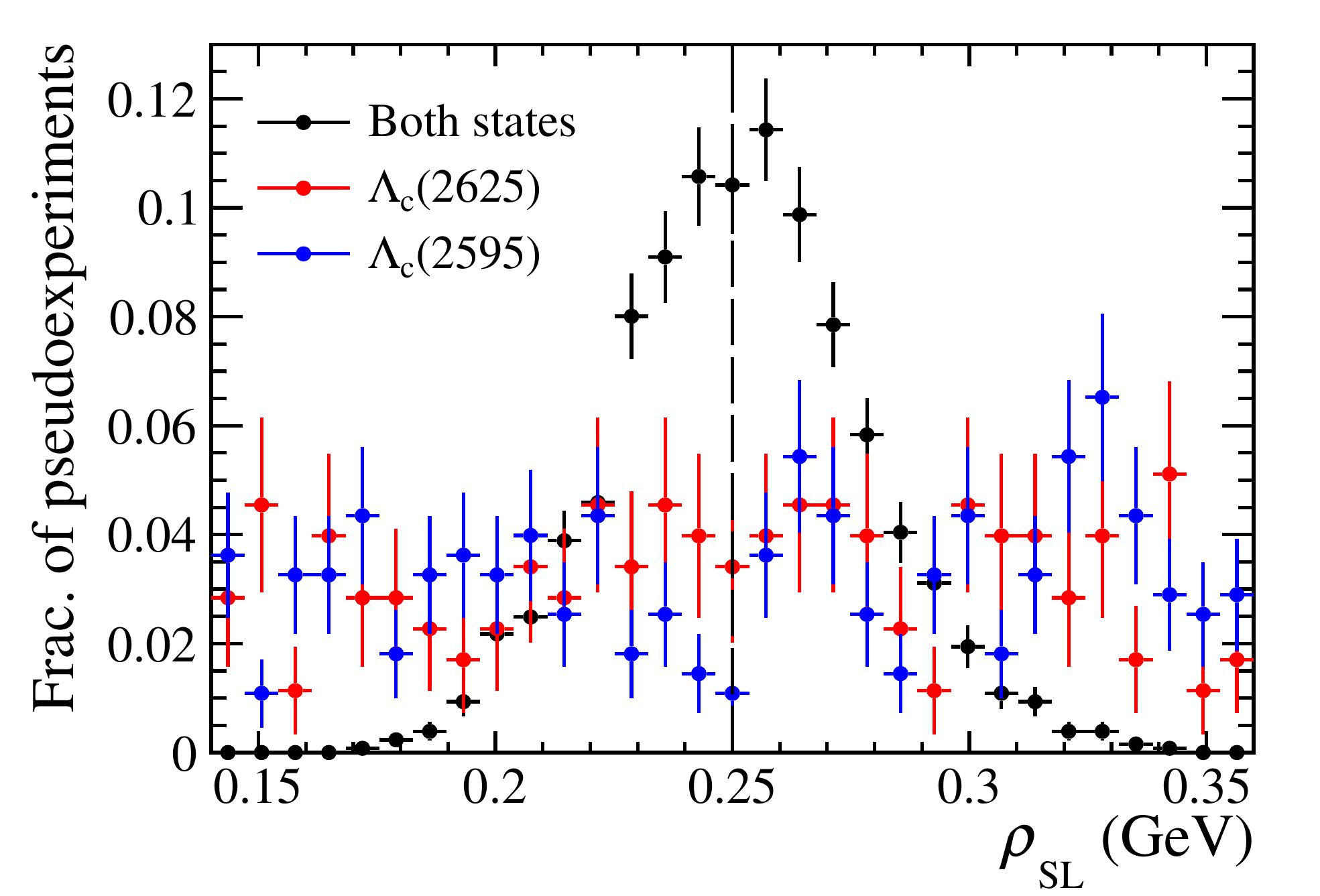}
        \includegraphics[width=0.32\linewidth]{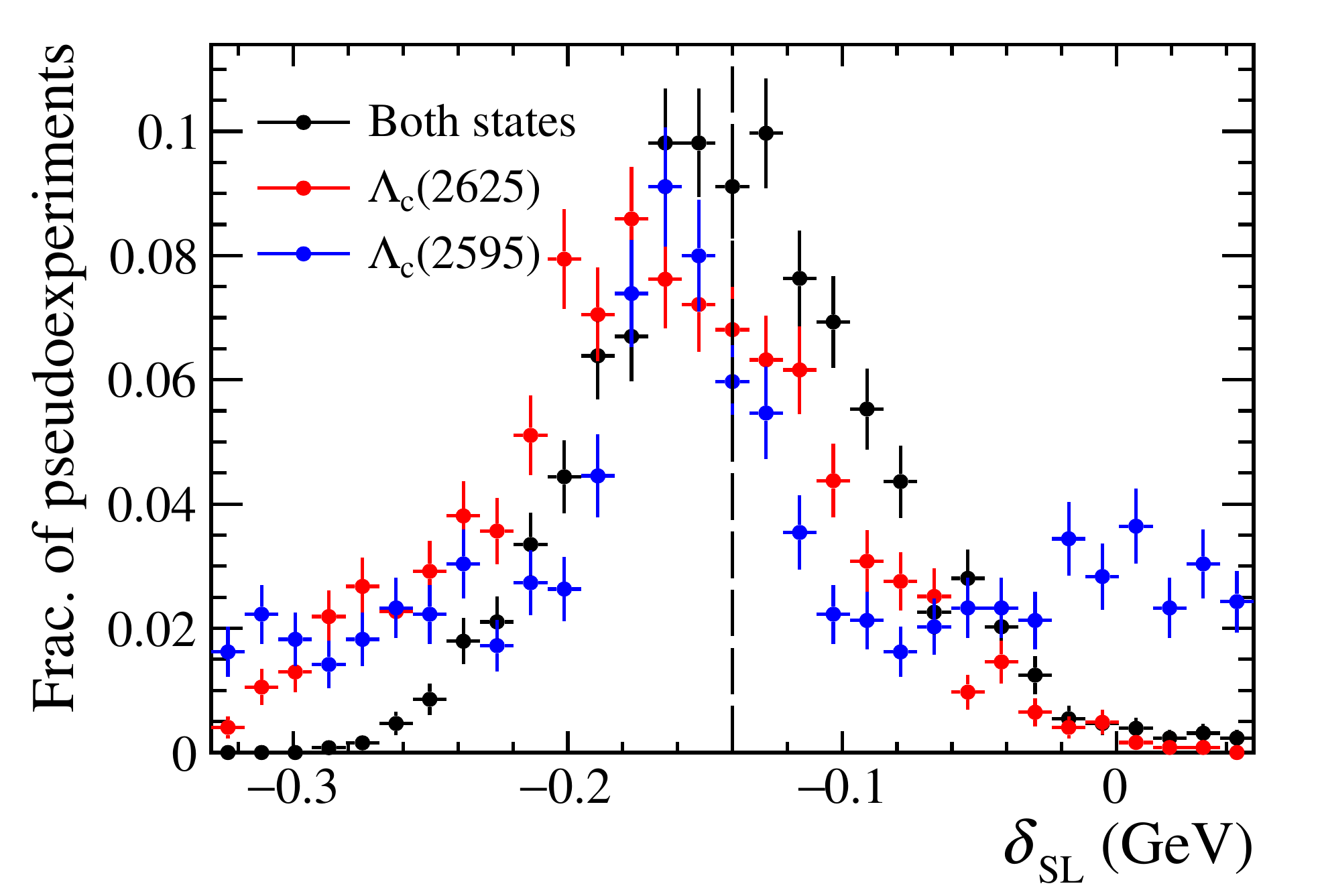}
	\caption{
		Distribution of the IW parameters as fitted from an ensemble of pseudo-experiments. The distributions are shown for the cases when one of the two \LamCstBoth states is fitted, as well as the combination of both. The dashed lines indicate the numerical values of the parameters used to generate the pseudoexperiments.
		\label{fig:states}
	}
\end{figure}


For the purpose of this analysis we fix the two HQE parameters
$\bar{\Lambda} = \mLamB - m_b$ and $\bar{\Lambda}^\prime = \mLamCst - m_c$ in the fits.\footnote{For upcoming experimental analyses, however, we recommend to let these parameters float in order to reflect theoretical ambiguities in their definitions. The
concrete window should reflect the definition of the heavy-quark mass used in the fit.}

We start by fitting the one-dimensional \qq distribution of the \GoodDecay decay,
\BadDecay decay or a combination thereof.
We generate about 300 pseudoexperiments for each parametrisation and benchmark points, and for each pseudoexperiment we generate 50000 \GoodDecay and 20000 \BadDecay events, corresponding to the expected size of the LHCb dataset at the end of the LHC Run II. 
The resulting one-dimensional distributions of the form factor parameters are shown in Fig.~\ref{fig:states} for the benchmark point described in \refeq{bp1}. All benchmark points yield similar results.
When fitting a single decay mode, we find that there is a degeneracy between 
the two slope parameters $\rho$ and $\rho_{\text{SL}}$ due to a strong correlation that is positive for the 
\GoodDecay decay and negative for the \BadDecay decay. Only by combining both states in a single fit can the interference 
between the positive and negative correlation break this degeneracy. 

\begin{figure}[tb]
	\centering
	\includegraphics[width=0.32\linewidth]{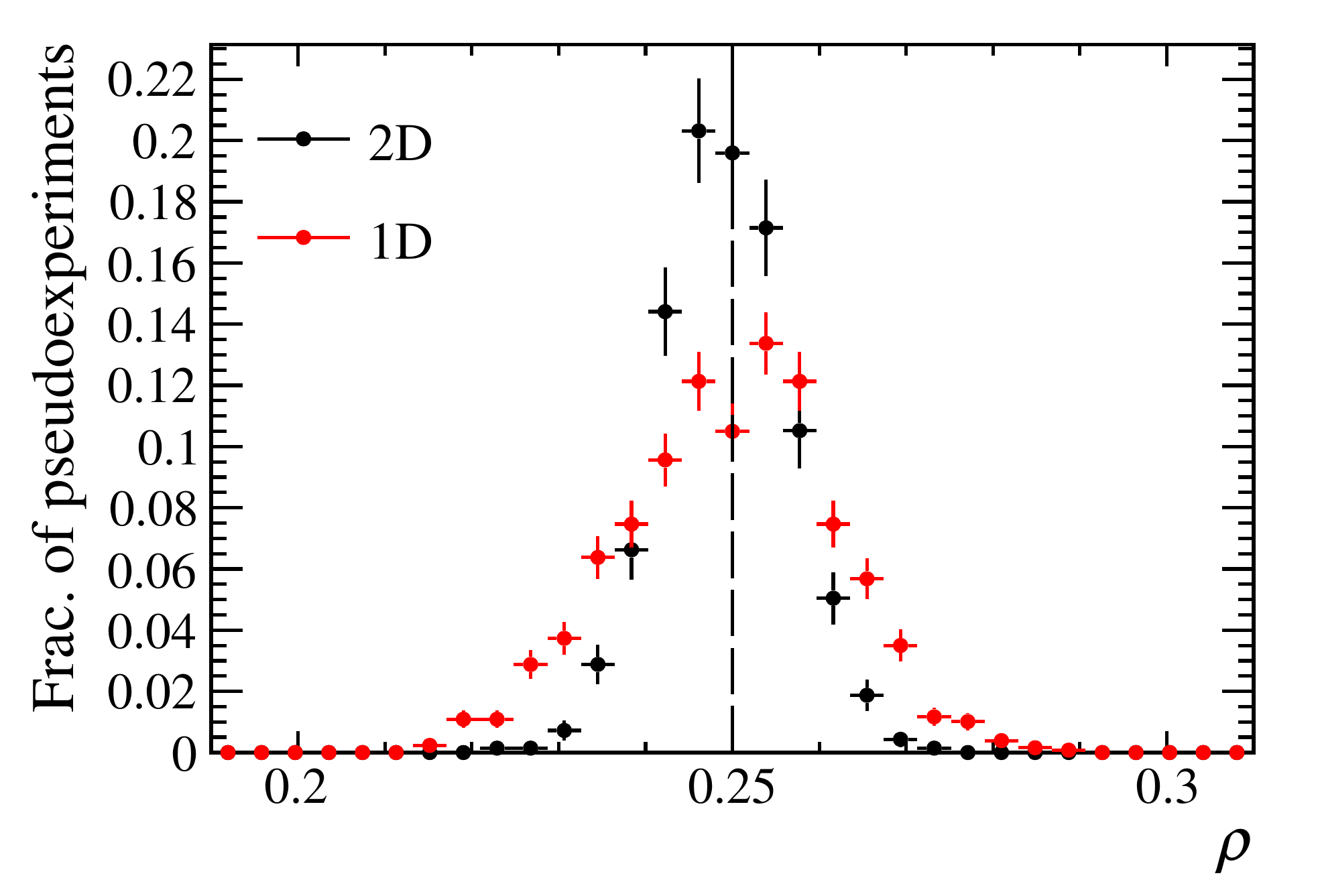}
	\includegraphics[width=0.32\linewidth]{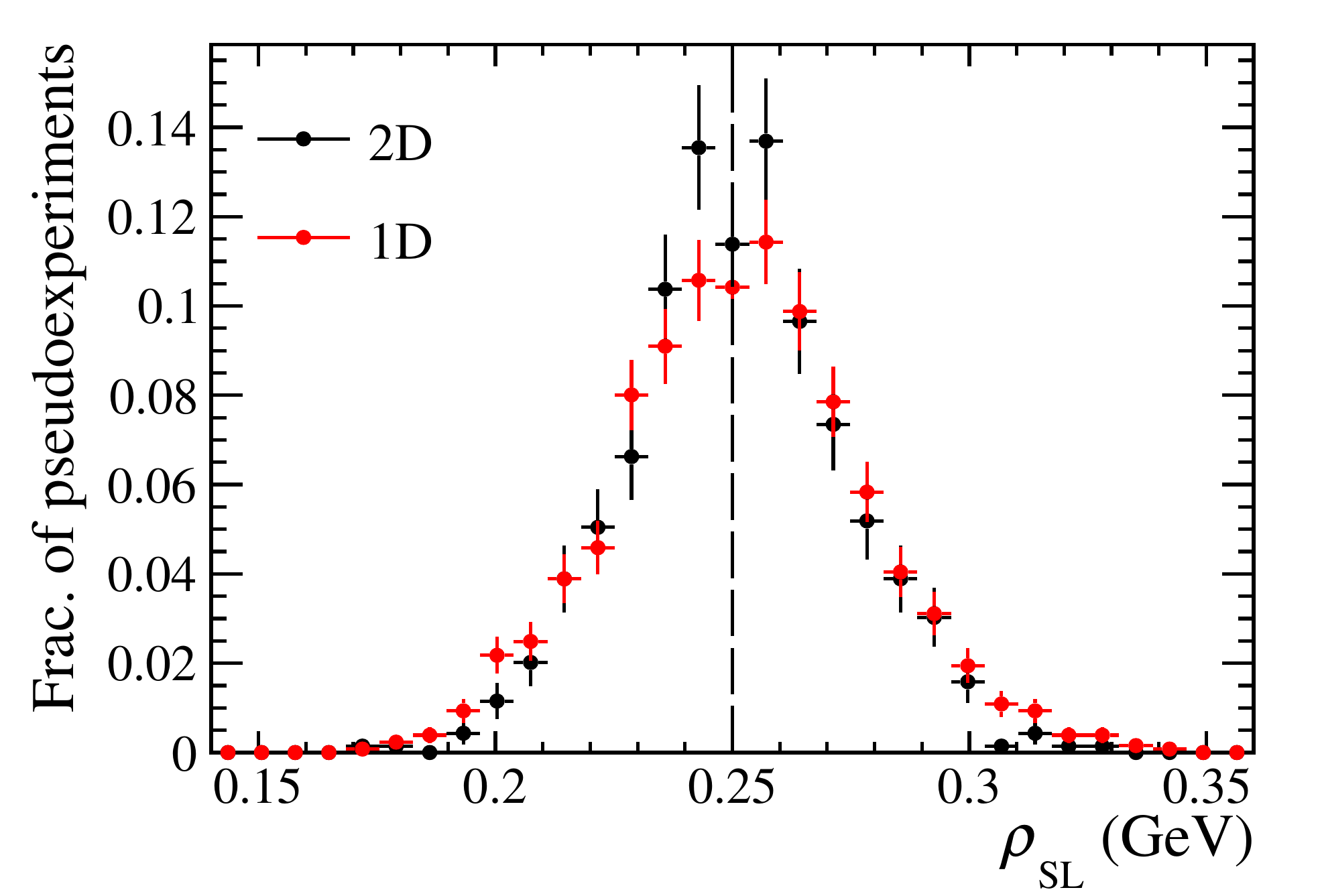}
        \includegraphics[width=0.32\linewidth]{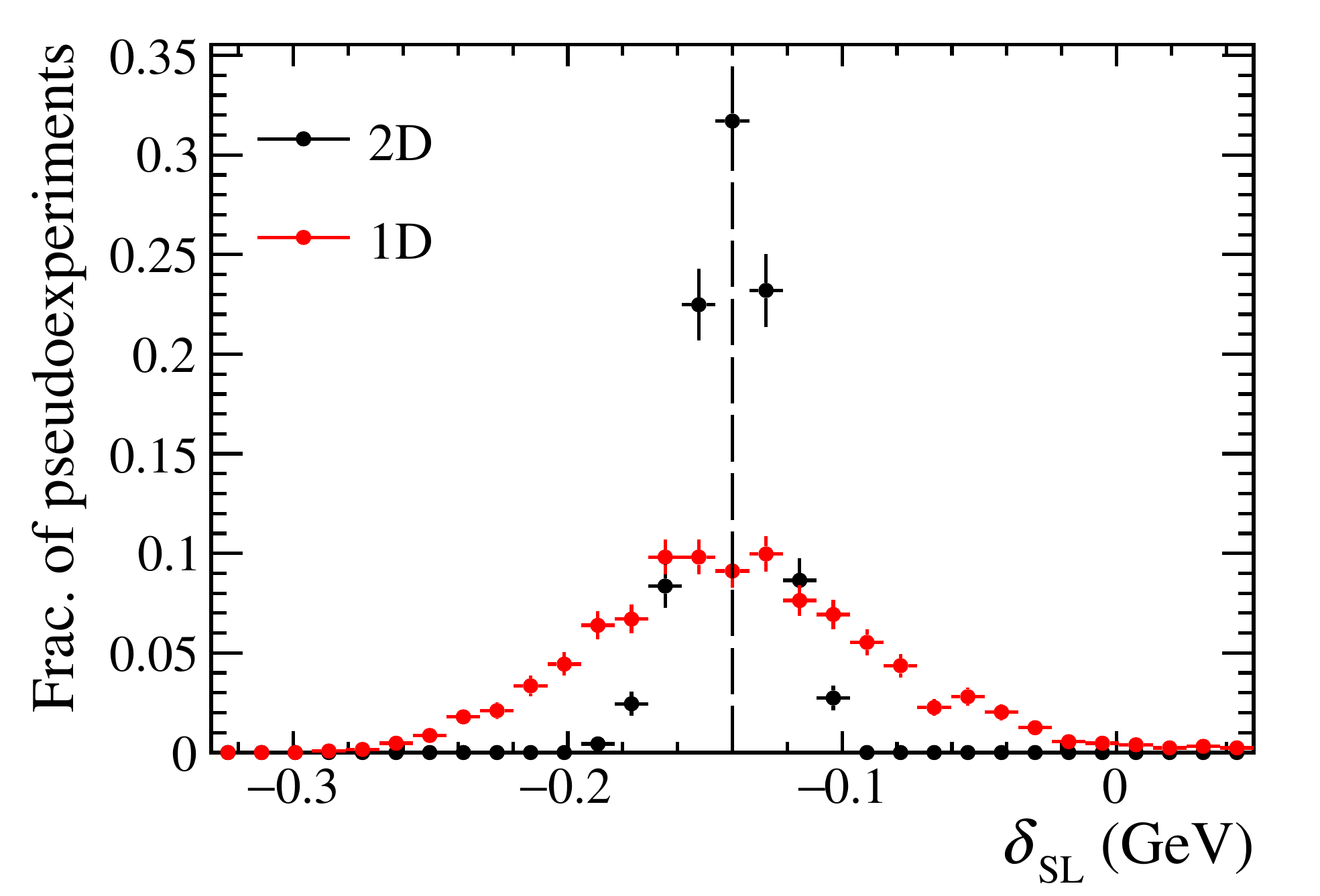}
	\caption{
		Distribution of the IW parameters as fitted from an ensemble of pseudo-experiments. The sensitivity is shown for fits to both the one-dimensional \qq and two-dimensional  $\qq\times \costhetal$ distributions. The dashed lines indicate the numerical values of the parameters used to generate the pseudoexperiments.
		\label{fig:params}
	}
\end{figure}
 
In order to maximise the sensitivity to all three form factor parameters and make full use of the LHCb dataset, we investigate fits to the two-dimensional \qq and \costhetal.
The resulting one-dimensional and two-dimensional distributions of the parameters are shown in Appendix~\ref{app:sens}.
A comparison between the distributions of the IW parameters for the one- and two-dimensional fits is shown in Fig.~\ref{fig:params}. 
The results show that a two-dimensional fit improves the precision on all three parameters with reduced correlations between them, as shown in Fig~\ref{fig:details}.
This strongly motivates a full two-dimensional fit to both \LamCstBoth states simultaneously for any future LHCb analysis to give the best possible precision on the form factor parameters.

\subsection{Projected precision on the $R_{\Lambda_c^{*}}$ predictions}

\begin{figure}[tb]
\centering
\includegraphics[width=0.6\linewidth]{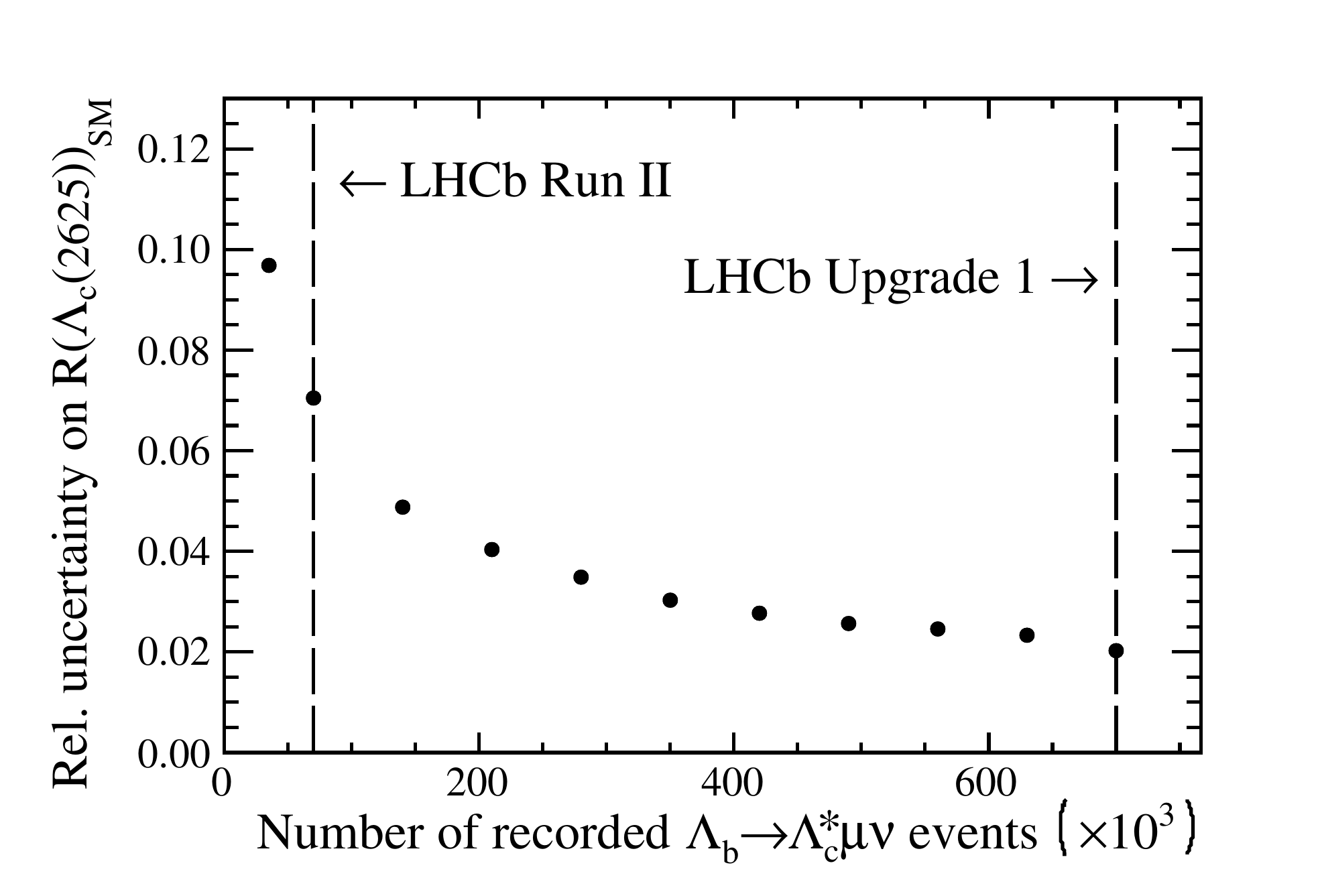}
\caption{
Expected theoretical precision of the $R_{\LamCst[2625]}$ ratio as a function of the amount of $\LamB\to\LamCstBoth\mu\nu$ events recorded by the LHCb experiment. The yields expected at the end of the LHC Run II and after the LHCb upgrade 1 are highlighted by the vertical lines.
\label{fig:result}
}
\end{figure}

Finally, by using the expected precision on the form factors, one can calculate the precision on the ratio $R_{\Lambda_{c}^{*}}$, 
which denotes both the $R_{\LamCst[2595]}$ and $R_{\LamCst[2625]}$ ratios as they are derived from the same parameters 
and therefore have similar uncertainties.
We carry out our study for each of the two paramatrisations of the IW functions given in Sec.~\ref{sec:parametrization}, and each of the common benchmark points defined in Sec.~\ref{sec:benchmark}.
The precision as a function of the luminosity collected by the LHCb experiment is 
shown in Fig.~\ref{fig:result}, where in order to
be conservative and ensure the legibility of our results we only show the worst case of our studies.
Assuming the exponential model\footnote{With exponential model we indicate the exponential parametrisation described in \refsec{parametrization} together with the benchmark points obtained in \refsec{benchmark}.} describes the data well, a statistical precision of $\sim$ 7\% can be expected from run I$+$II data. A
reduction to $\sim$\ 2\% can be expected after upgrade 1 of the LHCb detector. For the linear model, we find in general smaller uncertainties than for the exponential model. Our estimates for the uncertainties ignore power suppressed terms in the HQET expansion
and experimental systematic uncertainties, which could become relevant at that level of precision.

Similar to what has been done in the literature for $R_{D^{*}}$, we can estimate the impact of the dominant 
unknown $1/m_c^2$ corrections to the HQET relations on the theory predictions for the $R_{\Lambda_c^*}$.
Following the discussion \cite{Bigi:2017jbd}, we wish to separate the term involving the timelike form factors from the term
that can be taken directly from data on the semimuonic decay mode. We therefore decompose
\begin{equation}
    \frac{\text{d}\Gamma^{(\tau)}_J}{\text{d} q^2} = \frac{\text{d}\Gamma^{(\tau,1)}_J}{\text{d} q^2} + \frac{\text{d}\Gamma^{(\tau,2)}_J}{\text{d} q^2}
\end{equation}
in two contributions
\begin{align}
    \frac{\text{d}\Gamma^{(\tau,1)}_J}{\text{d} q^2}
        & = \frac{1}{3}\, \left(1 - \frac{m_\tau^2}{q^2}\right)^2\, \left(2 + \frac{m_\tau^2}{q^2}\right)\, \frac{\text{d}\Gamma^{(\ell)}}{q^2}\bigg|_{m_\ell \to 0}\,,\\
    \frac{\text{d}\Gamma^{(\tau,2)}_{J}}{\text{d} q^2}
        & = \begin{cases}
     \phantom{2}\,\Gamma_0^{(\tau)}\, \left[|f_{1/2,t}|^2\, \frac{m_\tau^2}{q^2}\left(\mLamB - \mLamCst\right)^2 + |g_{1/2,t}|^2\, \frac{m_\tau^2}{q^2}\left(\mLamB + \mLamCst\right)^2\right] & J=1/2\\[\smallskipamount]
              2 \,\Gamma_0^{(\tau)}\, \left[|F_{1/2,t}|^2\, \frac{m_\tau^2}{q^2}\left(\mLamB - \mLamCst\right)^2 + |G_{1/2,t}|^2\, \frac{m_\tau^2}{q^2}\left(\mLamB + \mLamCst\right)^2\right] & J=3/2
        \end{cases}\,.
\end{align}
Note here that the $(\tau,1)$ terms are taken directly from data, while the $(\tau,2)$ terms rely on the HQET relations between the
form factors for theoretical predictions.
Correspondingly, we then decompose $R_{\Lambda_c^*} = R_{\Lambda_c^*,1} + R_{\Lambda_c^*,2}$ with
\begin{equation}
    R_{\Lambda_c^*(J),i} = \frac{
            \int_{m_\tau^2}^{(\mLamB - \mLamCst)^2} \text{d}q^2\, \frac{\text{d}\Gamma_J^{(\tau,i)}}{\text{d}q^2}
        }{
            \int_{m_\mu^2 }^{(\mLamB - \mLamCst)^2} \text{d}q^2\, \frac{\text{d}\Gamma_J^{(\mu   )}}{\text{d}q^2}
        }\,.
\end{equation}
We find that the relative contribution by the $(\tau,1)$ term is both dominant and stable under variation
of the slope parameters across our four benchmark points in the exponential model.
We find that
\begin{align}
    R_{\Lambda_c(2595),1} & \simeq 0.76 \cdot R_{\LamCst[2595]}\,, \qquad \text{ and} &
    R_{\Lambda_c(2625),1} & \simeq 0.77 \cdot R_{\LamCst[2625]}\,.
\end{align}
For a conservative estimate, we can assume that the $1/m_c^2$ contributions yield $30\%$ corrections
to the HQET relations as estimated in \cite{Bigi:2017jbd}. Consequently, we would face an inherent theory
uncertainty of $\sim 8\%$ for $R_{\Lambda_c(2595)}$ and up to $\sim 7\%$ for $R_{\Lambda_c(2625)}$.
\footnote{
    Switching the $b$ and $c$ quark mass schemes from the pole to the kinetic scheme
    yields a shift in $R_{\Lambda_c^*}$ by less then $4\%$. The scheme dependence,
    and therefore the values of the heavy-quark expansion parameters $\bar{\Lambda}$
    and $\bar{\Lambda}^\prime$ are presently inconsequential compared to the
    inherent $1/m_c^2$ uncertainty.

}
Given that projected statistical uncertainty in Fig.~\ref{fig:result} are of similar
size already with the full run II dataset, we come to the conclusion that our
theoretical uncertainty estimates strongly motivate dedicated lattice QCD
studies of the $\Lambda_b\to \Lambda_c^*$ form factors.

\section{Conclusion}
\label{sec:conclusion}

Motivated by the recent deviations in LFU in semileptonic $b\to s$ and $b\to c$ decays, we have provided the 
theoretical ingredients needed to constrain the theoretical uncertainty of the lepton universality ratios $R_{\LamCst[2595]}$ and $R_{\LamCst[2625]}$, 
collectively denoted as $R_{\Lambda_{c}^{*}}$. 

To this end, we have improved and extended upon the work in~\cite{Leibovich:1997az}. We provide a 
new definition of the hadronic form factors, convenient for the decay observables, and work out formulae for $\mathcal{O}(\alpha_s)$ corrections to HQE. 
We then propose a 
parameterisation of the Isgur-Wise function informed from previous studies on the ground 
state $\Lambda_b^0 \to \Lambda_c^{+}$ transition~\cite{Jenkins:1992se} and 
perform a zero recoil sum rule to provide a benchmark point for these parameters 
to be used in a study of the sensitivity to these parameters for a future analysis of LHCb data. Last but not least, we provide the finite lepton mass 
terms for the two double differential decay distributions.

We investigated the benefits of fitting the two-dimensional $q^{2}-\cos\theta_{l}$ distribution 
over fitting only the $q^{2}$ distribution, for either of the \LamCstBoth 
hadronic states and their combination. We find that fitting the 
angular information in addition to the $q^{2}$ spectrum is crucial to obtain sensitivity
 to the sub-leading Igsur-Wise function. In addition, we stress that a combined analysis of both \LamCstBoth 
 states is necessary to break the degeneracy between the slopes of the leading and 
 sub-leading Igsur-Wise functions. Finally, we show that by measuring
 the differential decay rate of $\Lambda_b^{0} \to \Lambda_{c}^{*+} \mu^{-} \bar{\nu}$, 
 small statistical uncertainty for a data driven determination of the $R_{\Lambda_{c}^{*}}$ ratios can be achieved.
Our results therefore motivate an LHCb analysis of the $\Lambda_b^{0} \to
\Lambda_{c}^{*+} \mu^{-} \bar{\nu}$ double-differential decay rate and the
subsequent experimental measurement of the $R_{\Lambda_{c}^{*}}$ ratios.  On
the other hand, we also demonstrate that the unknown $1/m^2$ terms in the form
factors' expansion produce at present an irreducible uncertainty that is of the
same order as the statistical uncertainty. This motivates further theoretical
studies of the form factors, e.g. from lattice QCD.

\acknowledgments

We thank Marcin Chrzaszcz, Gino Isidori, Zoltan Ligeti, and Nicola Serra for helpful and enjoyable discussions.
P.B.\ acknowledges support by the Bundesministerium f\"ur Bildung und Forschung (BMBF), and by the Deutsche Forschungsgemeinschaft (DFG) within Research Unit FOR 1873 (Quark Flavour Physics and Effective Field Theories).
D.v.D.\ gratefully acknowledges partial support by the Swiss National Science Foundation (SNF) under contract
200021-159720, and by the Deutsche Forschungsgemeinschaft (DFG) within the Emmy Noether programme under grant
DY 130/1-1 and through the DFG Collaborative Research Center 110 ``Symmetries and the Emergence of Structure in QCD''.
The work of M.B.\ and P.O.\ is supported in part by the Swiss National Science Foundation (SNF) under contracts
200021-159720 and BSSGI0\_155990, respectively.
The authors gratefully acknowledge the compute and data resources provided by the Leibniz Supercomputing Centre.

\bibliography{references,LHCb-PAPER}

\appendix

\section{Details on the Rarita-Schwinger object}
\label{app:RSspinor}

We describe a $J^P = 3/2^-$ state by the spin-$3/2$ projection $u^\alpha$ of a generic
Rarita-Schwinger object $u^\alpha_\text{RS}(k, \eta) = \eta^\alpha u(k)$,
\begin{equation}
\begin{aligned}
    u_{(3/2)}^\alpha(k, \eta, s_c)
        & = \left[\eta^\alpha - \frac{1}{3}\left(\gamma^\alpha + \frac{k^\alpha}{\mLamCst}\right)\slashed{\eta}\right] u(k, s_c)\\
        & = \left[g^\alpha{}_\beta - \frac{1}{3}\left(\gamma^\alpha + \frac{k^\alpha}{\mLamCst}\right) \gamma_\beta\right]\, u_\text{RS}^\beta(k, \eta(\lambda), s_c)\\
        & \equiv \left[P_{3/2}\right]^{\alpha}{}_\beta\, u_\text{RS}^\beta(k, \eta(\lambda), s_c)\,.
\end{aligned}
\end{equation}
In the above, $u(k, s_c)$ denotes a spin-$1/2^+$ spinor of four momentum $k$ and
rest-frame helicity $s_c = \pm 1/2$, and $\eta$ denotes a polarisation vector with $J^P = 1^-$. Likewise, we can also characterise the $J^{P}=1/2^-$ state in term of the projection onto the spin-$1/2$ component as:
\begin{align}
u^{\alpha}_{(1/2)}(k,\eta,s_c)
&=\frac{1}{3}\left[\gamma^{\alpha}+\frac{k^{\alpha}}{\mLamCst}\right]\slashed{\eta} \ u(k,s_c) \\
&=\frac{1}{3}\left[\gamma^{\alpha}+\frac{k^{\alpha}}{\mLamCst}\right]\gamma_{\beta} \ u_{\mathrm{RS}}^{\beta}(k,\eta(\lambda),s_c)  \\
       & \equiv \left[P_{1/2}\right]^{\alpha}{}_\beta\, u_\text{RS}^\beta(k, \eta(\lambda), s_c)\,.
       \end{align}
The Rarita-Schwinger object fulfills the equation of motion
\begin{equation}
    \left[i \eps_{\mu\alpha\beta\sigma} \gamma^5 \gamma^\mu k^\sigma + i m\,\sigma_{\alpha\beta}\right] u^\beta(k) = 0\,.
\end{equation}

By virtue of the equations of motions, the following identities hold
\begin{align}
    k^\alpha u^{\mathrm{RS}}_\alpha(k, \eta, s_c) & = 0 = \eta(t)^\alpha u_\alpha^{\mathrm{RS}}(k, \eta, s_c)\,
    \end{align}
 while for the spin $3/2$ projection $u_\alpha$ of a Rarita-Schwinger object, the following relations are also true:
    \begin{align}
    \gamma^\alpha u^{(3/2)}_\alpha(k, \eta, s_c) & = 0\,,\\
    -i \sigma^{\alpha\beta} \, u^{(3/2)}_\alpha(k, \eta, s_c) & = u_{(3/2)}^\beta(k, \eta, s_c)\,.
\end{align}

The completeness relation for the $3/2$ spinor read
\begin{equation}
    \sum_{\lambda('),s_c(')} u_{3/2}^\alpha(k, \eta(\lambda), s_c) \bar{u}_{3/2}^{\alpha'}(k, \eta(\lambda'), s_c')
    = (\slashed{k} + \mLamCst) \left[-g^{\alpha\alpha'} + \frac{k^\alpha k^{\alpha'}}{\mLamCst^2} + \frac{1}{3} \left(\gamma^\alpha - \frac{k^\alpha}{\mLamCst}\right)\left(\gamma^{\alpha'} + \frac{k^{\alpha'}}{\mLamCst}\right)\right] \, ,
\end{equation}
while for the $1/2$ spinor we have:
\begin{equation}
  \sum_{\lambda('),s_c(')} u_{1/2}^\alpha(k, \eta(\lambda), s_c) \bar{u}_{1/2}^{\alpha'}(k, \eta(\lambda'), s_c')
    = -\frac{1}{3}(\slashed{k}+\mLamCst)\left(\gamma^{\alpha}-\frac{k^\alpha}{\mLamCst}\right)\left(\gamma^{\alpha'}+\frac{k^{\alpha'} }{\mLamCst}\right)
\end{equation}

\section{Details on the form factor definitions}
\label{app:ff-details}
The spin structures $\Gamma^{\alpha\mu}_{J,i}$ that contribute to the transition $\Lambda_b\to\Lambda_c^{*}$ are listed in the following. \\ 
For the final state $\LamCst[2595]$ and for the vector current ($J=V$) we find :
\begin{equation}
\begin{aligned}
\label{eq:FF-12-basis-V}
    \Gamma_{V,(1/2,t)}^{\alpha\mu} & =  \frac{\sqrt{4 \mLamB \mLamCst}}{\sqrt{s_+}}\frac{2 \mLamCst}{\sqrt{s_ +s_-}} p^\alpha\,\frac{\mLamB - \mLamCst}{\sqrt{q^2}} \, \frac{q^\mu}{\sqrt{q^2}}\,,\\
\Gamma_{V,(1/2,0)}^{\alpha\mu}     & = \frac{\sqrt{4 \mLamB \mLamCst}}{\sqrt{s_-}} \frac{2 \mLamCst}{\sqrt{s_ +s_-}} p^\alpha\,\frac{\mLamB + \mLamCst}{s_+}\left[(p + k)^\mu - \frac{\mLamB^2 - \mLamCst^2}{q^2} q^\mu\right]\,,\\
\Gamma_{V,(1/2,\perp)}^{\alpha\mu} & = \frac{\sqrt{4 \mLamB \mLamCst}}{\sqrt{s_-}} \frac{2 \mLamCst}{\sqrt{s_ +s_-}} p^\alpha\,\left[\gamma^\mu - \frac{2 \mLamCst}{s_+} p^\mu - \frac{2 \mLamB}{s_+} k^\mu\right]\,,
\end{aligned}
\end{equation}
while for the axialvector current ($J=A$) we obtain:
\begin{equation}
\begin{aligned}
\label{eq:FF-12-basis-A}
    \Gamma_{A,(1/2,t)}^{\alpha\mu} & =  \frac{\sqrt{4 \mLamB \mLamCst}}{\sqrt{s_-}}\frac{2 \mLamCst}{\sqrt{s_ +s_-}} p^\alpha\,\frac{\mLamB + \mLamCst}{\sqrt{q^2}} \, \frac{q^\mu}{\sqrt{q^2}}\,,\\
\Gamma_{A,(1/2,0)}^{\alpha\mu}     & =  \frac{\sqrt{4 \mLamB \mLamCst}}{\sqrt{s_+}}\frac{2 \mLamCst}{\sqrt{s_ +s_-}} p^\alpha\,\frac{\mLamB - \mLamCst}{s_-}\left[(p + k)^\mu - \frac{\mLamB^2 - \mLamCst^2}{q^2} q^\mu\right]\,,\\
\Gamma_{A,(1/2,\perp)}^{\alpha\mu} & =  \frac{\sqrt{4 \mLamB \mLamCst}}{\sqrt{s_+}}\frac{2 \mLamCst}{\sqrt{s_ +s_-}} p^\alpha\,\left[\gamma^\mu + \frac{2 \mLamCst}{s_-} p^\mu - \frac{2 \mLamB}{s_-} k^\mu\right]\,.
\end{aligned}
\end{equation}
In the case of the final state $\LamCst[2625]$, for the vector current ($J=V$) we obtain:
\begin{equation}
\begin{aligned}
\label{eq:FF-32-basis-V}
    \Gamma_{V,(1/2,t)}^{\alpha\mu} & = \frac{\sqrt{4 \mLamB \mLamCst}}{\sqrt{s_+}}\, \frac{2 \mLamCst}{\sqrt{s_ +s_-}} p^\alpha\,\frac{\mLamB - \mLamCst}{\sqrt{q^2}} \, \frac{q^\mu}{\sqrt{q^2}}\,,\\
\Gamma_{V,(1/2,0)}^{\alpha\mu}     & = \frac{\sqrt{4 \mLamB \mLamCst}}{\sqrt{s_-}}\, \frac{2 \mLamCst}{\sqrt{s_ +s_-}} p^\alpha\,\frac{\mLamB + \mLamCst}{s_+}\left[(p + k)^\mu - \frac{\mLamB^2 - \mLamCst^2}{q^2} q^\mu\right]\,,\\
\Gamma_{V,(1/2,\perp)}^{\alpha\mu} & = \frac{\sqrt{4 \mLamB \mLamCst}}{\sqrt{s_-}}\, \frac{2 \mLamCst}{\sqrt{s_ +s_-}} p^\alpha\,\left[\gamma^\mu - \frac{2 \mLamCst}{s_+} p^\mu - \frac{2 \mLamB}{s_+} k^\mu\right]\,,\\
\Gamma_{V,(3/2,\perp)}^{\alpha\mu} & = \frac{\sqrt{4 \mLamB \mLamCst}}{\sqrt{s_-}}\, \frac{-4 i \eps^{\alpha\mu p k}}{\sqrt{s_+ s_-}} \gamma_5 + \Gamma_{V,(1/2,\perp)}\,,
\end{aligned}
\end{equation}
while for the axialvector current ($J=A$) we use
\begin{equation}
\begin{aligned}
\label{eq:FF-32-basis-A}
    \Gamma_{A,(1/2,t)}^{\alpha\mu} & = \frac{\sqrt{4 \mLamB \mLamCst}}{\sqrt{s_-}}\, \frac{2 \mLamCst}{\sqrt{s_ +s_-}} p^\alpha\,\frac{\mLamB + \mLamCst}{\sqrt{q^2}} \, \frac{q^\mu}{\sqrt{q^2}}\,,\\
\Gamma_{A,(1/2,0)}^{\alpha\mu}     & = \frac{\sqrt{4 \mLamB \mLamCst}}{\sqrt{s_+}}\, \frac{2 \mLamCst}{\sqrt{s_ +s_-}} p^\alpha\,\frac{\mLamB - \mLamCst}{s_-}\left[(p + k)^\mu - \frac{\mLamB^2 - \mLamCst^2}{q^2} q^\mu\right]\,,\\
\Gamma_{A,(1/2,\perp)}^{\alpha\mu} & = \frac{\sqrt{4 \mLamB \mLamCst}}{\sqrt{s_+}}\, \frac{2 \mLamCst}{\sqrt{s_ +s_-}} p^\alpha\,\left[\gamma^\mu + \frac{2 \mLamCst}{s_-} p^\mu - \frac{2 \mLamB}{s_-} k^\mu\right]\,,\\
\Gamma_{A,(3/2,\perp)}^{\alpha\mu} & = \frac{\sqrt{4 \mLamB \mLamCst}}{\sqrt{s_+}}\, \frac{-4 i \eps^{\alpha\mu p k}}{\sqrt{s_+ s_-}} \gamma_5 - \Gamma_{A,(1/2,\perp)}\,.
\end{aligned}
\end{equation}
Note that we adopted the convention $\eps^{0123} = - \eps_{0123} = +1$ for the Levi-Civita tensor.\\

In the above a recurring term fulfills
\begin{equation}
    \bar{u}_\alpha(k) \frac{-2 \mLamCst}{\sqrt{s_ +s_-}} p^\alpha = \bar{u}_\alpha(k) \eta^\alpha(0)\,.
\end{equation}
To conclude, we also provide the matching between our form factor definitions and the ones in \cite{Leibovich:1997az}:
\begin{equation}
\label{eq:conversion-LS}
\begin{aligned}
F_{1/2,t}(q^2(w))=& \  +\frac{\sqrt{w-1}}{\sqrt{2}(r-1)}(w+1)\left[(r-1)l_{V_1}+(rw-1)l_{V_2}+(r-w)l_{V_3}-l_{V_4}\right]\,,\\
F_{1/2,0}(q^2(w))=& \ +\frac{\sqrt{w+1}}{\sqrt{2}(1+r)} \left[(r+1)(w-1)l_{V_1}+(w^2-1)(r l_{V_2}+l_{V_3})+(w-r)l_{V_4}\right] \,,\\
F_{1/2,\perp}(q^2(w))=& \  -\frac{\sqrt{w+1}}{2\sqrt{2}} \left[2 (1-w)l_{V_1}+l_{V_4}\right]\,,\\
F_{3/2,\perp}(q^2(w))=& \ -\frac{\sqrt{w+1}}{2\sqrt{2}}\, l_{V_4}\,,\\
G_{1/2,t}(q^2(w))=& \  +\frac{\sqrt{w+1}}{\sqrt{2}(r+1)}(w-1)\left[(r+1)l_{A_1}+(rw-1)l_{A_2}+(r-w)l_{A_3}-l_{A_4}\right]\,,\\
G_{1/2,0}(q^2(w))=& \ +\frac{\sqrt{w-1}}{\sqrt{2}(1-r)} \left[(r-1)(w+1)l_{A_1}+(w^2-1)(r l_{A_2}+l_{A_3})+(w-r)l_{A_4}\right]\,,\\
G_{1/2,\perp}(q^2(w))=& \ -\frac{\sqrt{w-1}}{2\sqrt{2}} \left[-2 (1+w)l_{A_1}+l_{A_4}\right]\,,\\
G_{3/2,\perp}(q^2(w))=& \ +\frac{\sqrt{w-1}}{2\sqrt{2}}\, l_{A_4}\,,\\
\end{aligned}
\end{equation}

with $r=\mLamCst/\mLamB$. \\
We worked out the matching between our convention and \cite{Leibovich:1997az} also for the form factors of $\Lambda_b\to\LamCst[2595]$ transitions. This is slightly more involved since our approach and the approach of \cite{Leibovich:1997az} for the spin $1/2^{-}$ projection of the Rarita-Schwinger object differ. We find it convenient to use:
\begin{equation}
\label{eq:matching_12_spinor}
\sum_{\lambda'_c, s'_c} C_{\lambda'_c, s_c'}^{1/2, s_c}\, \bar{u}_{\alpha}^{(1/2)}(k, \eta(\lambda'_c), s'_c) p^{\alpha}= -\frac{1}{\sqrt{3}}\bar{u}(k,s_c)\gamma^5 \left(\frac{1}{\mLamCst}k\cdot q+\slashed{q}\right)\,,
\end{equation}
with the $C_{\lambda'_c, s'_c}^{1/2, s_c}$ being the Clebsch-Gordan coefficients for $j_1 \oplus j_2 = 1 \oplus 1/2$
angular momentum.
Using \refeq{matching_12_spinor}, the matching between our form factors for the $\Lambda_b\to\LamCst[2595]$ transition and the ones in \cite{Leibovich:1997az} reads:
\begin{equation}
\begin{aligned}
f_{1/2,t}(q^2(w))=& \ +\sqrt{\frac{3}{2}}\frac{\sqrt{w-1}}{r-1}\left[(r+1) d_{V_1} +(rw-1)d_{V_2}+(r-w)d_{V_3}\right]\,,\\
f_{1/2,0}(q^2(w))=& \ +\sqrt{\frac{3}{2}}\frac{\sqrt{w+1}}{r+1}\left[(r-1)d_{V_1}+(w-1)(r d_{V_2}+d_{V_3})\right] \,,\\
f_{1/2,\perp}(q^2(w))=& \   -\sqrt{\frac{3}{2}}\sqrt{w+1} \,d_{V_1}\,,\\
g_{1/2,t}(q^2(w))=& \ +\sqrt{\frac{3}{2}}\frac{\sqrt{w+1}}{r+1}\left[(r-1) d_{A_1} +(rw-1)d_{A_2}+(r-w)d_{A_3}\right]\,,\\
g_{1/2,0}(q^2(w))=& \  +\sqrt{\frac{3}{2}}\frac{\sqrt{w-1}}{r-1}\left[(r+1)d_{A_1}+(w+1)(r d_{A_2}+d_{A_3})\right] \,,\\
g_{1/2,\perp}(q^2(w))=& \   -\sqrt{\frac{3}{2}}\sqrt{w-1} \,d_{A_1}\,.
\end{aligned}
\end{equation}
\section{Helicity Amplitudes}
\label{app:helamp}
\subsection{$1/2^+\rightarrow 1/2^-$}
For the scalar current, defined as
\begin{equation}
    h^\alpha_S(s_b, s_c, \lambda_c) \equiv \bar{u}^\alpha(k, \eta(\lambda_c), s_c) u(p, s_b)\,,
\end{equation}
we find the following non vanishing terms:
\begin{align}
    \frac{1}{ \sqrt{2}}   h^\alpha_S(-1/2, -1/2, +1) =h^\alpha_S(-1/2, +1/2, 0) & = \frac{\sqrt{2}}{3} \sqrt{s_+} \eta^{*\alpha}(+1) \,,\\
 \frac{1}{ \sqrt{2}} h^\alpha_S(+1/2, +1/2, -1) = h^\alpha_S(+1/2, -1/2, 0) & = \frac{\sqrt{2}}{3} \sqrt{s_+} \eta^{*\alpha}(-1) \,,\\
  -  h^\alpha_S(+1/2, -1/2,+1) = \sqrt{2} h^\alpha_S(+1/2, +1/2, 0) & = \frac{\sqrt{2}}{3} \sqrt{s_+} \eta^{*\alpha}(0) \,,\\
  -  h^\alpha_S(-1/2, +1/2,-1) = \sqrt{2} h^\alpha_S(-1/2, -1/2, 0) & = \frac{\sqrt{2}}{3} \sqrt{s_+} \eta^{*\alpha}(0) \,.
\end{align}
For the pseudoscalar current, defined as
\begin{equation}
    h^\alpha_P(s_b, s_c, \lambda_c) \equiv \bar{u}^\alpha(k, \eta(\lambda_c), s_c) \gamma_5 u(p, s_b)\,,
\end{equation}
one finds:
\begin{align}
\frac{1}{\sqrt{2}} h^\alpha_P(-1/2, -1/2, +1) = -h^\alpha_P(-1/2, +1/2,  0) & = + \frac{\sqrt{2}}{3} \sqrt{s_-} \eta^{*\alpha}(+1) \,,\\
 -\frac{1}{\sqrt{2}} h^\alpha_P(+1/2, +1/2, -1) = h^\alpha_P(+1/2, -1/2,  0) & = +\frac{\sqrt{2}}{3} \sqrt{s_-} \eta^{*\alpha}(-1) \,,\\
    h^\alpha_P(+1/2, -1/2, +1) =- \sqrt{2}h^\alpha_P(+1/2, +1/2,  0) & = +\frac{\sqrt{2}}{3} \sqrt{s_-} \eta^{*\alpha}(0) \,,\\
   - h^\alpha_P(-1/2, +1/2, -1) = \sqrt{2}h^\alpha_P(-1/2, -1/2,  0) & = + \frac{\sqrt{2}}{3} \sqrt{s_-} \eta^{*\alpha}(0) \,.
\end{align}

For the vector current
\begin{equation}
    h^\alpha_{V,\lambda_q}(s_b, s_c, \lambda_c)
        \equiv \bar{u}^\alpha(k, \eta(\lambda_c), s_c) \slashed{\eps}^*(\lambda_q) u(p, s_b)\,,
\end{equation}
we identify
\begin{equation}
    h^\alpha_{V,t}(s_b, s_c, \lambda_c) = \frac{\mLamB - \mLamCst}{\sqrt{q^2}} h^\alpha_{S}(s_b, s_c, \lambda_c)\,.
\end{equation}
For the transverse polarisation we find:
\begin{align}
-\frac{1} {\sqrt{2} }h^\alpha_{V,-1}(+1/2, -1/2, +1) = h^\alpha_{V,-1}(+1/2, +1/2, 0) & = + \frac{2}{3} \sqrt{s_-} \eta^{*\alpha}(+1) \,,\\
  -\frac{1}{\sqrt{2}} h^\alpha_{V,+1}(-1/2, +1/2, -1) = h^\alpha_{V,+1}(-1/2, -1/2, 0) & = + \frac{2}{3} \sqrt{s_-} \eta^{*\alpha}(-1) \,,\\
    h^\alpha_{V,+1}(-1/2, -1/2, +1) =- \sqrt{2} h^\alpha_{V,+1}(-1/2, +1/2, 0) & = +\frac23 \sqrt{s_-} \eta^{*\alpha}(0) \,, \\
    h^\alpha_{V,-1}(+1/2, +1/2, -1) = -\sqrt{2} h^\alpha_{V,-1}(+1/2, -1/2, 0) & = +\frac23 \sqrt{s_-} \eta^{*\alpha}(0) \,.
\end{align}
For the longitudinal polarisation we find:
\begin{align}
  \frac{1}{\sqrt{2}} h^\alpha_{V, 0}(-1/2, -1/2, +1) = -h^\alpha_{V, 0}(-1/2, +1/2, 0) & = \frac{\sqrt{2}}{3} \frac{\mLamB + \mLamCst}{\sqrt{q^2}} \sqrt{s_-} \eta^{*\alpha}(+1) \,, \\
\frac{1}{\sqrt{2}} h^\alpha_{V, 0}(+1/2, +1/2, -1) = -h^\alpha_{V, 0}(+1/2, -1/2, 0) & = \frac{\sqrt{2}}{3} \frac{\mLamB + \mLamCst}{\sqrt{q^2}} \sqrt{s_-} \eta^{*\alpha}(-1) \,, \\
   - h^\alpha_{V, 0}(+1/2, -1/2, +1) =\sqrt{2} h^\alpha_{V, 0}(+1/2, +1/2, 0) & = \frac{\sqrt{2}}{3} \frac{\mLamB + \mLamCst}{\sqrt{q^2}} \sqrt{s_-} \eta^{*\alpha}(0) \,, \\
    -h^\alpha_{V, 0}(-1/2, +1/2, -1) = \sqrt{2} h^\alpha_{V, 0}(-1/2, -1/2, 0) & = \frac{\sqrt{2}}{3} \frac{\mLamB + \mLamCst}{\sqrt{q^2}} \sqrt{s_-} \eta^{*\alpha}(0) \,.
\end{align}
Similarly for the axialvector current
\begin{equation}
    h^\alpha_{A,\lambda_q}(s_b, s_c, \lambda_c)
        \equiv \bar{u}^\alpha(k, \eta(\lambda_c), s_c) \slashed{\eps}^*(\lambda_q) \gamma_5 u(p, s_b)\,,
\end{equation}
we identify
\begin{equation}
    h^\alpha_{A,t}(s_b, s_c, \lambda_c) = -\frac{\mLamB + \mLamCst}{\sqrt{q^2}} h^\alpha_{P}(s_b, s_c, \lambda_c)\,.
\end{equation}
For the transverse polarisation we find
\begin{align}
\frac{1}{\sqrt{2}} h^\alpha_{A,-1}(+1/2, -1/2, +1) =- h^\alpha_{A,-1}(+1/2, +1/2, 0) & = + \frac{2}{3} \sqrt{s_+} \eta^{*\alpha}(+1) \,,\\
-\frac{1}{ \sqrt{2} }h^\alpha_{A,+1}(-1/2, +1/2, -1) = h^\alpha_{A,+1}(-1/2, -1/2, 0) & = +\frac{2}{3} \sqrt{s_+} \eta^{*\alpha}(-1) \,,\\
    h^\alpha_{A,+1}(-1/2, -1/2, +1) = -\sqrt{2} h^\alpha_{A,+1}(-1/2, +1/2, 0) & = + \frac23 \sqrt{s_+} \eta^{*\alpha}(0) \,, \\
    -h^\alpha_{A,-1}(+1/2, +1/2, -1) = \sqrt{2} h^\alpha_{A,-1}(+1/2, -1/2, 0) & = + \frac23 \sqrt{s_+} \eta^{*\alpha}(0) \,.
\end{align}
For the longitudinal polarisation we find
\begin{align}
-\frac{1}{\sqrt{2}} h^\alpha_{A, 0}(-1/2, -1/2, +1) = h^\alpha_{A, 0}(-1/2, +1/2, 0) & = + \frac{\sqrt{2}}{3} \frac{\mLamB - \mLamCst}{\sqrt{q^2}} \sqrt{s_+} \eta^{*\alpha}(+1) \,, \\
\frac{1}{\sqrt{2}} h^\alpha_{A, 0}(+1/2, +1/2, -1) =- h^\alpha_{A, 0}(+1/2, -1/2, 0) & = + \frac{\sqrt{2}}{3} \frac{\mLamB - \mLamCst}{\sqrt{q^2}} \sqrt{s_+} \eta^{*\alpha}(-1) \,, \\
  -  h^\alpha_{A, 0}(+1/2, -1/2, +1) =\sqrt{2}h^\alpha_{A, 0}(+1/2, +1/2, 0) & = + \frac{\sqrt{2}}{3} \frac{\mLamB - \mLamCst}{\sqrt{q^2}} \sqrt{s_+} \eta^{*\alpha}(0) \,,\\
    h^\alpha_{A, 0}(-1/2, +1/2, -1) = -\sqrt{2} h^\alpha_{A, 0}(-1/2, -1/2, 0) & = + \frac{\sqrt{2}}{3} \frac{\mLamB - \mLamCst}{\sqrt{q^2}} \sqrt{s_+} \eta^{*\alpha}(0) \,.
\end{align}

Using the above expressions, we can now list the helicity amplitudes for the transition $\Lambda_{b}\to\LamCst[2595]$.
For the vector current we find the following non-zero helicity amplitudes:
\begin{align}
\label{eq:helamp12-vector-first}
    +\mathcal{A}^{(1/2)}_V(+1/2, +1/2,  0) = +\mathcal{A}^{(1/2)}_V(-1/2, -1/2,  0)
        & = -\sqrt{\frac{1}{3}} f_{1/2, 0} \frac{\mLamB +\mLamCst}{\sqrt{q^2}}  \sqrt{4\mLamB\mLamCst}\,,\\
    +\mathcal{A}^{(1/2)}_V(+1/2, +1/2,  t) = +\mathcal{A}^{(1/2)}_V(-1/2, -1/2,  t)
        & = -\sqrt{\frac{1}{3}} f_{1/2, t} \frac{\mLamB - \mLamCst}{\sqrt{q^2}}  \sqrt{4\mLamB\mLamCst}\,,\\
\label{eq:helamp12-vector-last}
    +\mathcal{A}^{(1/2)}_V(+1/2, -1/2, -1) = +\mathcal{A}^{(1/2)}_V(-1/2, +1/2, +1)
        & = -\sqrt{\frac{2}{3}} f_{1/2,\perp} \sqrt{4\mLamB\mLamCst} \, .
\end{align}
For the axialvector current we find similarly
\begin{align}
\label{eq:helamp12-axialvector-first}
    +\mathcal{A}^{(1/2)}_A(+1/2, +1/2,  0) = -\mathcal{A}^{(1/2)}_A(-1/2, -1/2,  0)
        & = -\sqrt{\frac{1}{3}} g_{1/2, 0} \frac{\mLamB - \mLamCst}{\sqrt{q^2}} \sqrt{4\mLamB\mLamCst}\,,\\
    +\mathcal{A}^{(1/2)}_A(+1/2, +1/2,  t) = -\mathcal{A}^{(1/2)}_A(-1/2, -1/2,  t)
        & = -\sqrt{\frac{1}{3}} g_{1/2, t} \frac{\mLamB + \mLamCst}{\sqrt{q^2}}\sqrt{4\mLamB\mLamCst}\,,\\
\label{eq:helamp12-axialvector-last}
    +\mathcal{A}^{(1/2)}_A(+1/2, -1/2, -1) = -\mathcal{A}^{(1/2)}_A(-1/2, +1/2, +1)
        & = +\sqrt{\frac{2}{3}} g_{1/2,\perp} \sqrt{4\mLamB\mLamCst}\,.
\end{align}

In the heavy quark expansion, if we use \refeq{hqet_vect_current} for the vector current, we calculated the following helicitity amplitudes:
\begin{align}
    \mathcal{A}_V(+1/2, +1/2,  0) &= +\mathcal{A}_V(-1/2, -1/2,  0) = -\sqrt{\frac{1}{3}} \, \frac{\mLamB + \mLamCst}{\sqrt{q^2}} \, \frac{\sqrt{ s_+}}{\mLamB \mLamCst} \, \bigg\{\bigg[s_-\left(C_1(\bar w)+\frac{s_+(C_2(\bar w)\mLamCst+C_3(\bar w)\mLamB)}{2\mLamB\mLamCst(\mLamB+\mLamCst)}\right) \notag \\
    &+\frac{\mLamB-\mLamCst}{\mLamB+\mLamCst}\left(\frac{\mLamB^2-\mLamCst^2+q^2}{2\mLamB}\bar\Lambda-\frac{\mLamB^2-\mLamCst^2-q^2}{2\mLamCst}\bar\Lambda'\right)\bigg]\zeta-2(\mLamB-\mLamCst)\zeta_{\text{SL}}\bigg\}\,, \label{eq:hqe_hel_ampl_vect_first12} \\
    \mathcal{A}_V(+1/2, +1/2,  t) &= +\mathcal{A}_V(-1/2, -1/2,  t) = -\sqrt{\frac{1}{3}} \, \frac{\mLamB - \mLamCst}{\sqrt{q^2}} \, \frac{\sqrt{s_-}}{\mLamB \mLamCst} \,\bigg\{\bigg[C_1(\bar w)s_+ \notag \\
  &+  \frac{\mLamB+\mLamCst}{\mLamB-\mLamCst}\bigg(\frac{\mLamB^2-\mLamCst^2+q^2}{2\mLamB}\left(\bar\Lambda+\frac{C_2(\bar w)s_+}{\mLamB+\mLamCst}\right) \notag \\
    &-\frac{\mLamB^2-\mLamCst^2-q^2}{2\mLamCst}\left(\bar\Lambda'-\frac{C_3(\bar w)s_+}{\mLamB+\mLamCst}\right)\bigg)\bigg]\zeta-2\frac{(\mLamB+\mLamCst)^2}{\mLamB-\mLamCst}\zeta_{\text{SL}}\bigg\}\,,\\
    \mathcal{A}_V(+1/2, -1/2, +1) &= +\mathcal{A}_V(-1/2, +1/2, -1) = -\sqrt{\frac{2}{3}} \, \frac{\sqrt{s_+}}{\mLamB \mLamCst} \, \bigg\{\bigg[C_1(\bar w)s_{-}+\frac{3\mLamB^2+\mLamCst^2-q^2}{2\mLamB} \bar\Lambda \notag \\
    -&\frac{\mLamB^2+3\mLamCst^2-q^2}{2\mLamCst}\bar\Lambda'\bigg]\zeta-2\mLamB\zeta_{\text{SL}}\bigg\}\,,     \label{eq:hqe_hel_ampl_vect_last12}
\end{align}
while for the axial vector current in \refeq{hqet_axial_current} we obtain:
\begin{align}
    \mathcal{A}_A(+1/2, +1/2,  0) =& -\mathcal{A}_A(-1/2, -1/2,  0)  = -\sqrt{\frac{1}{3}} \, \frac{\mLamB - \mLamCst}{\sqrt{q^2}} \, \frac{\sqrt{s_- }}{\mLamB \mLamCst} \, \bigg\{\bigg[s_{+}\left(C_1(\bar w)-\frac{s_-(C_2(\bar w)\mLamCst+C_3(\bar w)\mLamB)}{2\mLamB\mLamCst(\mLamB-\mLamCst)}\right) \notag \\ 
    &+\frac{\mLamB+\mLamCst}{\mLamB-\mLamCst}\left(\frac{\mLamB^2-\mLamCst^2+q^2}{2\mLamB}\bar\Lambda-\frac{\mLamB^2-\mLamCst^2-q^2}{2\mLamCst}\bar\Lambda'\right)\bigg]\zeta-2(\mLamB+\mLamCst)\zeta_{\text{SL}}\bigg\}\,, \label{eq:hqe_hel_ampl_axial_first12}  \\
    \mathcal{A}_A(+1/2, +1/2,  t) = &-\mathcal{A}_A(-1/2, -1/2,  t)  = -\sqrt{\frac{1}{3}} \, \frac{\mLamB + \mLamCst}{\sqrt{q^2}} \, \frac{\sqrt{s_+ }}{\mLamB \mLamCst} \, \bigg\{\bigg[C_1(\bar w)s_{-} \notag \\
    &+\frac{\mLamB-\mLamCst}{\mLamB+\mLamCst}\bigg(\frac{\mLamB^2-\mLamCst^2+q^2}{2\mLamB}\left(\bar\Lambda-\frac{C_2(\bar w)s_+}{\mLamB-\mLamCst}\right) \notag \\
    &-\frac{\mLamB^2-\mLamCst^2-q^2}{2\mLamCst}\left(\bar\Lambda'+\frac{C_3(\bar w)s_+}{\mLamB+\mLamCst}\right)\bigg) \bigg]\zeta-2\frac{(\mLamB-\mLamCst)^2}{\mLamB+\mLamCst}\zeta_{\text{SL}}\bigg\}\,,\\    
    \mathcal{A}_A(+1/2, -1/2, +1) = &-\mathcal{A}_A(-1/2, +1/2, -1)  = \sqrt{\frac{2}{3}} \, \frac{\sqrt{s_- }}{\mLamB \mLamCst} \, \bigg\{\bigg[C_1(\bar w)s_{+} \notag \\
    &+\frac{3\mLamB^2+\mLamCst^2-q^2}{2\mLamB}\bar\Lambda-\frac{\mLamB^2+3\mLamCst^2-q^2}{2\mLamCst}\bar\Lambda'\bigg]\zeta+2\mLamB\zeta_{\text{SL}}\bigg\}\,.   \label{eq:hqe_hel_ampl_axial_last12}
\end{align}

\subsection{$1/2^+\rightarrow 3/2^-$}
We list here the $\Lambda_{b}\to\LamCst[2625]$ helicity amplitudes for various currents.
For the scalar current
\begin{equation}
    h^\alpha_S(s_b, s_c, \lambda_c) \equiv \bar{u}^\alpha(k, \eta(\lambda_c), s_c) u(p, s_b)
\end{equation}
one finds the non-vanishing helicity amplitudes as follows:
\begin{align}
    \frac{\sqrt{2}}{3} h^\alpha_S(+1/2, +1/2,+1) = \sqrt{2} h^\alpha_S(-1/2, -1/2, +1) = h^\alpha_S(-1/2, +1/2, 0) & = \frac{\sqrt{2}}{3} \sqrt{s_+} \eta^{*\alpha}(+1) \,,\\
    \frac{\sqrt{2}}{3} h^\alpha_S(-1/2, -1/2,-1) = \sqrt{2} h^\alpha_S(+1/2, +1/2, -1) = h^\alpha_S(+1/2, -1/2, 0) & = \frac{\sqrt{2}}{3} \sqrt{s_+} \eta^{*\alpha}(-1) \,,\\
    h^\alpha_S(+1/2, -1/2,+1) = \frac{1}{\sqrt{2}} h^\alpha_S(+1/2, +1/2, 0) & = \frac{\sqrt{2}}{3} \sqrt{s_+} \eta^{*\alpha}(0) \,,\\
    h^\alpha_S(-1/2, +1/2,-1) = \frac{1}{\sqrt{2}} h^\alpha_S(-1/2, -1/2, 0) & = \frac{\sqrt{2}}{3} \sqrt{s_+} \eta^{*\alpha}(0) \,.
\end{align}

For the pseudoscalar current
\begin{equation}
    h^\alpha_P(s_b, s_c, \lambda_c) \equiv \bar{u}^\alpha(k, \eta(\lambda_c), s_c) \gamma_5 u(p, s_b)
\end{equation}
one finds similarly:
\begin{align}
    -\frac{\sqrt{2}}{3} h^\alpha_P(+1/2, +1/2, +1) = \sqrt{2} h^\alpha_P(-1/2, -1/2, +1) = h^\alpha_P(-1/2, +1/2,  0) & = + \frac{\sqrt{2}}{3} \sqrt{s_-} \eta^{*\alpha}(+1) \,,\\
    -\frac{\sqrt{2}}{3} h^\alpha_P(-1/2, -1/2, -1) = \sqrt{2} h^\alpha_P(+1/2, +1/2, -1) = h^\alpha_P(+1/2, -1/2,  0) & = - \frac{\sqrt{2}}{3} \sqrt{s_-} \eta^{*\alpha}(-1) \,,\\
    h^\alpha_P(+1/2, -1/2, +1) = \frac{1}{\sqrt{2}} h^\alpha_P(+1/2, +1/2,  0) & = - \frac{\sqrt{2}}{3} \sqrt{s_-} \eta^{*\alpha}(0) \,,\\
    h^\alpha_P(-1/2, +1/2, -1) = \frac{1}{\sqrt{2}} h^\alpha_P(-1/2, -1/2,  0) & = + \frac{\sqrt{2}}{3} \sqrt{s_-} \eta^{*\alpha}(0) \,.
\end{align}

For the vector current we investigate
\begin{equation}
    h^\alpha_{V,\lambda_q}(s_b, s_c, \lambda_c)
        \equiv \bar{u}^\alpha(k, \eta(\lambda_c), s_c) \slashed{\eps}^*(\lambda_q) u(p, s_b)\,,
\end{equation}
and identify
\begin{equation}
    h^\alpha_{V,t}(s_b, s_c, \lambda_c) = \frac{\mLamB - \mLamCst}{\sqrt{q^2}} h^\alpha_{S}(s_b, s_c, \lambda_c)\,.
\end{equation}
For the transverse polarisations we find:
\begin{align}
   - \frac{\sqrt{2}}{3} h^\alpha_{V,+1}(-1/2, +1/2, +1) = \sqrt{2} h^\alpha_{V,-1}(+1/2, -1/2, +1) = h^\alpha_{V,-1}(+1/2, +1/2, 0) & = - \frac{2}{3} \sqrt{s_-} \eta^{*\alpha}(+1) \,,\\
    \frac{\sqrt{2}}{3} h^\alpha_{V,-1}(+1/2, -1/2, -1) = \sqrt{2} h^\alpha_{V,+1}(-1/2, +1/2, -1) = h^\alpha_{V,+1}(-1/2, -1/2, 0) & = - \frac{2}{3} \sqrt{s_-} \eta^{*\alpha}(-1) \,,\\
    h^\alpha_{V,+1}(-1/2, -1/2, +1) = \frac{1}{\sqrt{2}} h^\alpha_{V,+1}(-1/2, +1/2, 0) & = - \frac23 \sqrt{s_-} \eta^{*\alpha}(0) \,, \\
    h^\alpha_{V,-1}(+1/2, +1/2, -1) = \frac{1}{\sqrt{2}} h^\alpha_{V,-1}(+1/2, -1/2, 0) & = - \frac23 \sqrt{s_-} \eta^{*\alpha}(0) \,.
\end{align}
For the longitudinal polarisation we find
\begin{align}
    \frac{\sqrt{2}}{3} h^\alpha_{V, 0}(+1/2, +1/2, +1) = \sqrt{2} h^\alpha_{V, 0}(-1/2, -1/2, +1) = h^\alpha_{V, 0}(-1/2, +1/2, 0) & = \frac{\sqrt{2}}{3} \frac{\mLamB + \mLamCst}{\sqrt{q^2}} \sqrt{s_-} \eta^{*\alpha}(+1) \,, \\
    \frac{\sqrt{2}}{3} h^\alpha_{V, 0}(-1/2, -1/2, -1) = \sqrt{2} h^\alpha_{V, 0}(+1/2, +1/2, -1) = h^\alpha_{V, 0}(+1/2, -1/2, 0) & = \frac{\sqrt{2}}{3} \frac{\mLamB + \mLamCst}{\sqrt{q^2}} \sqrt{s_-} \eta^{*\alpha}(-1) \,, \\
    h^\alpha_{V, 0}(+1/2, -1/2, +1) = \frac{1}{\sqrt{2}} h^\alpha_{V, 0}(+1/2, +1/2, 0) & = \frac{\sqrt{2}}{3} \frac{\mLamB + \mLamCst}{\sqrt{q^2}} \sqrt{s_-} \eta^{*\alpha}(0) \,, \\
    h^\alpha_{V, 0}(-1/2, +1/2, -1) = \frac{1}{\sqrt{2}} h^\alpha_{V, 0}(-1/2, -1/2, 0) & = \frac{\sqrt{2}}{3} \frac{\mLamB + \mLamCst}{\sqrt{q^2}} \sqrt{s_-} \eta^{*\alpha}(0) \,.
\end{align}
For the axialvector current we investigate
\begin{equation}
    h^\alpha_{A,\lambda_q}(s_b, s_c, \lambda_c)
        \equiv \bar{u}^\alpha(k, \eta(\lambda_c), s_c) \slashed{\eps}^*(\lambda_q) \gamma_5 u(p, s_b)\,,
\end{equation}
and identify
\begin{equation}
    h^\alpha_{A,t}(s_b, s_c, \lambda_c) = -\frac{\mLamB + \mLamCst}{\sqrt{q^2}} h^\alpha_{P}(s_b, s_c, \lambda_c)\,.
\end{equation}
For the transverse polarisations we find:
\begin{align}
    - \frac{\sqrt{2}}{3} h^\alpha_{A,+1}(-1/2, +1/2, +1) = \sqrt{2} h^\alpha_{A,-1}(+1/2, -1/2, +1) = h^\alpha_{A,-1}(+1/2, +1/2, 0) & = + \frac{2}{3} \sqrt{s_+} \eta^{*\alpha}(+1) \,,\\
    - \frac{\sqrt{2}}{3} h^\alpha_{A,-1}(+1/2, -1/2, -1) = \sqrt{2} h^\alpha_{A,+1}(-1/2, +1/2, -1) = h^\alpha_{A,+1}(-1/2, -1/2, 0) & = - \frac{2}{3} \sqrt{s_+} \eta^{*\alpha}(-1) \,,\\
    h^\alpha_{A,+1}(-1/2, -1/2, +1) = \frac{1}{\sqrt{2}} h^\alpha_{A,+1}(-1/2, +1/2, 0) & = - \frac23 \sqrt{s_+} \eta^{*\alpha}(0) \,, \\
    h^\alpha_{A,-1}(+1/2, +1/2, -1) = \frac{1}{\sqrt{2}} h^\alpha_{A,-1}(+1/2, -1/2, 0) & = + \frac23 \sqrt{s_+} \eta^{*\alpha}(0) \,.
\end{align}
For the longitudinal polarisation we find
\begin{align}
    - \frac{\sqrt{2}}{3} h^\alpha_{A, 0}(+1/2, +1/2, +1) = \sqrt{2} h^\alpha_{A, 0}(-1/2, -1/2, +1) = h^\alpha_{A, 0}(-1/2, +1/2, 0) & = - \frac{\sqrt{2}}{3} \frac{\mLamB - \mLamCst}{\sqrt{q^2}} \sqrt{s_+} \eta^{*\alpha}(+1) \,, \\
    - \frac{\sqrt{2}}{3} h^\alpha_{A, 0}(-1/2, -1/2, -1) = \sqrt{2} h^\alpha_{A, 0}(+1/2, +1/2, -1) = h^\alpha_{A, 0}(+1/2, -1/2, 0) & = + \frac{\sqrt{2}}{3} \frac{\mLamB - \mLamCst}{\sqrt{q^2}} \sqrt{s_+} \eta^{*\alpha}(-1) \,, \\
    h^\alpha_{A, 0}(+1/2, -1/2, +1) = \frac{1}{\sqrt{2}} h^\alpha_{A, 0}(+1/2, +1/2, 0) & = + \frac{\sqrt{2}}{3} \frac{\mLamB - \mLamCst}{\sqrt{q^2}} \sqrt{s_+} \eta^{*\alpha}(0) \,,\\
    h^\alpha_{A, 0}(-1/2, +1/2, -1) = \frac{1}{\sqrt{2}} h^\alpha_{A, 0}(-1/2, -1/2, 0) & = - \frac{\sqrt{2}}{3} \frac{\mLamB - \mLamCst}{\sqrt{q^2}} \sqrt{s_+} \eta^{*\alpha}(0) \,.
\end{align}

For the vector current we find the following non-zero helicity amplitudes:
\begin{align}
\label{eq:helamp32-vector-first}
    +\mathcal{A}^{(3/2)}_V(+1/2, +3/2, +1) = +\mathcal{A}^{(3/2)}_V(-1/2, -3/2, -1)
        & = -2 F_{3/2,\perp} \sqrt{4\,\mLamB\,\mLamCst}\,,\\
    +\mathcal{A}^{(3/2)}_V(+1/2, +1/2,  0) = +\mathcal{A}^{(3/2)}_V(-1/2, -1/2,  0)
        & = +\sqrt{\frac{2}{3}} F_{1/2, 0} \frac{\mLamB + \mLamCst}{\sqrt{q^2}} \sqrt{4\,\mLamB\,\mLamCst}\,,\\
    +\mathcal{A}^{(3/2)}_V(+1/2, +1/2,  t) = +\mathcal{A}^{(3/2)}_V(-1/2, -1/2,  t)
        & = +\sqrt{\frac{2}{3}} F_{1/2, t} \frac{\mLamB - \mLamCst}{\sqrt{q^2}} \sqrt{4\,\mLamB\,\mLamCst}\,,\\
\label{eq:helamp32-vector-last}
    +\mathcal{A}^{(3/2)}_V(+1/2, -1/2, -1) = +\mathcal{A}^{(3/2)}_V(-1/2, +1/2, +1)
        & = -\frac{2}{\sqrt{3}} F_{1/2,\perp} \sqrt{4\,\mLamB\,\mLamCst}\,.
\end{align}
For the axialvector current we find similarly
\begin{align}
\label{eq:helamp32-axialvector-first}
    +\mathcal{A}^{(3/2)}_A(+1/2, +3/2, +1) = -\mathcal{A}^{(3/2)}_A(-1/2, -3/2, -1)
        & = -2 G_{3/2,\perp} \sqrt{4\,\mLamB\,\mLamCst}\,,\\
    +\mathcal{A}^{(3/2)}_A(+1/2, +1/2,  0) = -\mathcal{A}^{(3/2)}_A(-1/2, -1/2,  0)
        & = +\sqrt{\frac{2}{3}} G_{1/2, 0} \frac{\mLamB - \mLamCst}{\sqrt{q^2}} \sqrt{4\,\mLamB\,\mLamCst}\,,\\
    +\mathcal{A}^{(3/2)}_A(+1/2, +1/2,  t) = -\mathcal{A}^{(3/2)}_A(-1/2, -1/2,  t)
        & = +\sqrt{\frac{2}{3}} G_{1/2, t} \frac{\mLamB + \mLamCst}{\sqrt{q^2}} \sqrt{4\,\mLamB\,\mLamCst}\,,\\
\label{eq:helamp32-axialvector-last}
    +\mathcal{A}^{(3/2)}_A(+1/2, -1/2, -1) = -\mathcal{A}^{(3/2)}_A(-1/2, +1/2, +1)
        & = +\frac{2}{\sqrt{3}} G_{1/2,\perp} \sqrt{4\,\mLamB\,\mLamCst}\,.
\end{align}
In the heavy quark expansion, the helicity amplitudes related to the vector current eq.~(\ref{eq:hqet_vect_current}) read
\begin{align}
    \mathcal{A}_V(+1/2, +3/2, +1) =& +\mathcal{A}_V(-1/2, -3/2, -1) = + 2\frac{\sqrt{s_+}}{\mLamB}\zeta_{\text{SL}}\,,  \label{eq:hqe_hel_ampl_vect_first32}\\
    \mathcal{A}_V(+1/2, +1/2,  0) =& +\mathcal{A}_V(-1/2, -1/2,  0) = +\sqrt{\frac{2}{3}} \, \frac{\mLamB + \mLamCst}{\sqrt{q^2}} \, \frac{\sqrt{s_+}}{\mLamB \mLamCst} \, \bigg\{\bigg[s_{-}\left(C_1(\bar w)+\frac{s_+(C_2(\bar w)\mLamCst+C_3(\bar w)\mLamB)}{2\mLamB\mLamCst(\mLamB+\mLamCst)}\right) \notag \\
    &+\frac{\mLamB - \mLamCst}{\mLamB + \mLamCst}\left(\frac{\mLamB^2-\mLamCst^2+q^2}{2\mLamB}\bar\Lambda-\frac{\mLamB^2-\mLamCst^2-q^2}{2\mLamCst}\bar\Lambda'\right)\bigg]\zeta+(\mLamB-\mLamCst)\zeta_{\text{SL}}\bigg\}\,,\\
    \mathcal{A}_V(+1/2, +1/2,  t) = &+\mathcal{A}_V(-1/2, -1/2,  t) = +\sqrt{\frac{2}{3}} \, \frac{\mLamB - \mLamCst}{\mLamB \mLamCst} \, \frac{\sqrt{s_-}}{\sqrt{q^2}} \,\bigg\{\bigg[s_{+} \notag  \\
    &+\frac{\mLamB +\mLamCst}{\mLamB - \mLamCst}\left(\frac{\mLamB^2-\mLamCst^2+q^2}{2\mLamB}(\bar\Lambda+\frac{C_2(\bar w)s_+}{\mLamB +\mLamCst})-\frac{\mLamB^2-\mLamCst^2-q^2}{2\mLamCst}(\bar\Lambda'-\frac{C_3(\bar w)s_+}{\mLamB +\mLamCst})\right)\bigg]\zeta \notag \\
    &+\frac{(\mLamB+\mLamCst)^2}{\mLamB-\mLamCst}\zeta_{\text{SL}}\bigg\}\,,\\
    \mathcal{A}_V(+1/2, -1/2, -1) = &+\mathcal{A}_V(-1/2, +1/2, +1) = -\sqrt{\frac{4}{3}} \, \frac{\sqrt{s_+}}{\mLamB \mLamCst} \, \bigg\{\bigg[s_{-}C_1(\bar w)-\frac{3\mLamB^2+\mLamCst^2-q^2}{2\mLamB}\bar\Lambda \notag \\
    &+\frac{\mLamB^2+3\mLamCst^2-q^2}{2\mLamCst}\bar\Lambda'\bigg]\zeta+\mLamB\zeta_{\text{SL}}\bigg\}\,, \label{eq:hqe_hel_ampl_vect_last32}
\end{align}
while for the axial vector current eq.~(\ref{eq:hqet_axial_current}), we obtain
\begin{align}
    \mathcal{A}_A(+1/2, +3/2, -1) =& -\mathcal{A}_A(-1/2, -3/2, +1)  = 2\frac{\sqrt{s_-}}{\mLamB}\zeta_{\text{SL}}\,,   \label{eq:hqe_hel_ampl_axial_first32} \\
    \mathcal{A}_A(+1/2, +1/2,  0) =& -\mathcal{A}_A(-1/2, -1/2,  0)  = +\sqrt{\frac{2}{3}} \, \frac{\mLamB - \mLamCst}{\sqrt{q^2}} \, \frac{\sqrt{s_-}}{\mLamB \mLamCst} \,\bigg\{\bigg[s_{+}\left(C_1(\bar w)-\frac{s_-(C_2(\bar w)\mLamCst+C_3(\bar w)\mLamB)}{2\mLamB\mLamCst(\mLamB+\mLamCst)}\right) \notag\\
    &+\left(\frac{\mLamB^2-\mLamCst^2+q^2}{2\mLamB}\bar\Lambda-\frac{\mLamB^2-\mLamCst^2-q^2}{2\mLamCst}\bar\Lambda'\right)\frac{(\mLamB+\mLamCst)}{\mLamB-\mLamCst}\bigg]\zeta+(\mLamB+\mLamCst)\zeta_{\text{SL}}\bigg\}\,,\\
    \mathcal{A}_A(+1/2, +1/2,  t) =& -\mathcal{A}_A(-1/2, -1/2,  t)  =+\sqrt{\frac{2}{3}} \, \frac{\mLamB + \mLamCst}{\sqrt{q^2}} \, \frac{\sqrt{s_+}}{\mLamB \mLamCst} \,\bigg\{\bigg[s_{-} \notag \\
    &+\left(\frac{\mLamB^2-\mLamCst^2+q^2}{2\mLamB}\bar(\Lambda-\frac{C_2(\bar w) s_-}{\mLamB-\mLamCst})-\frac{\mLamB^2-\mLamCst^2-q^2}{2\mLamCst}(\bar\Lambda'+\frac{C_3(\bar w) s_-}{\mLamB-\mLamCst})\right)\frac{(\mLamB-\mLamCst)}{\mLamB+\mLamCst}\bigg]\zeta \notag \\
    &+\frac{(\mLamB-\mLamCst)^2}{\mLamB+\mLamCst}\zeta_{\text{SL}}\bigg\}\,,\\
    \mathcal{A}_A(+1/2, -1/2, +1) =& -\mathcal{A}_A(-1/2, +1/2, -1)  = \mb{+}\sqrt{\frac{4}{3}} \, \frac{\sqrt{s_-}}{\mLamB \mLamCst} \, \bigg\{\bigg[ s_{+}C_1(\bar w)+\frac{3\mLamB^2+\mLamCst^2-q^2}{2\mLamB} \bar\Lambda \notag \\
   &- \frac{\mLamB^2+3\mLamCst^2-q^2}{2\mLamCst} \bar\Lambda'\bigg]\zeta+\mLamB\zeta_{\text{SL}}\bigg\}\,. \label{eq:hqe_hel_ampl_axial_last32}
\end{align}

\section{Details on the Kinematics}
\label{app:kin}

We choose the $z$ axis along the flight direction of the $\Lambda_c^{*}$. Thus, in the
rest frame of the $\LamB$ (B-RF) one has
\begin{align}
    p^\mu\big|_\text{B-RF} & = (m_{\LamB}, 0, 0, 0)\,,\\
    q^\mu\big|_\text{B-RF} & = (q^0, 0, 0, -|\vec{q}\,|)\,,\\
    k^\mu\big|_\text{B-RF} & = (m_{\LamB} - q^0, 0, 0, +|\vec{q}\,|)\,.
\end{align}
We chose to describe the decay through the invariants $q^2$ and obtain
\begin{align}
        q^0\big|_{B-RF} & = \frac{\mLamB^2 - \mLamCst^2 + q^2}{2 \mLamB}\,, &
|\vec{q}\,|\big|_{B-RF} & = \frac{\sqrt{\lambda(\mLamB^2, \mLamCst^2, q^2)}}{2 \mLamB}\,,
\end{align}
where $\lambda$ is the usual K\"all\'en function.

The description of the $\Lambda_c^{*}$ involves a spin-$1$ polarisation vector $\eta(m)$ along the
positive $z$ direction. According to \cite{Haber:1994pe} we can use
\begin{align}
    \eta(\pm)|_{B-RF} & = (0, \mp 1, - i, 0) / \sqrt{2}\,,\\
    \eta(0)|_{B-RF}   & = (|\vec{q}\,|, 0, 0, \mLamB - q^0) / \mLamCst\,.
\end{align}

In order to facilitate the calculation we introduce artificial polarisation vectors
$\eps(n)$ which fulfill the following relations:
\begin{align}
    \eps(n) \cdot q & = 0\qquad n=\pm, 0\\
    \eps(n) \cdot \eps^\dagger(n') & = g_{nn'}\qquad g_{nn'} = \text{diag}(+, -, -, -)\text{ for }n,n'=t,+,-,0\\
    \eps(n)_\mu \eps^\dagger(n')_\nu g_{nn'} & = g_{\mu\nu}\,.
\end{align}
Within the $\ell\nu$ rest frame these relations are fulfilled by the set
\begin{align}
    \eps^\mu(t)\big|_{\ell\nu-RF}   & = (1, 0, 0, 0)\,,\\
    \eps^\mu(\pm)\big|_{\ell\nu-RF} & = (0, \pm 1, - i, 0) / \sqrt{2}\,,\\
    \eps^\mu(0)\big|_{\ell\nu-RF}   & = (0, 0, 0, -1)\,.
\end{align}
Using a boost along $z$, one obtains in the
$B$ rest frame
\begin{align}
    \eps^\mu(t)\big|_{B-RF}   & = (q^0, 0, 0, -|\vec{q}\,|) / \sqrt{q^2} = q^\mu / \sqrt{q^2}\,,\\
    \eps^\mu(0)\big|_{B-RF}   & = (+|\vec{q}\,|, 0, 0, -q_0) / \sqrt{q^2}\,,
\end{align}
while the $\eps(\pm)$ remain invariant under that boost. Comments are due on the choice
of the polarisation vectors, especially the signs of $\eps^z(0)$ as well as $\eps^y(\pm)$.
These haven been adopted to obtain longitudinal and right-handed/left-handed polarisation
of the $\ell\nu$ system, which moves along the \emph{negative} $z$-axis. The phase
convention is as in \cite{Haber:1994pe}.

\section{Explicit Spinor Representations}
\label{app:spinors}

In the course of the calculations we need to use explicit representations of spinors
for an arbitrary momentum and fixed helicity in their rest frame. In the chiral representation of Dirac spinors,
one obtains for a $u$ spinor with momentum $p^\mu$,
\begin{equation}
    p^\mu = (p^0, |\vec{p}|\sin\theta\cos\phi, |\vec{p}|\sin\theta\sin\phi, |\vec{p}|\cos\theta),
\end{equation}
with $p^2 = m^2$ and helicity $h = \pm 1/2$ in their respective rest frames \cite{Haber:1994pe}
\begin{align}
    u(p, h=+1/2) & = \frac{\gamma^0}{\sqrt{2 (p^0 + m)}} \left[\begin{matrix}
    +(p^0 + m - |\vec{p}|) & \cos(\theta/2) & \\
    +(p^0 + m - |\vec{p}|) & \sin(\theta/2) & \exp(+i \phi)\\
    +(p^0 + m + |\vec{p}|) & \cos(\theta/2) & \\
    +(p^0 + m + |\vec{p}|) & \sin(\theta/2) & \exp(+i \phi)
    \end{matrix}\right]\\
    u(p, h=-1/2) & = \frac{\gamma^0}{\sqrt{2 (p^0 + m)}} \left[\begin{matrix}
    -(p^0 + m + |\vec{p}|) & \sin(\theta/2) & \exp(-i\phi)\\
    +(p^0 + m + |\vec{p}|) & \cos(\theta/2) & \\
    -(p^0 + m - |\vec{p}|) & \sin(\theta/2) & \exp(-i\phi)\\
    +(p^0 + m - |\vec{p}|) & \cos(\theta/2) &
    \end{matrix}\right]\,.
\end{align}

\section{Formulae}
\label{app:formulae}

For the Levi-Civita tensor we use the convention
\begin{equation}
\eps^{0123} = - \eps_{0123} = +1\,.
\end{equation}
In this convention one has
\begin{align}
\tr \gamma^\mu \gamma^\nu \gamma^\rho \gamma^\sigma \gamma_5 & = -4i \eps^{\mu\nu\rho\sigma}\\
\eps^{\alpha\beta\mu\nu} \eps_{\alpha\beta\rho\sigma} & = -2(\delta^\mu_\rho \delta^\nu_\sigma - \delta^\mu_\sigma \delta^\nu_\rho)\\
\sigma_{\mu\nu}\gamma_5 & = \frac{i}{2}\eps_{\mu\nu\alpha\beta}\sigma^{\alpha\beta}
\end{align}

\section{Additional material on the sensitivity study}\label{app:sens}

\begin{figure}[tbph]
	\centering
	\includegraphics[width=0.32\linewidth]{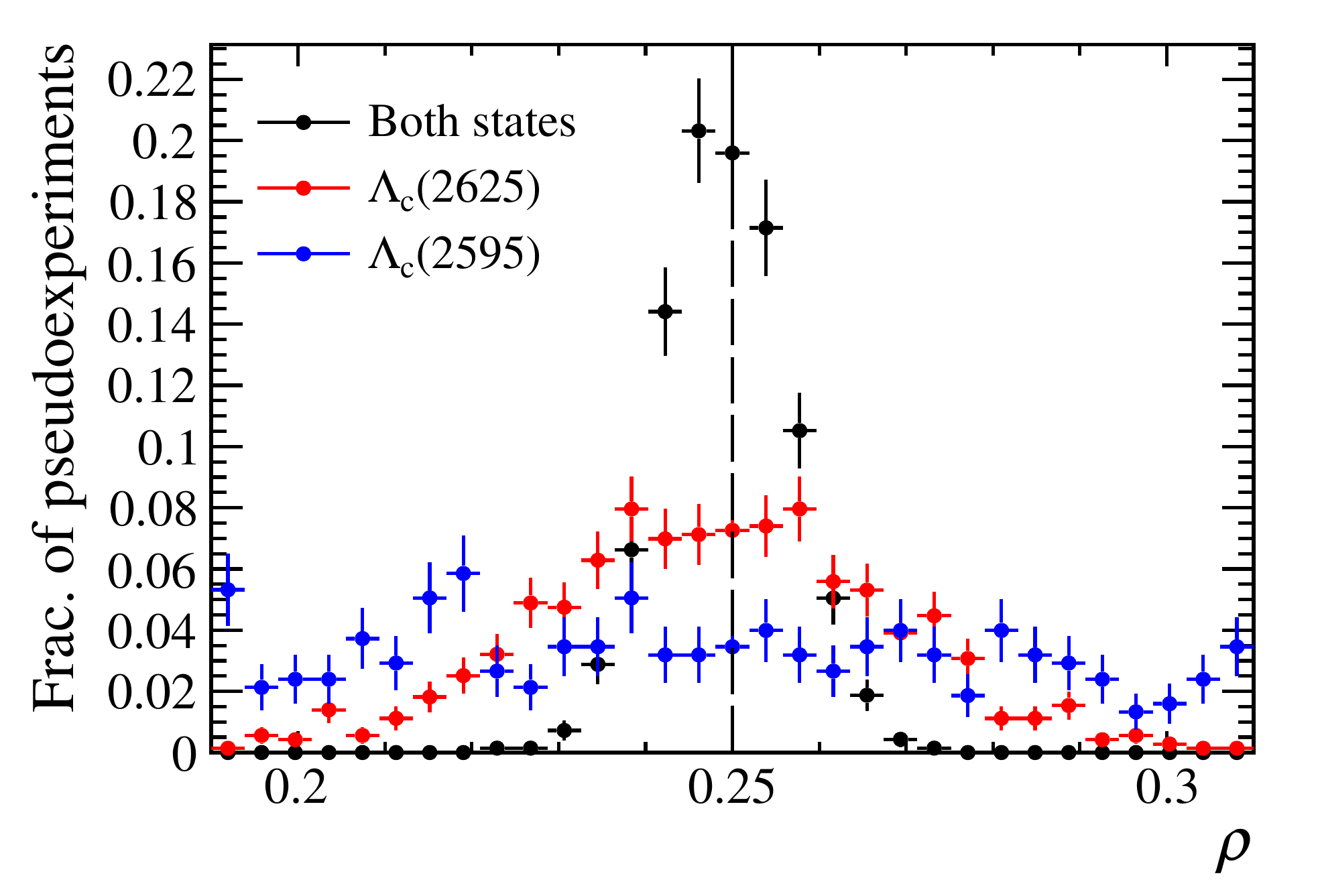}
	\includegraphics[width=0.32\linewidth]{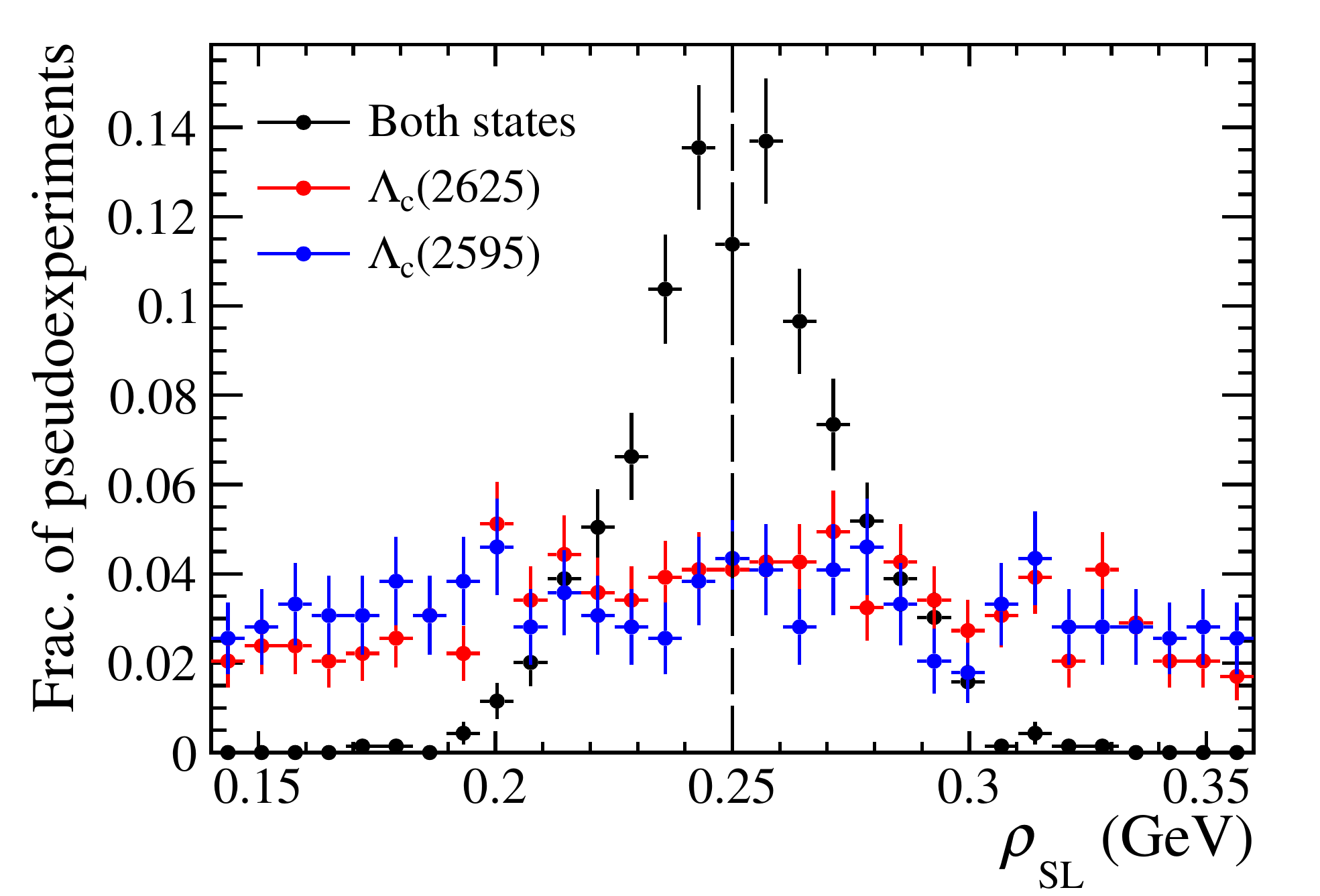}
        \includegraphics[width=0.32\linewidth]{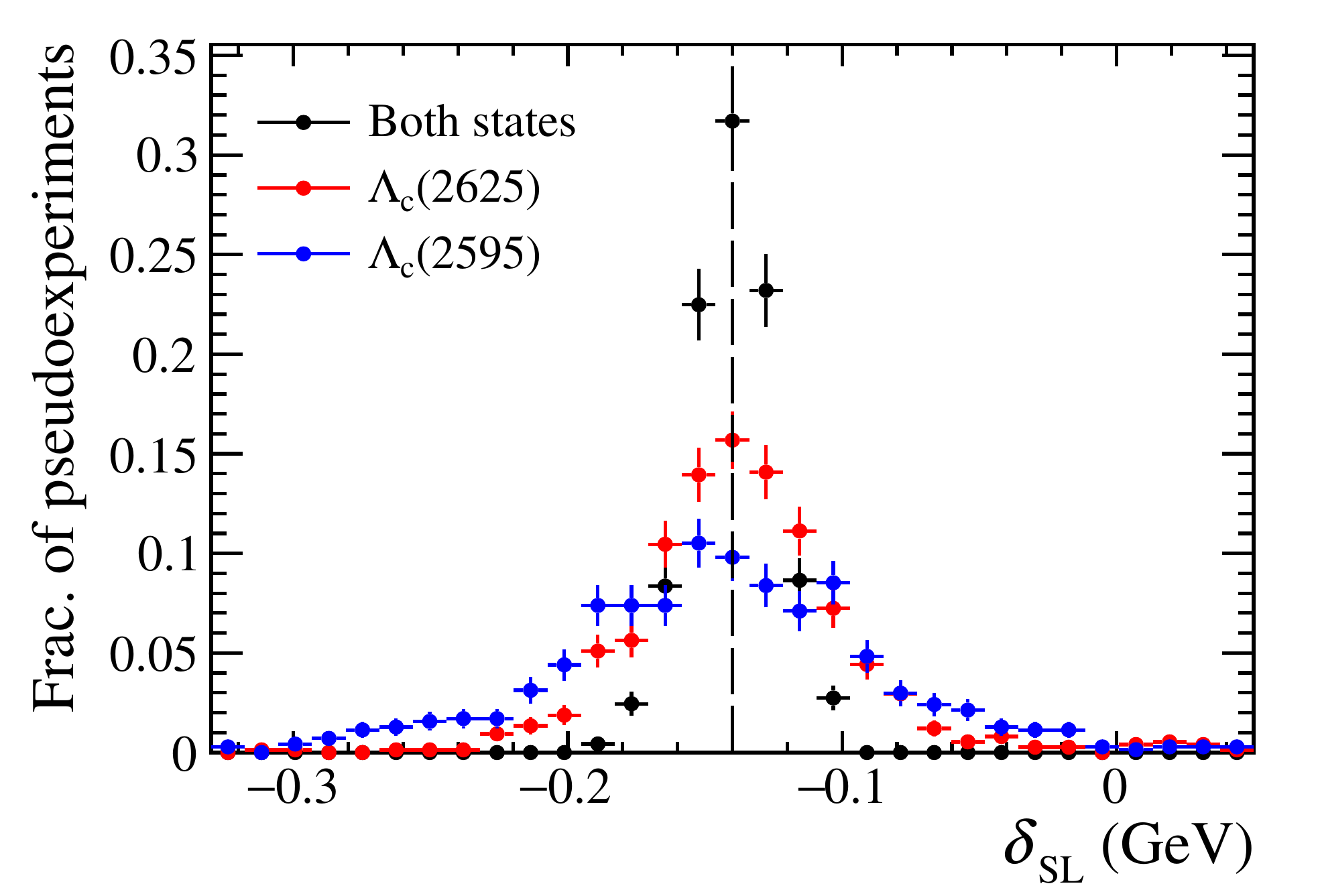}
	\caption{
		Distribution of the Isgur-Wise parameters as fitted from an ensemble of pseudo-experiments. The distributions are shown for the cases when one of the two \LamCstBoth states is fitted, as well as the combination of both. Both \qq and \costhetal are fitted simultaneously.
		\label{fig:params2d}
	}
\end{figure}

\begin{figure}[tbph]
    \centering
    \includegraphics[width=.32\linewidth]{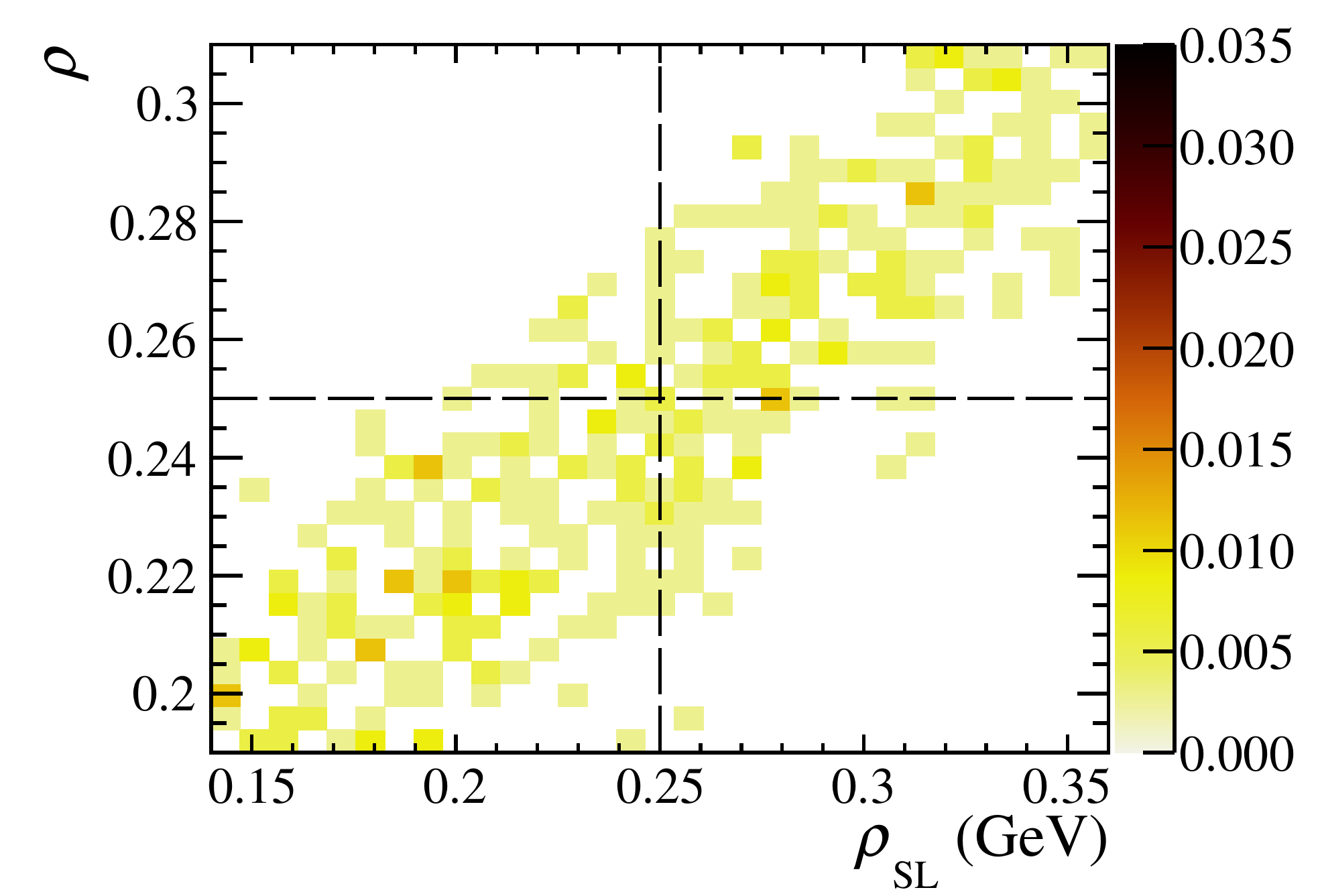}
    \includegraphics[width=.32\linewidth]{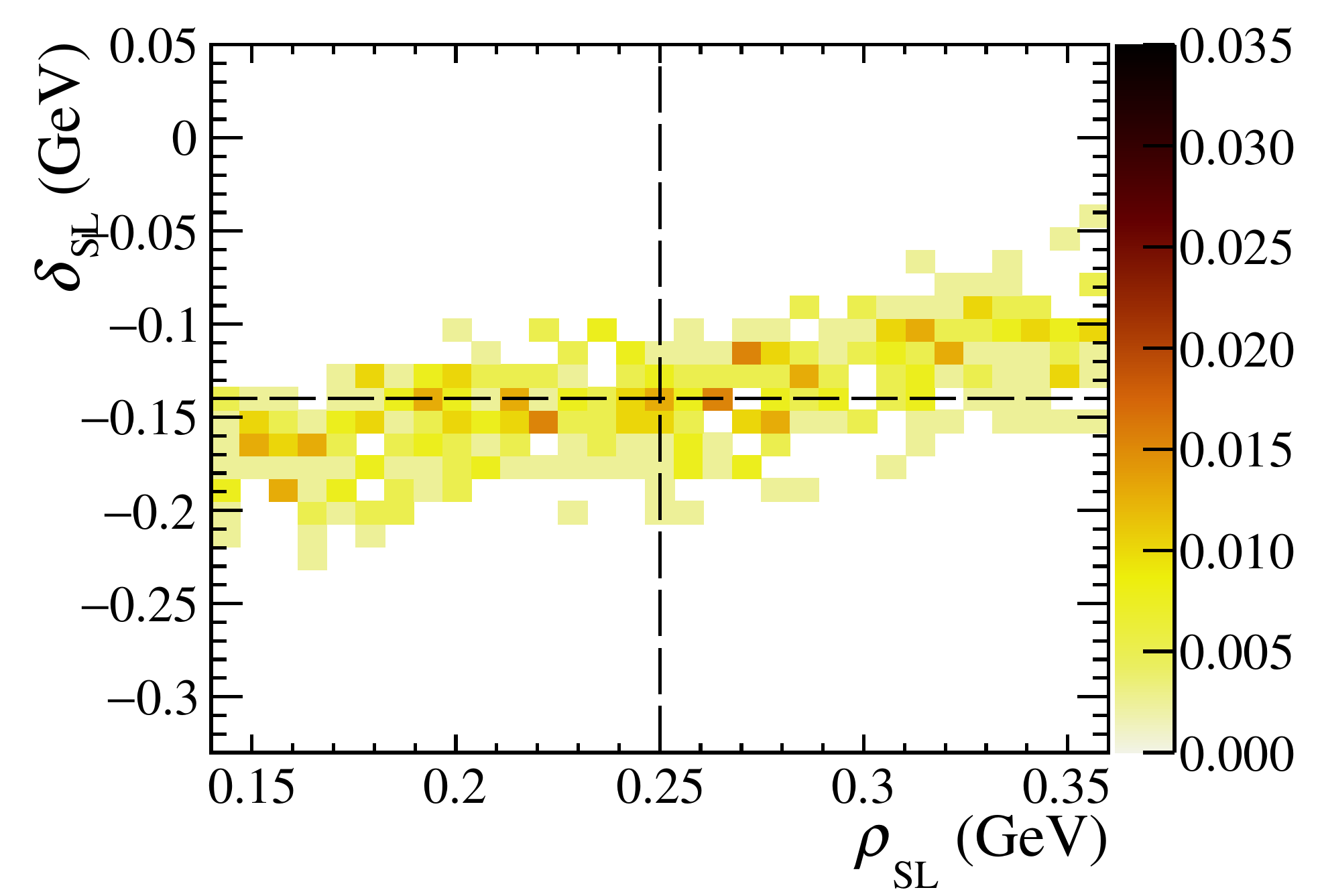}
    \includegraphics[width=.32\linewidth]{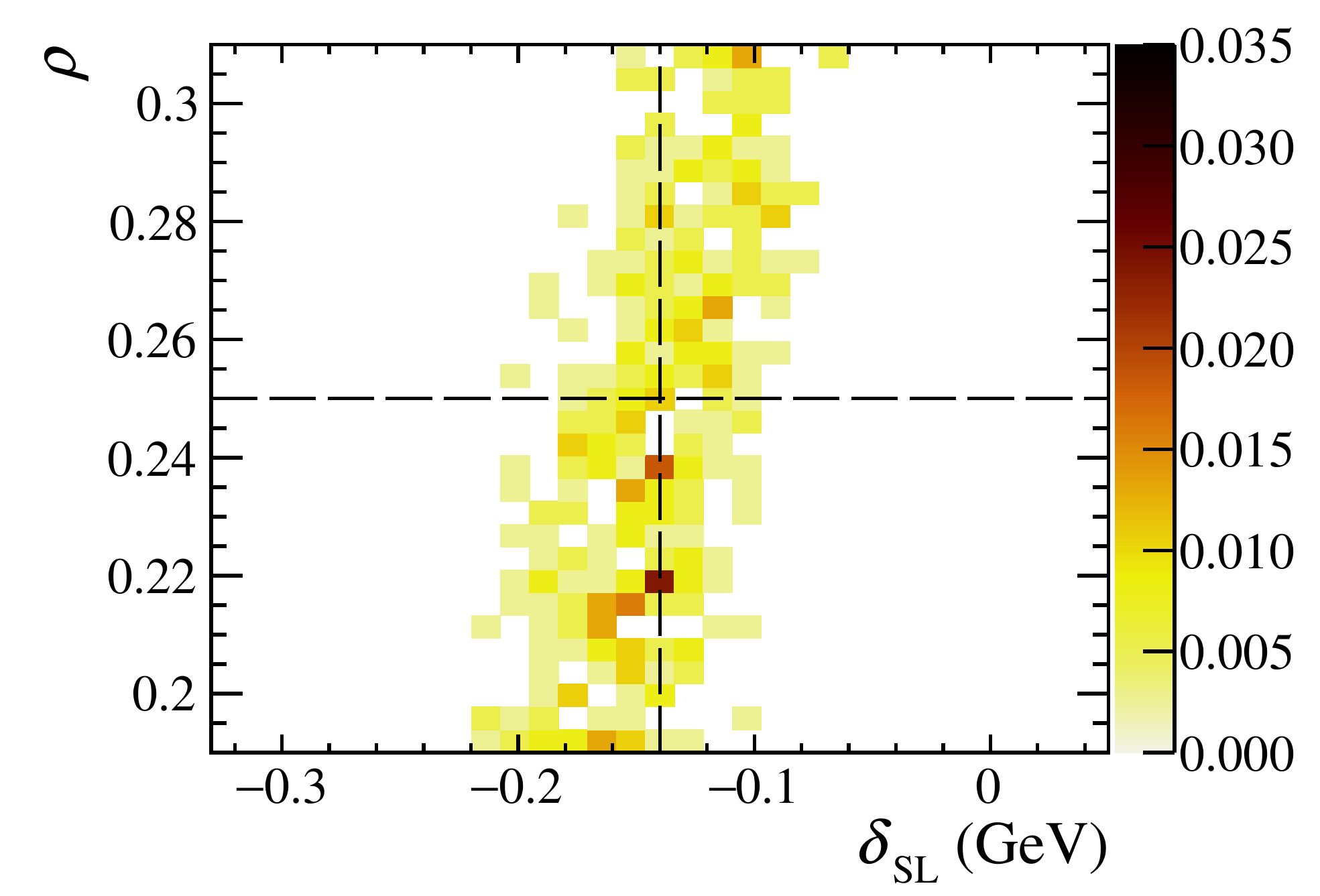}\\
    \includegraphics[width=.32\linewidth]{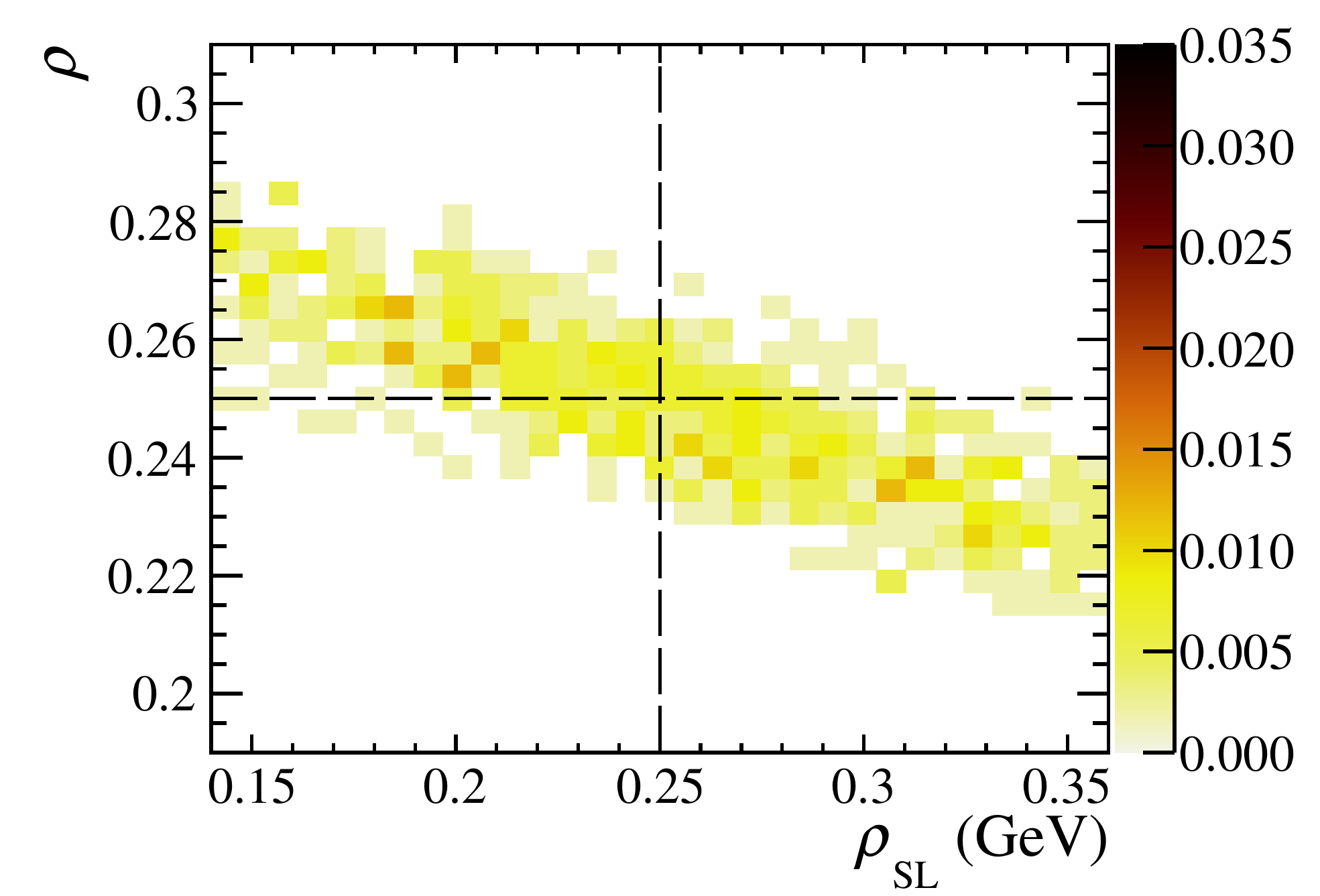}
    \includegraphics[width=.32\linewidth]{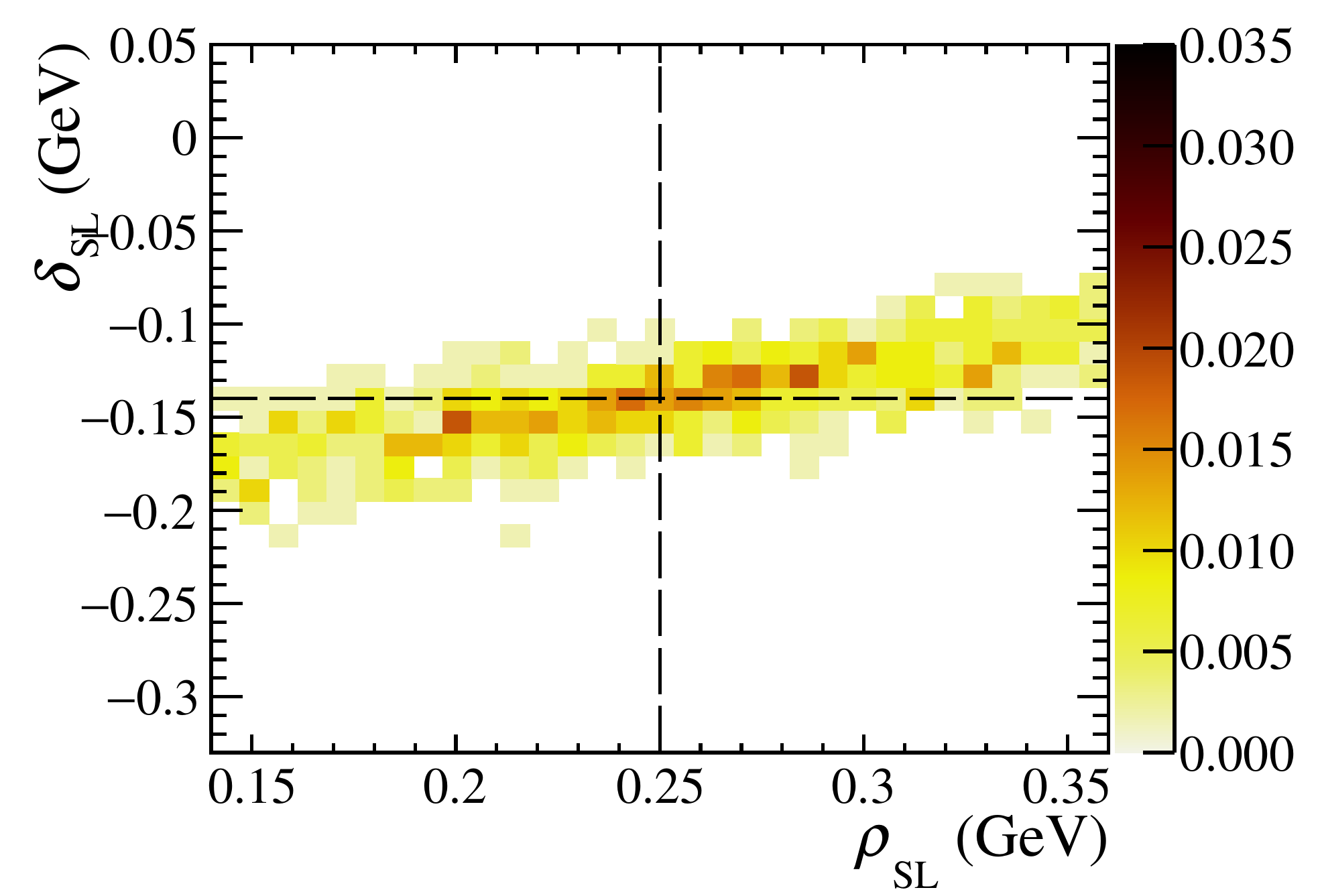}
    \includegraphics[width=.32\linewidth]{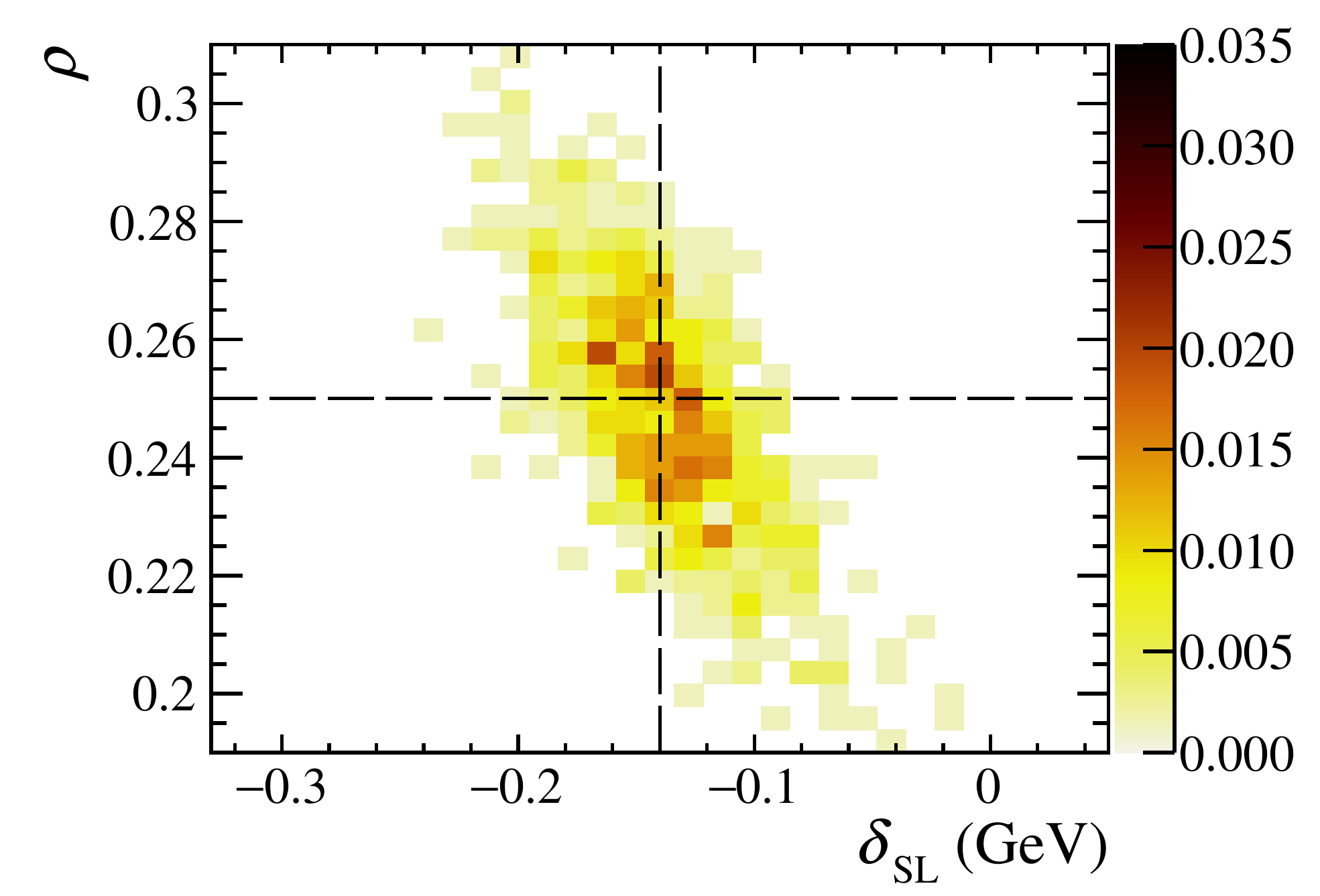}\\
    \includegraphics[width=.32\linewidth]{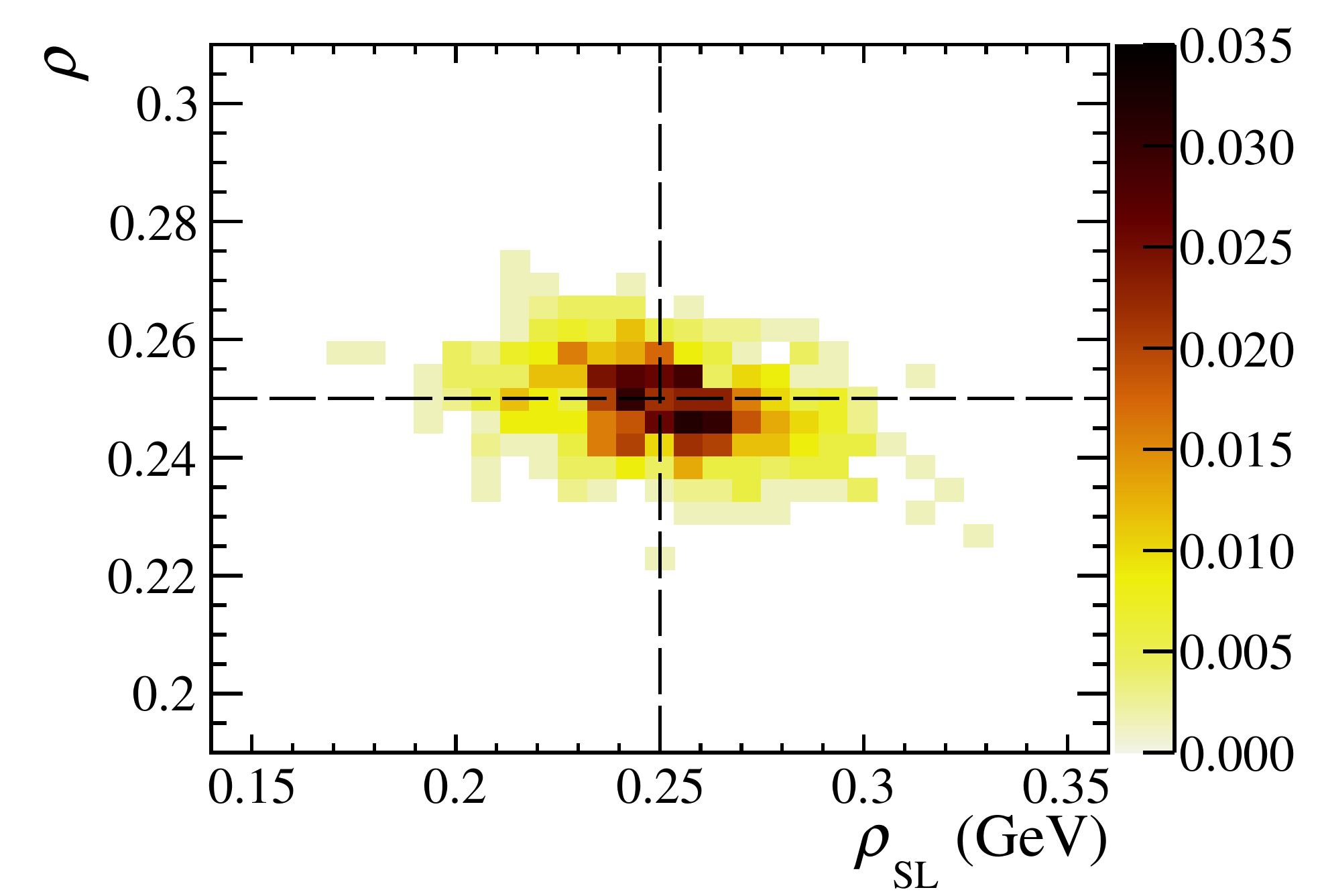}
    \includegraphics[width=.32\linewidth]{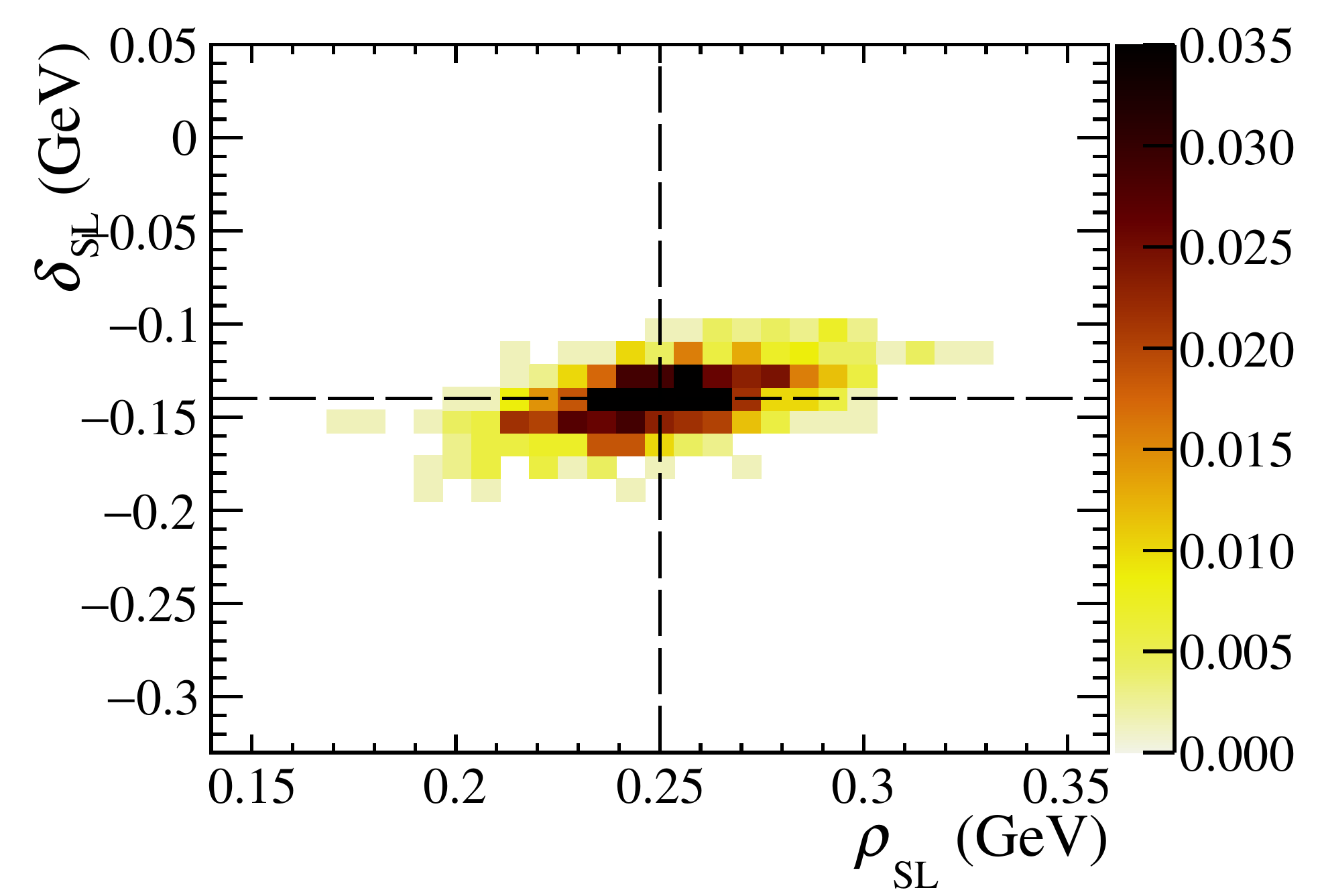}
    \includegraphics[width=.32\linewidth]{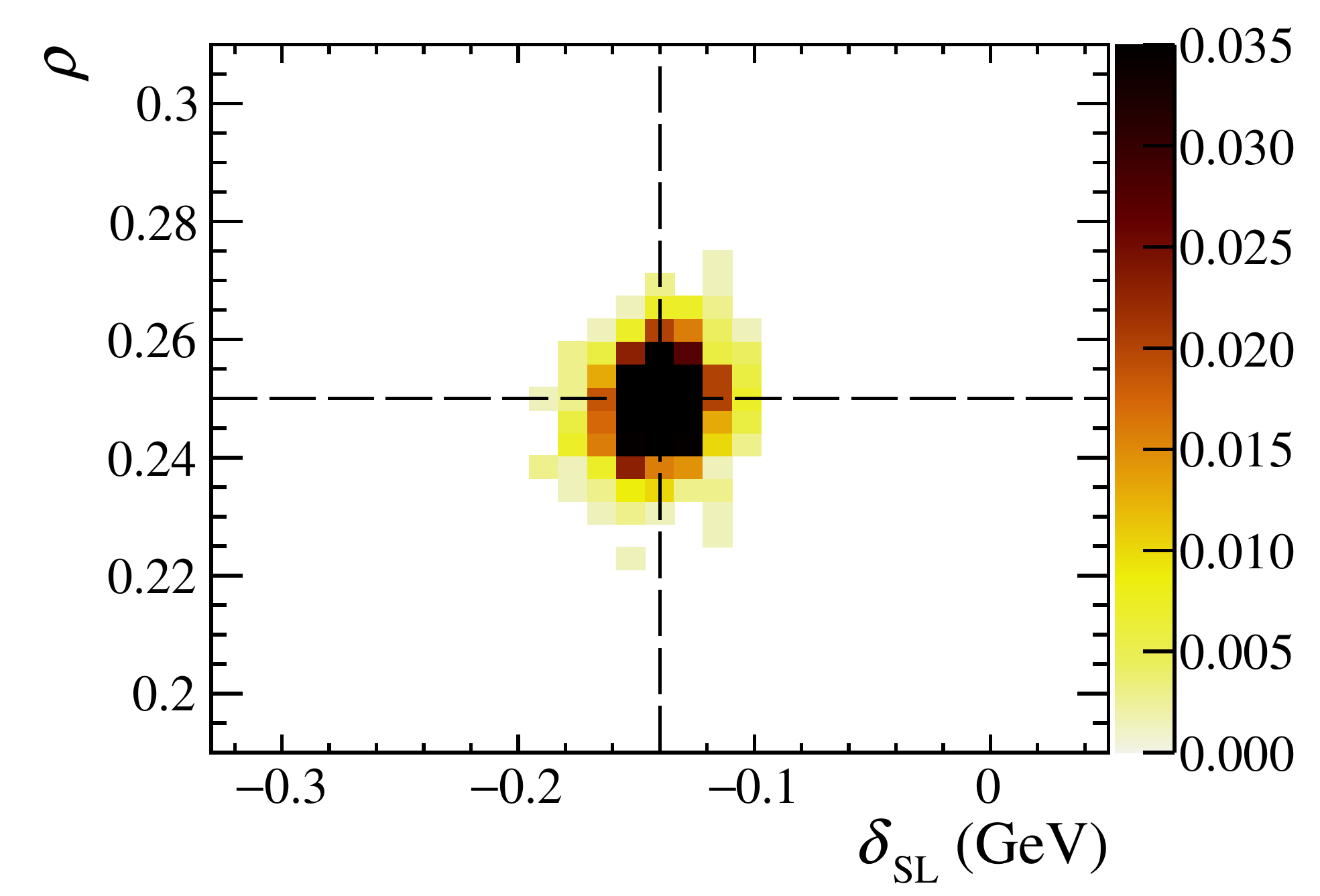}
    \caption{Two-dimensional distributions of the Isgur-Wise parameters as fitted from an ensemble of pseudoexperiments. Both \qq and \costhetal are fitted simultaneously. Only simulated \LamCst[2595] and \LamCst[2625] data are used for the pseudoexperiments shown in the first and second row, respectively. Both states are fitted in the pseudoexperiments shown in the third row. The dashed lines indicate the numerical values of the parameters used to generate the pseudoexperiments.}\label{fig:details}
\end{figure}

We show in Fig.~\ref{fig:params2d} the distributions of the Isgur-Wise parameters resulting from a two-dimensional fit to both \qq and \costhetal, comparing ensembles of pseudo-experiments using only the \LamCst[2595], only the \LamCst[2625], or both.
In Fig.~\ref{fig:details} we investigate the correlations between the Isgur-Wise parameters resulting from a two-dimensional fit to \qq and \costhetal of the three sets of pseudo-experiments. In particular, the leftmost plots demonstrate how only a simultaneous fit to both \LamCstBoth states can solve the degeneracy between the two slope parameters. Moreover, both \LamCst[2595] and \LamCst[2625] data sets are individually sensitive to the $\delta_{\text{SL}}$ parameter, but a simultaneous fit provides much better precision.

\end{document}